%% file: sajovic-2026-mesh-csg.tex
\documentclass{egpubl}
\pdfoutput=1 % force pdflatex on arXiv (png figures)
\JournalSubmission
\copyrightTextRunPag{} % LOCAL COPY ONLY: hide the "submitted to ..." page footer; delete before submission
\copyrightTextTitPag{} % LOCAL COPY ONLY: hide the "submitted to ..." title-page footer; delete before submission
\makeatletter % LOCAL COPY ONLY: blank the title-page "Volume xx (200y)..." header/footer; delete before submission
\def\ps@titlepage{\let\@mkboth\@gobbletwo
  \def\@oddhead{}\def\@evenhead{}\def\@oddfoot{}\def\@evenfoot{}}
\makeatother

\usepackage[T1]{fontenc}
\usepackage{mathptmx} % Times-like Adobe substitute; replace with dfadobe for camera-ready

\usepackage{cite}
\BibtexOrBiblatex
\electronicVersion
\PrintedOrElectronic
\ifpdf \usepackage[pdftex]{graphicx} \pdfcompresslevel=9
\else \usepackage[dvips]{graphicx} \fi
\usepackage{egweblnk}

\usepackage{amsmath,amssymb}
\usepackage{booktabs}
\usepackage{array}
\usepackage{algorithm}
\usepackage{algpseudocode}
\usepackage{tikz}
\usepackage{tikz-3dplot}
\usetikzlibrary{calc}
\usepackage{pgfplots}
\pgfplotsset{compat=1.16}
\usepgfplotslibrary{fillbetween}
\usepackage{pifont}

\definecolor{tfteal}{HTML}{00D5BE}
\definecolor{tfrose}{HTML}{C8467A}
\definecolor{tfink}{HTML}{1F2937}

\definecolor{cOne}{HTML}{4C78A8}
\definecolor{cTwo}{HTML}{F58518}
\definecolor{cThree}{HTML}{54A24B}
\definecolor{cFour}{HTML}{8E6BBE}
\definecolor{cFive}{HTML}{E45756}
\definecolor{cSix}{HTML}{8C564B}
\definecolor{tfgrey}{HTML}{9CA3AF}
\definecolor{tfsoft}{HTML}{F3F4F6}
\definecolor{tfbg}{HTML}{ECECEE} % plate background for rendered figures

\newcommand{\chcomment}[2][]{}

\newcommand{\chdeleted}[2][]{}

\title[trueform for mesh CSG]%
      {trueform: fast and robust mesh CSG\\
       via topological aggregation}

\author[\v{Z}. Sajovic et al.]
{\parbox{\textwidth}{\centering \v{Z}iga Sajovic$^{1}$\qquad Dejan Knez$^{1}$
        }
        \\
{\parbox{\textwidth}{\centering $^1$Polydera, \texttt{\{ziga.sajovic, dejan.knez\}@polydera.com}
       }
}
}

\begin{document}
\teaser{%
  \centering
  \setlength{\fboxsep}{0pt}%
  \colorbox{tfbg}{\includegraphics[width=0.49\linewidth]{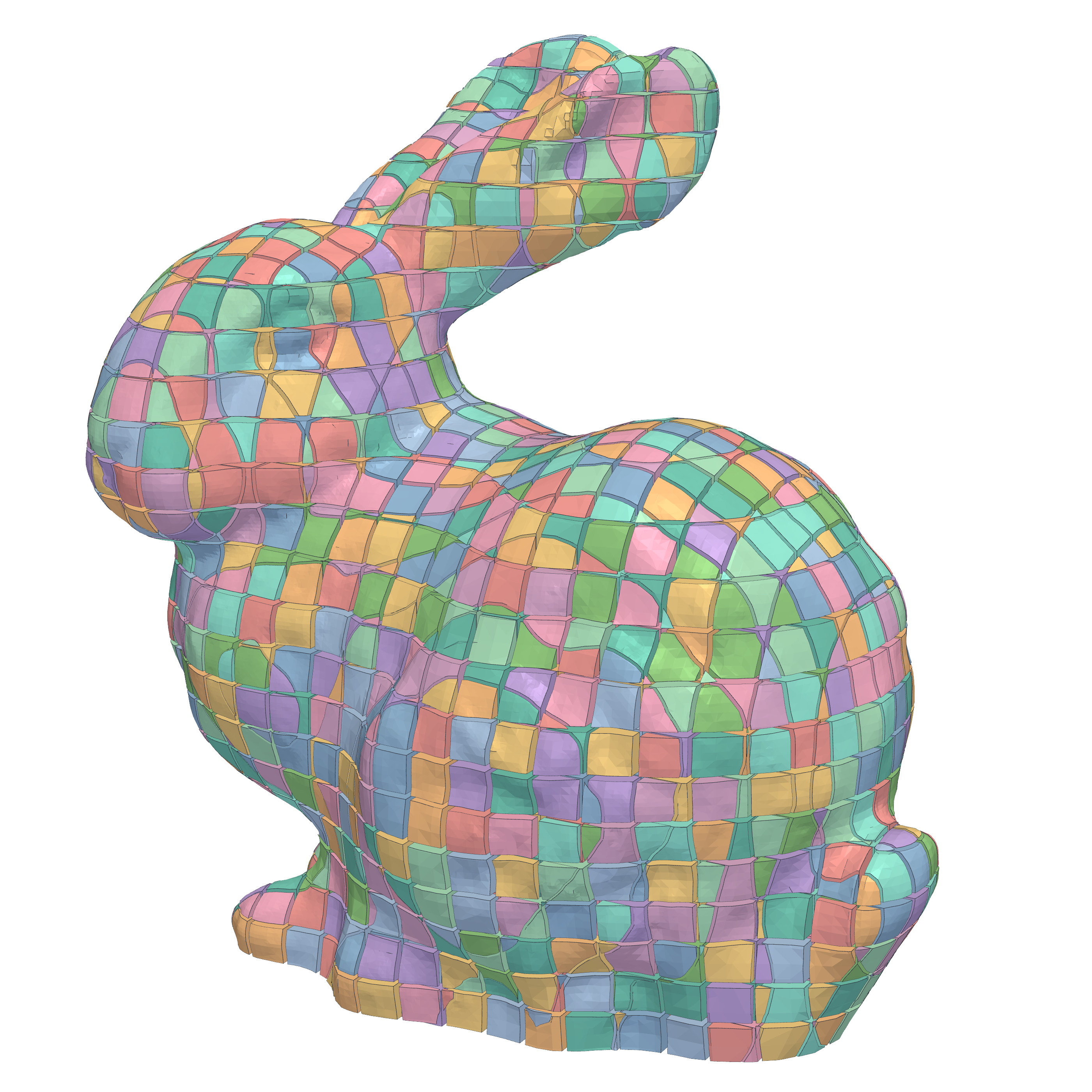}}\hfill
  \colorbox{tfbg}{\includegraphics[width=0.49\linewidth]{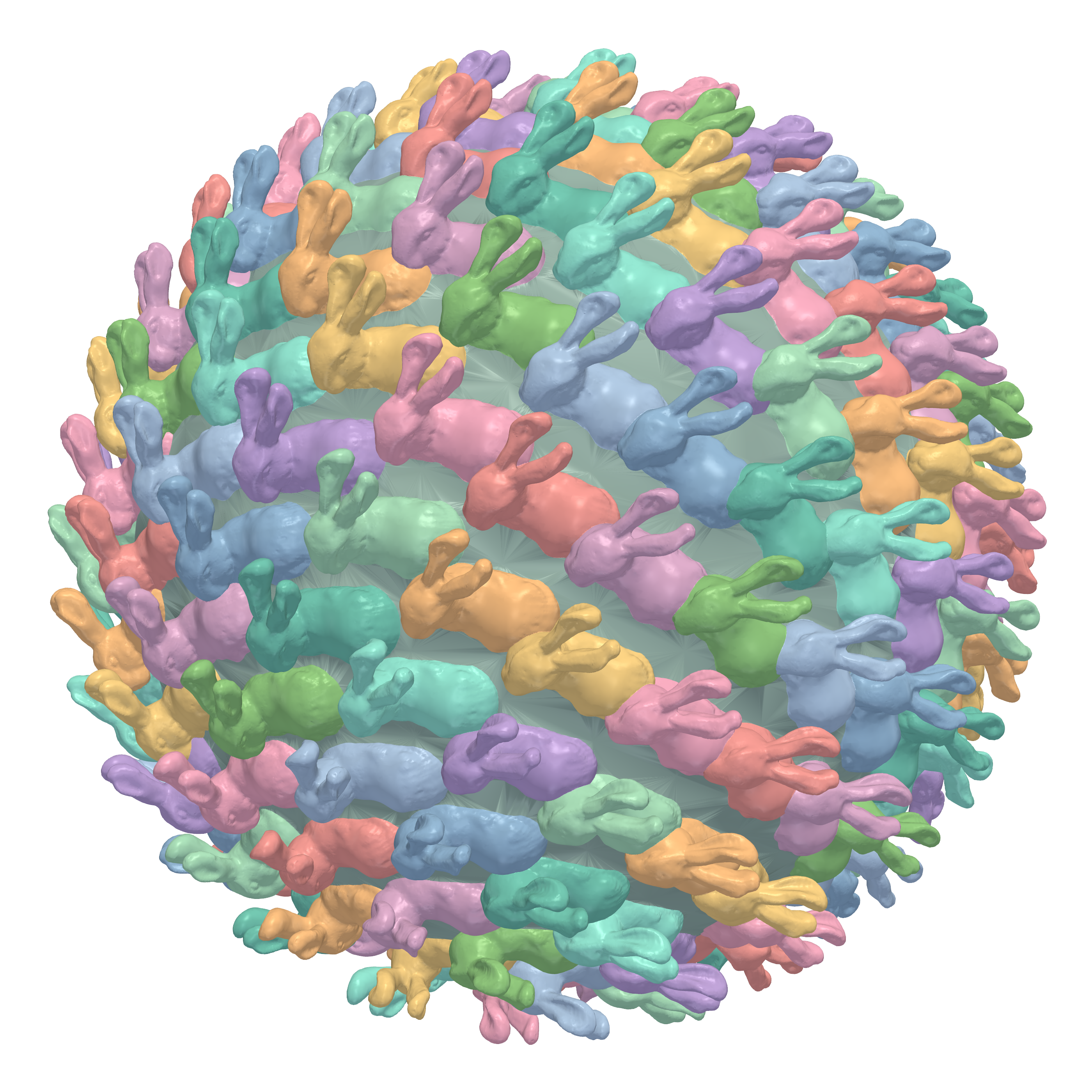}}
  \caption{\textbf{Volumetric domains and \emph{N}-ary booleans.} \emph{Left}: a Stanford bunny arranged against $60$ axis-aligned cutting planes ($20$ per axis) and split into its $3459$ interior volumetric cells, each a separate watertight solid (coloured arbitrarily). \emph{Right}: the union of a sphere with $200$ Fibonacci-placed bunnies (sphere $\cup$ bunnies) --- $22$M input triangles arranged in $0.5$\,s --- a $201$-operand boolean with every face coloured by the operand it came from. Each panel is a single locally exact arrangement build (Section~\ref{sec:results-performance}).}
  \label{fig:teaser}%
}

\maketitle

\begin{abstract}
Mesh CSG output is consumed in floating point: however exact the computation, every emitted coordinate is materialised --- rounded to a representable position --- and the next stage can observe crossings and orderings the exact result never had. Only index-based topology survives materialisation. We keep it exact: within the build, the arrangement's radial structure is ordered by exact predicates on the original input planes --- exact without exact constructions --- and where a decision spans faces, the intended answer is recovered by topological aggregation: a majority vote over the disagreeing geometric observations within their topological unit.

We compute the arrangement locally with integer-exact predicates, every stage a graph problem on graphs it never explicitly constructs. Pairwise intersections are classified into five canonical types (VV, VE, VF, EE, EF), each cut face is arranged in its own plane, and a two-level identity keeps the result consistent across faces with no global structure. The arrangement and its domain partition are built once and queried arbitrarily often: a boolean of any arity is a per-domain bit test, volumetric regions read straight off the partition, and open surfaces --- declared as oriented sheets --- cut volumes through the same algebra. The method is implemented in the header-only \emph{trueform} library, in C\texttt{++} with Python and TypeScript bindings. Compared to prior art, it produces valid, watertight output while running up to two orders of magnitude faster, and stays interactive in the browser.
%-------------------------------------------------------------------------
%  ACM CCS 2012 (placeholder --- to be refined with final categories)
\end{abstract}

\section{Introduction}
\label{sec:intro}

Constructive Solid Geometry (CSG) on triangulated meshes --- encompassing the arrangement of surfaces and the boolean operations that select among them --- serves as a foundational tool in geometric modelling, computer-aided design, scientific simulation, and computer graphics. Over the past decades, robust pipelines have been developed, with recent advances enabling interactive performance. In these downstream applications, the output of a CSG pipeline is rarely the final product; it is typically passed to subsequent stages, such as remeshing, finite-element analysis, or file-format conversion. Critically, this output is represented in floating-point arithmetic. Composing CSG stages is therefore hard on two fronts at once: constructing new geometric elements --- new vertices, for instance --- with exact or inexact representations of their embedding, and controlling the topology between those elements as it passes through the pipeline.

For the geometry, exact predicates have received attention for decades, and several kernels extend exactness through chained constructions~\cite{Shewchuk1997,Attene2020,Trettner2022,Levy2025}. When computations reach the limits of the floating-point representation, the classical trade-off appears: exact but expensive arbitrary-precision kernels against filtered or approximated but fast ones. Existing techniques push the exact representation as far as possible through the pipeline before either meeting a numerical failure or switching to a costlier representation. Yet exactness inside a stage is not the missing piece. However exact the computation, its result must be \emph{materialised} --- committed to a representable position, whether IEEE double, an integer grid, or a downstream format --- to leave one stage and enter the next, and the guarantee stops there. Pushing exactness further postpones this materialisation; it does not remove it.

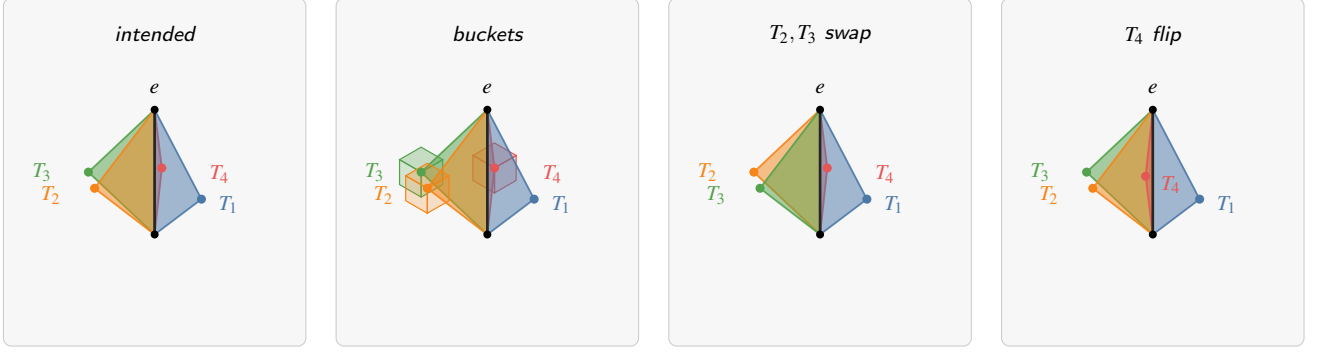
\begin{figure*}[t]
  \centering
  \input{diagrams/observed_wedge.tex}
  \caption{\textbf{The observed wedge.} \emph{Intended}: four triangles $T_1, T_2, T_3, T_4$ meeting at a non-manifold edge $e$. $T_2$ and $T_3$ are near-coplanar; $T_4$ is near-degenerate with apex close to the edge axis. The intended radial order is $T_1, T_2, T_3, T_4$. \emph{Buckets}: whenever a vertex is quantised, snap-rounded, or format-converted, it lands at one of the nearby representable values in its bucket. $T_2$ and $T_3$ are close enough that their buckets overlap; $T_4$'s bucket crosses the plane through the edge axis. The remaining panels show two observed manifestations. \emph{$T_2$, $T_3$ swap}: when the two apices fall in the bucket overlap, the \texttt{orient3d} test ordering them can return the opposite sign, swapping them in the observed order. \emph{$T_4$ flip}: when $T_4$'s apex lands on the other side of the edge plane, $T_4$'s position around the edge axis flips. Bucket size is exaggerated for visibility; the same ambiguity arises wherever bucket perturbations can flip an \texttt{orient3d} sign, including at the endpoints of $e$, where they redefine the edge axis itself. Within the build the radial order never reads these coordinates --- it is decided on the original planes (Section~\ref{sec:graphs}); the wedge shown here is what any consumer of the \emph{materialised} arrangement observes.}
  \label{fig:observed_wedge}
\end{figure*}

Besides the geometry, there is the topology --- and the topology is what survives. Connected components, the ring of faces around a non-manifold edge, the incidences the user models: these are decided by index adjacency, and they are the same combinatorial objects at any precision, on either side of materialisation. What does not survive are the \emph{relations} between them. The radial order of faces around a non-manifold edge, or the nesting of one connected component inside another, is a single fact of the intended geometry, yet each is read through exact \texttt{orient3d} evaluations on coordinates that materialisation has already perturbed, so readings of the same relation can disagree (Figure~\ref{fig:observed_wedge}). Inside our own build the radial order escapes this entirely --- it is decided by exact predicates on the original input planes, never on a constructed coordinate (Section~\ref{sec:graphs}) --- but any consumer of the materialised arrangement, composed pipelines included, faces the disagreement. Our answer is to make exact only what survives --- the topology --- and to treat every geometric answer as an observation of an intended geometry. Each kind of decision has a natural aggregation scope, its \emph{topological unit}: a component for orientation, a relation between components for merging, a volumetric domain for classification. Within a unit the intended geometry admits one answer, so a weighted-majority vote across the unit's observations recovers it. What a CSG stage can still guarantee once its output is materialised, it turns out, is the topology.

\begin{figure}[t]
  \centering
  \input{diagrams/box_rectangle_split.tex}
  \caption{Mesh arrangements often divide a volume into regions of interest. When those regions are cut parametrically, primitives end up coplanar by construction. Here, a box is split into sub-domains $D_1, D_2$ by an internal rectangle $R$ whose vertices sit at midpoints of cube edges; the four intersection points (and edges) are shown in red. SoS removes the coplanarity by deterministic but arbitrary perturbation: each red vertex lands infinitesimally inside or outside the box, yielding $2^4 = 16$ possible intersection configurations depending on index order. Only one of these (all four perturbed outside the box) produces the two intended volumetric domains.}
  \label{fig:sos_box_divider}
\end{figure}
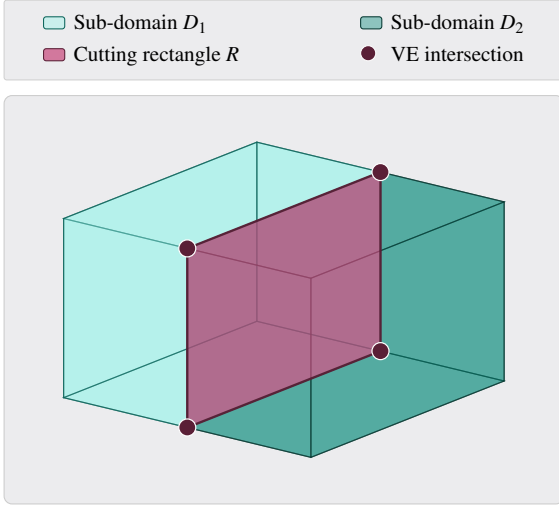

The contacts the user modelled --- coincident vertices, on-edge points, coplanar faces --- are classified exactly and survive intact, never blurred by perturbation. For the same reason, symbolic perturbation (Simulation of Simplicity~\cite{EdelsbrunnerMucke1990}) is excluded from the arrangement: where contact is intended, perturbation decides topology arbitrarily (Figure~\ref{fig:sos_box_divider}). Non-manifold input is first-class data: flaps, T-junctions, and isolated edges are units to aggregate over, not exceptions to clean up before processing.

Booleans are classically defined on solids, and an open surface has no inside. Yet cutting by open surfaces is everyday work: a geological horizon splits a volume into strata, a free-form section halves a part. Because classification here is a per-domain bitvector rather than an inside test, open surfaces enter the same algebra directly: declared as \emph{sheets} --- oriented separators whose bit reads ``behind the normal'' --- they cut volumes through the ordinary boolean operators (Figure~\ref{fig:sheets}).

The result is a composable CSG stage: an arrangement computed locally --- per face pair, per face plane, on a bounded integer kernel --- with no global exact structure to maintain, classified by votes into volumetric domains. The build is paid once and any expression over the operands then reads off per domain (Section~\ref{sec:selection}), at interactive rates, in C\texttt{++}, Python, and the browser. We position the method against prior work next, and state its contributions in \S\ref{sec:contributions}.

\subsection{Related work}
\label{sec:related}

\textbf{Geometric exactness.} The arithmetic-kernel literature has converged on a small set of approaches. Shewchuk's adaptive floating-point expansions~\cite{Shewchuk1997} compute a predicate's sign exactly by accumulating non-overlapping floating-point components, refining only when the filter cascade fails. The Filtered Predicate Generator (FPG)~\cite{MeyerPion2008} generates such filtered predicates from polynomial expressions. CGAL's lazy rationals~\cite{CGAL2023} offer the same exactness through arbitrary-precision evaluation. Indirect predicates~\cite{Attene2020} take a different route: they store constructed points as recipes (line-plane intersections, triple-plane intersections) and fold them into the predicate's polynomial. Cherchi's mesh arrangement~\cite{Cherchi2020} is built on this. EMBER's plane-based representation~\cite{Trettner2022,NehringWirxel2021}, after Campen and Kobbelt~\cite{Campen2010Intersections,Campen2010Boundary}, uses fixed-width 256-bit integer arithmetic with no filter cascade, delivering exact CSG at interactive rates. L\'evy's recent CSG pipeline~\cite{Levy2025} implements two exact kernels --- Shewchuk expansions and multi-precision floats --- and compares them, with the multi-precision kernel the robust default. Extending exactness from one evaluation to chained constructions is where the cost concentrates: expansions can reach tens of thousands of components in degenerate cases~\cite{Levy2025}, and fixed-width representations must widen at each composition or change representation. Interval arithmetic~\cite{Bronnimann1998} provides the cheap filter all of these depend on. A separate strategy resolves degeneracies rather than computing through them: Simulation of Simplicity (SoS)~\cite{EdelsbrunnerMucke1990} perturbs inputs symbolically so predicates never return zero, eliminating degenerate cases by construction. The perturbation is deterministic but arbitrary, and not topologically exact (\S\ref{sec:intro}). Trueform sits between Attene and EMBER: a fixed three-width integer ladder (T0$\to$T1$\to$T2), no dynamic allocation, no fallback kernel. Degeneracies are classified exactly through the five-type discipline of Section~\ref{sec:predicates}, not perturbed away.

\textbf{Materialisation.} Once exact coordinates materialise to IEEE double, new intersections may appear that were not in the exact result. Prior work has taken two responses. The first keeps exactness inside the system as long as possible: EMBER's bounded-width plane representation~\cite{Trettner2022,NehringWirxel2021} guarantees exactness without leaving its kernel, and L\'evy's pipeline~\cite{Levy2025} keeps exact coordinates throughout (Shewchuk expansions or multi-precision floats). The second rounds back to a tractable representation at materialisation: 3D snap rounding~\cite{DevillersLazardLenhart2018,Valque2019}, geometric rounding for meshes~\cite{MilenkovicSacks2019}, and Cherchi's iterated cast-to-float heuristic~\cite{Cherchi2020} minimise the damage the round-trip causes. Neither response removes it: EMBER's own snap-rounding step at materialisation~\cite{Trettner2022} and L\'evy explicitly setting the float conversion aside as an open problem~\cite{Levy2025} mark materialisation as intrinsic to the use case, not a defect of any particular kernel. Trueform takes a third path: the observed disagreements materialisation introduces are inputs to topological aggregation, not noise to be hidden; the intended consistency of each topological unit recovers the answer.

\textbf{Topology and classification.} Prior approaches to classification differ in where the geometric decision lives. Zhou's pipeline~\cite{Zhou2016} decides globally, building the arrangement with CGAL exact predicates and flooding labels through the resulting topology. Cherchi's arrangement~\cite{Cherchi2020} does the same with indirect predicates and ear-cut triangulation. L\'evy~\cite{Levy2025} also floods globally but over a richer representation: the Weiler 3-map~\cite{Weiler1985} in the language of combinatorial maps~\cite{Lienhardt1988}, with operand-membership bit-vectors as the labels. Cherchi's 2022 interactive variant~\cite{Cherchi2022} localises: a per-patch ray cast decides inside/outside one face patch at a time. B\"ohm and Runge~\cite{BohmRunge2025} localise further, sorting incident triangles angularly around each non-manifold edge (a floating-point angular sort, degenerate elements excluded by precondition) and extracting domains per edge --- each edge read in isolation, with nothing to reconcile edges whose readings disagree. Generalized winding numbers~\cite{JacobsonKavanSorkine2013} take an orthogonal path: inside/outside is a scalar field evaluated at any query point, tolerant of non-manifold input but indifferent to combinatorial topology. Trueform places the decision at topological-unit scope (\S\ref{sec:intro}): a component or relation that carries disagreeing observations is resolved by a majority vote within it, and domains inherit the result.

\textbf{Sequences of booleans.} A model is rarely one boolean: it is a CSG tree, or a family of queries against the same operands. The classical route evaluates the tree pairwise, materialising every intermediate: CGAL's Nef polyhedra~\cite{Nef1978,Hachenberger2007} compose this way robustly at heavy cost, and OpenSCAD~\cite{OpenSCAD} inherits the pattern. BSP merging~\cite{NaylorAmanatidesThibault1990} composes trees rather than meshes, at the price of the splitting blow-up. QuickCSG~\cite{Douze2017QuickCSG} evaluates an $N$-ary expression in a single pass over the final arrangement, in floating point and explicitly outside degenerate configurations. EMBER~\cite{Trettner2022} evaluates CSG trees inside its exact plane representation, never materialising an intermediate mesh; Manifold~\cite{Manifold} evaluates operation trees lazily, keeping every intermediate as a manifold solid. Zhou~\cite{Zhou2016} and L\'evy~\cite{Levy2025} carry per-operand inclusion bits through one global arrangement, so a single expression needs no intermediates at all. Trueform takes the bit-vector route and splits build from query: the arrangement and its classification are computed once, and any expression over the operands --- any arity, arbitrarily many expressions --- is a per-domain bit test against the same build (Section~\ref{sec:selection}); a sequence costs one arrangement, not one per operation. The three strata of Figure~\ref{fig:sheets} are three such expressions read from one build.

\textbf{Application context.} Mesh CSG drives demanding workflows in geomodelling~\cite{Caumon2003,Pellerin2017}, interactive mechanical modelling~\cite{Tymms2016}, and CSG for fabrication~\cite{Garg2016}. Geomodelling's defining cut is by open horizon surfaces, yet mesh-boolean pipelines require solid operands --- Zhou's inputs are piecewise-constant winding-number meshes~\cite{Zhou2016}, Manifold's are manifold solids~\cite{Manifold} --- so the workflow falls to dedicated trimming machinery~\cite{Pellerin2017}; trueform admits open surfaces as first-class boolean operands (Section~\ref{sec:sheets}). All these workflows share the same constraint from \S\ref{sec:intro}: CSG output is consumed by another stage in floating point, and that stage cannot tolerate topology drift.

\subsection{Contributions}
\label{sec:contributions}

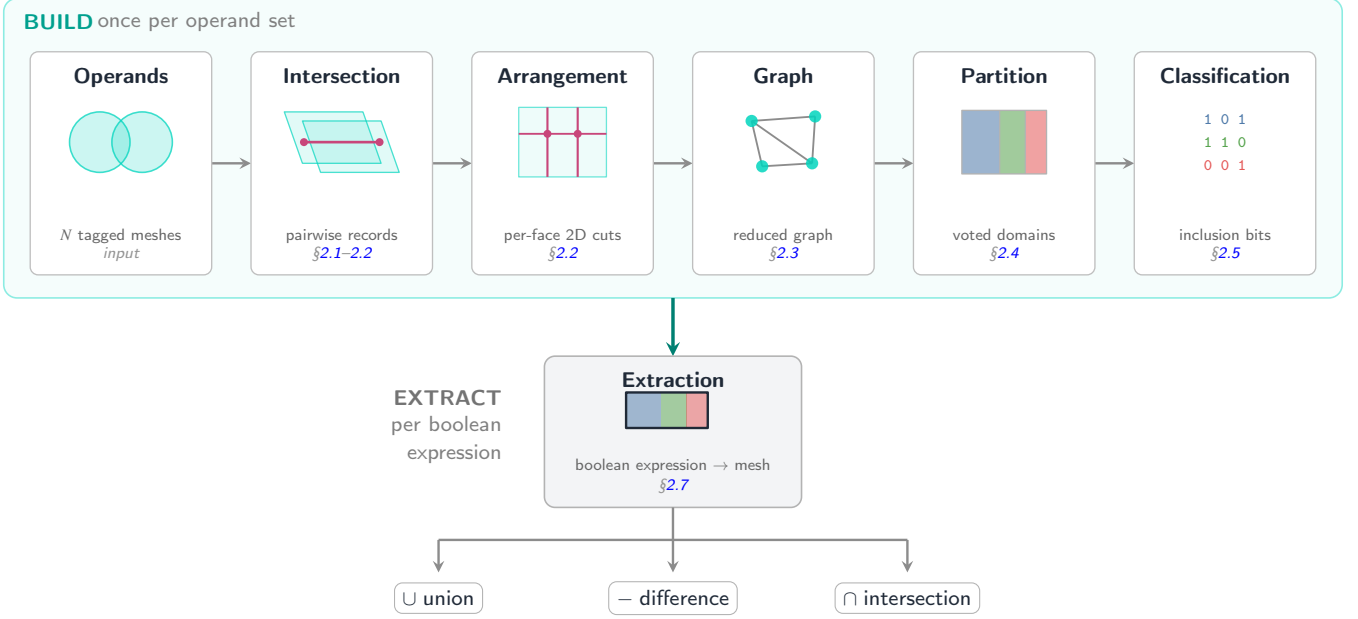
\begin{figure*}[t]
  \centering
  \input{diagrams/algorithm_pipeline.tex}
  \caption{\textbf{Pipeline overview.} The \textbf{build} is a single forward pass over the $N$ operands (Section~\ref{sec:algorithm}): pairwise intersection emits tagged records, each cut face is arranged in its local plane, the cut sub-faces and reused manifold links form the implicit reduced graph, topological-scope votes partition it into domains, and each domain is classified by an inclusion bitvector. The build is paid once. \textbf{Extraction} then answers any boolean expression over the operands by selecting domains and triangulating their boundary (Section~\ref{sec:selection}); a whole family of booleans is read from one build, the amortisation measured in Section~\ref{sec:results-performance}.}
  \label{fig:pipeline}
\end{figure*}

The method is one pipeline, but its contributions separate along four subtasks: the predicates that classify contact, the construction that arranges it, the classification that absorbs uncertainty, and the sequences of booleans one build answers.

\begin{enumerate}
\item \textbf{Predicates: exact contact classification without perturbation.}
Pairwise intersections are classified into five canonical types (VV, VE, VF, EE, EF) by exact predicates on a single bounded integer kernel, before any point is constructed and without Simulation of Simplicity~\cite{EdelsbrunnerMucke1990} (Section~\ref{sec:predicates}). Coincident vertices, on-edge points, and coplanar faces survive classification intact, and an optional tolerance band leaves the five-type partition invariant.

\item \textbf{Construction: a locally computed arrangement with no global structure.}
Each cut face is arranged in its own 2D plane (Section~\ref{sec:arrangement}, Figure~\ref{fig:arrangement_pipeline}), and a two-level identity scheme --- a topological identity naming each point by the primitives that meet there, plus a geometric merge for distinct identities landing at one coordinate --- keeps the arrangement consistent across faces without escalating precision. The radial order around every constructed intersection edge is decided by exact sign tests on the original input planes alone --- the constructed edge lies in every incident plane, so no materialised coordinate enters the comparison, and two faces tie exactly when their planes truly coincide. The arrangement's radial structure is exact without exact constructions --- the guarantee L\'evy obtains by escalating to exact homogeneous coordinates~\cite{Levy2025}. Every stage operates on graphs that are never explicitly constructed: flat arrays indexed by lookup, whose reduced graph carries the combinatorial content of L\'evy's Weiler 3-map~\cite{Weiler1985,Levy2025} with no half-edge structure to maintain (Section~\ref{sec:graphs}). The same path handles two meshes, $N$ meshes, and polygon-soup self-arrangement, with non-manifold flaps, T-junctions, and isolated edges first-class.

\item \textbf{Classification: uncertainty aggregated across topological units.}
Where a 3D decision spans faces --- the orientation of a component, the nesting of one connected component inside another, the radial order at a non-manifold edge once the arrangement is consumed materialised or the input sheets are inconsistently wound --- individual exact \texttt{orient3d} observations that should agree can disagree. Each is resolved within its topological unit by a weighted-majority vote, the maximum-a-posteriori estimate of the unit's intended outcome (Section~\ref{sec:aggregation}, Appendix~\ref{app:bayesian}). The resulting bitvector admits open surfaces as first-class operands, declared \emph{sheets} (Section~\ref{sec:sheets}, Figure~\ref{fig:sheets}).

\item \textbf{Sequences: one build, arbitrarily many booleans.}
Classification yields a domain partition carrying per-domain operand-inclusion bitvectors, so a boolean expression of any arity is a per-domain bit test (Section~\ref{sec:selection}). The arrangement is paid once; a whole family of CSG queries --- a tree, a sweep of parameters, an interactive session --- reads from the same build (Figure~\ref{fig:teaser}).
\end{enumerate}

The pipeline ships as a header-only C\texttt{++} library with Python (nanobind) and TypeScript (WebAssembly) bindings (Section~\ref{sec:implementation}). The same exact kernel runs on x86, ARM, and in the browser at interactive speeds, and the codebase is freely available for academic and research use~\cite{trueform2025}. To our knowledge, no prior CSG library is interactive in the browser at these scales.

\subsection{Algorithm overview}
\label{sec:overview}

The pipeline is a single forward pass --- arrangement, then topological aggregation, then output --- built once and reused across CSG queries (Figure~\ref{fig:pipeline}).

\textbf{Arrangement.} Stage one intersects every candidate face pair in parallel (Sections~\ref{sec:predicates},~\ref{sec:arrangement}). Each intersection point's topological identity is the pair of input primitives that produced it (one of VV, VE, VF, EE, EF): same identity means same point by construction. A geometric merge additionally unifies distinct identities landing at the same coordinate. Each intersection edge is then defined by its two endpoint identities, independent of any containing face. Stage two computes each cut face's 2D arrangement independently in its local plane. Every point created here is registered on the canonical intersection edges it lies on, and therefore on every face containing them. Identification proceeds at the same two levels. Topologically: stage-one identities carry through, and a new crossing is named by its face triple --- the faces meeting there. Each face containing a crossing detects it locally through a different pair of canonical edges; the triple is the name those observations share. Where coplanar contact makes a triple ambiguous --- one face pair meeting in several canonical edges --- the canonical edge pair refines it (\S\ref{sec:arrangement}). Geometrically: distinct identities at the same parametric position on a shared canonical edge merge. The arrangement is topologically consistent across faces despite being computed locally on each, without escalating precision. Edge adjacency is built only for the new cut sub-faces; non-intersected original faces reuse the input mesh's manifold-edge link directly. Together these form the implicit reduced graph that subsequent steps operate on.

\textbf{Topological aggregation.} When 3D decisions span faces, individual geometric observations of one intended fact may disagree --- orientation and nesting read materialised intersection points; the radial order, exact in-build, is perturbed when re-read from materialised geometry or presented inconsistently by the input. Components aggregate orientation votes. Relations between components aggregate wedge-merge votes; union-find on the reduced graph then propagates those decisions into a domain partition. Domains inherit inside/outside from the relations' seeds (Sections~\ref{sec:aggregation},~\ref{sec:classification}). Both votes are weighted-majority over exact \texttt{orient3d} observations. Disagreeing observations within a unit are resolved by majority; ambiguous ones are skipped, with faces inheriting their unit's label. Domains carrying no seeds fall back to exact SoS-perturbed segment-casting on the integer kernel.

\textbf{Output.} A boolean expression of any arity over the operands --- closed volumes or declared sheets (Section~\ref{sec:sheets}) --- selects domains by classification: each domain carries an inclusion bitvector, and selection evaluates the expression per domain. Selected uncut original faces pass through directly; selected cut sub-faces are Delaunay-triangulated into the output.

\section{The algorithm}
\label{sec:algorithm}

Everything below serves one invariant: combinatorics kept exact over coordinates that are merely grid-rounded. The kernel classifies contact before any point is constructed (\S\ref{sec:predicates}), the arrangement names every point it does construct (\S\ref{sec:arrangement}), the implicit graphs carry those names into topological units (\S\ref{sec:graphs}), and the votes over those units partition space into domains (\S\ref{sec:aggregation}).

\subsection{Topologically exact predicates on convex polygons}
\label{sec:predicates}

The kernel is a bounded integer ladder over snapped coordinates. Input floats are mapped onto an integer grid. Predicate expressions promote width as needed: subtractions cost one bit (T0$\to$T1), products of differences double width (T1$\times$T1$\to$T2). The ladder closes after T2. Every predicate the pipeline uses --- \texttt{orient2d}, \texttt{orient3d}, and the cross/dot work that feeds them --- is a polynomial of total degree at most three. The library exposes two ladders, $32{\to}64{\to}128$ and $64{\to}128{\to}256$ bits; the user picks based on input precision and resolution. The kernel is the substrate; Attene's indirect predicates~\cite{Attene2020} carry the same accounting via stored expressions, EMBER~\cite{Trettner2022} via fixed-width 256-bit planes. Trueform works in between, with a fixed promotion ladder of exactly known widths and no dynamic allocation.

Five canonical configurations carry the local topology forward: vertex-vertex (\texttt{VV}), vertex-edge (\texttt{VE}), vertex-face (\texttt{VF}), edge-edge (\texttt{EE}), and edge-face (\texttt{EF}). Each emitted intersection is exactly one of these five, named by the simplex pair that produced it. The same kind of simplex-pair taxonomy lives in L\'evy~\cite{Levy2025} as $(\sigma, \sigma')$ pairs over a triangle's seven open simplices; trueform generalises to convex polygons.

Convex polygons are the input scope because they fan-triangulate trivially online from any vertex $A_0$. There is no separate constrained Delaunay triangulation (CDT), no auxiliary index, no triangulation queue. The fan triangulation is what makes the rest of the section possible. Input polygons cannot be assumed to be in general position; three or more collinear vertices occur in practice. A fan triangle is the right unit of work. A non-degenerate triangle classifies cleanly via the zero-pattern below; a degenerate one (three collinear vertices) contributes nothing to that slice, and the loop advances. Non-convex face primitives are rare in mesh CSG inputs, so the convex scope captures the practical case at minimal cost.

The classifier evaluates three \texttt{orient3d} signs per fan triangle. For fan triangle $t = (A_0, A_{t+1}, A_{t+2})$ and segment $(D, E)$, the signs are
\[
\begin{aligned}
v_1 &= \mathrm{orient3d}(A_0,     A_{t+1}, D, E),\\
v_2 &= \mathrm{orient3d}(A_{t+1}, A_{t+2}, D, E),\\
v_3 &= \mathrm{orient3d}(A_0,     A_{t+2}, D, E).
\end{aligned}
\]
The zero-pattern decides the type (Table~\ref{tab:fan_zero_patterns}). No zeros and a sign disagreement on the fan triangle: \texttt{EF} (the segment crosses the face's interior). One zero on a real polygon edge: \texttt{EE} (the segment crosses a real edge at an interior point of both). Two zeros at a shared fan vertex: \texttt{VE} (the segment passes through a face vertex). Fan diagonals --- the edges $A_0 A_{t+1}$ for $t > 0$ and $A_0 A_{t+2}$ for $t + 3 < n$ --- are not real polygon edges. A zero on one of them is geometrically real but topologically a fan-interior point, classified as \texttt{EF}, never as \texttt{EE}. The real-edge predicate $(v_1 = 0 \land t = 0) \lor (v_2 = 0) \lor (v_3 = 0 \land t + 3 = n)$ is the only place the fan triangulation leaks structural information; the rest is rotation-symmetric.

\begin{table}[t]
\centering
\caption{Fan-triangle zero-pattern decision tree for the crossing-edge routine. On-plane primitives and vertex-in-face containment are handled by separate routines.}
\label{tab:fan_zero_patterns}
\begin{tabular*}{\columnwidth}{@{\extracolsep{\fill}}lr@{}}
\toprule
Pattern & Type \\
\midrule
No zeros, $v_1 = v_2 \ne v_3$  & \texttt{EF} \\
One zero on a real edge        & \texttt{EE} \\
One zero on a fan diagonal     & \texttt{EF} \\
Two zeros (face vertex)        & \texttt{VE} \\
\midrule
On-plane primitives            & \texttt{VV}, \texttt{VE}, \texttt{EE} \\
Vertex-in-face                 & \texttt{VF} \\
\bottomrule
\end{tabular*}
\end{table}

Two complementary routines run over this decomposition. The first processes edges that strictly cross the opposing plane, emitting \texttt{EF}, \texttt{EE}, or \texttt{VE} from the fan zero-pattern above. The second processes on-plane primitives --- any vertex whose plane-side sign is zero --- and emits \texttt{VV}, coplanar \texttt{VE}, or coplanar \texttt{EE}. A third routine emits \texttt{VF}. The three are not mutually exclusive: for a pair where some vertices cross and some touch the plane, all three may fire. The partition into types is still exact --- every real contact is emitted once, from one routine.

Degenerate polygons are not a special case. A chain of collinear vertices is a polygon that has shrunk to an edge. When the chain lies in the opposing plane, every vertex reads sign zero. The on-plane routine then catches all its contacts as the same \texttt{VV}/\texttt{VE}/\texttt{EE} primitives a real edge would produce. Convexity makes the geometry simple; the classifier makes the topology uniform. The five-type discipline preserves what SoS~\cite{EdelsbrunnerMucke1990} discards; the classifier perturbs nothing.

\paragraph*{Tolerance band.}
Users often want to declare a tolerance length within which near-coincident features in the input should be classified as coincident. The kernel accepts this length, but the predicates do not operate in length: \texttt{orient3d} returns a volume, \texttt{orient2d} an area. Each predicate therefore derives its own band per call from the user's length. The five-type partition is unchanged. The band is used only by the classifier: in stage one and in stage two's segment-arrangement phase (Section~\ref{sec:arrangement}). Region extraction and all subsequent stages run on exact predicates.

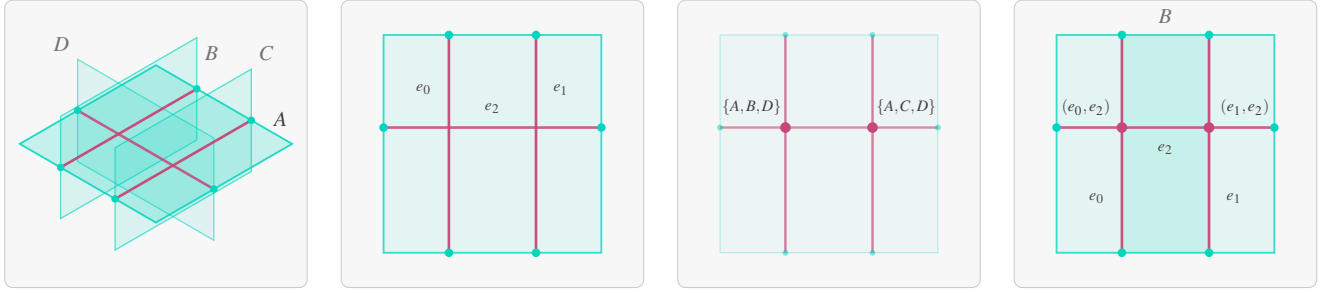
\begin{figure*}[t]
  \centering
  \input{diagrams/arrangement_pipeline.tex}
  \caption{\textbf{Two-stage local arrangement on face $A$}, which meets faces $B$, $C$, $D$. \emph{Left}: 3D configuration with three intersection edges and six boundary vertices on $A$. \emph{Middle left}: those edges in $A$'s local plane, labelled $e_0, e_1, e_2$ (the intersections of $A$ with $B$, $C$, $D$). \emph{Middle right}: the planar arrangement --- seven segments, two crossings, each named by its face triple, $\{A, B, D\}$ and $\{A, C, D\}$. Faces $B$ and $D$ detect the same crossings through different edge pairs; the triple is the name all observations share, so points computed independently on different faces unify without coordinate comparison. \emph{Right}: the coplanar exception --- a variant where $B$ is instead coplanar with $A$, meeting it in two contact edges $e_0, e_1$; a transversal $e_2$ crosses both at distinct points carrying the same triple $\{A, B, D\}$, which the canonical edge pairs $(e_0, e_2)$, $(e_1, e_2)$ keep distinct.}
  \label{fig:arrangement_pipeline}
\end{figure*}

\subsection{Two-stage local arrangement}
\label{sec:arrangement}

The arrangement is computed in two stages. Stage one intersects every candidate face pair in parallel, emitting tagged intersection records keyed by simplex pair on each face. Stage two arranges each cut face's records independently in its local 2D plane. Both stages follow the same two-level identification pattern (\S\ref{sec:overview}): a topological identity that establishes same-point-by-construction, plus a geometric merge that additionally unifies distinct identities landing at the same coordinate. The implementation exposes the arrangement either over a collection of meshes or over a polygon soup; the algorithm is the same, only the candidate set differs.

\subsubsection*{Stage one --- pairwise intersection records}

Candidate face pairs come from a parallel AABB-tree descent --- dual across two trees for the multi-mesh case, self over a single tree for polygon soup. Each surviving pair invokes the predicates of Section~\ref{sec:predicates}. The output is a flat tuple per intersection: which mesh, which polygon, which polygon on the other side, the simplex pair on each, the point identity. No edges, no per-face geometry yet --- just identities.

The stream is globally unique without any dedup pass. Two mechanisms ensure this. \emph{Within a pair}, the decision tree of Section~\ref{sec:predicates} fires exactly one type per real contact --- no overlapping sub-simplex queries like L\'evy's~\cite{Levy2025}, which require a \texttt{sort + unique} pass. \emph{Across pairs}, each vertex and edge of a mesh has one designated owner face, picked up front from the face-membership list and the manifold-edge link. During traversal, only the owner emits records for that primitive; the other faces that share it skip it. Together the two mechanisms keep every contact recorded exactly once, regardless of how many adjacent faces share the primitive.

Construction is decoupled from classification. \texttt{VV}, \texttt{VE}, and \texttt{VF} reuse input-vertex identities directly. Only \texttt{EE} and \texttt{EF} compute a new point, via barycentric-weighted \texttt{div\_round} on the integer endpoints. The point is a grid integer, but a rounded one. \texttt{div\_round} snaps the true intersection to the nearest T2 cell: exact as a grid coordinate, inexact as a point. Its topological identity is exact regardless --- the pair of input primitives that produced it, (edge, edge) for \texttt{EE}, (edge, face) for \texttt{EF}, a global key independent of where it was computed.

A parallel sort by key $(\text{tag}, \text{polygon}, \text{tag}', \text{polygon}')$ groups records. Offset-blocks on the prefix $(\text{tag}, \text{polygon})$ place every record in a contiguous slice belonging to one face. Stage two runs over these slices in parallel.

\subsubsection*{Stage two --- per-face planar arrangement}

Per-face processing has three phases: segment extraction, segment arrangements, and region extraction. The first two build a per-face segment graph; the third walks regions out of it.

All work happens in the local 2D plane of each face. Every intersection point --- stage-one outputs and the new crossings stage two produces --- carries a canonical identity and a 2D copy in each face that contains it. These per-face 2D copies let per-face arithmetic stay exact: coplanarity holds within each plane by construction, and the integer planar predicates of \S\ref{sec:predicates} apply without further constraint.

\paragraph*{Segment extraction.}
Extraction proceeds in two passes. The first inserts the face's intersection points into its base loop, ordered around the perimeter. The result is the cut base loop. The second walks the face's slice. The stage-one sort ordered records by $(\text{tag}, \text{polygon}, \text{tag}', \text{polygon}')$, so records sharing a cross-face context are already contiguous --- the grouping is the linear sequence itself. Each run classifies in $O(1)$ from its size: one record means no edge; two, a single intersection edge; three or more, a closed coplanar polygon-of-contact, which is a contiguous subset of the other polygon's cut base loop. The subset is traced linearly through the triplet identities of its points; no geometric check is needed. Each intersection edge has a canonical identity given by its two endpoint identities. The same canonical edge appears in every face that contains it; updates to the edge propagate to all containing faces by lookup.

\paragraph*{Segment arrangements.}
A per-face planar arrangement resolves crossings among the extracted segments. Each crossing produces a new intersection point, and its identity must be the same on every face that sees it. A crossing where three faces meet is detected independently on each of them: face $A$ as a crossing of $(e_{AB}, e_{AC})$, face $B$ as $(e_{AB}, e_{BC})$, face $C$ as $(e_{AC}, e_{BC})$. No pair of canonical edges is shared by all three observations; the face triple is, and it names the junction identically from every face (Figure~\ref{fig:arrangement_pipeline}). The triple fails in exactly one configuration: coplanar contact. There a single face pair meets in several canonical edges --- the boundary of its polygon-of-contact --- and one triple then spans several distinct crossings: an edge transversal to the contact crosses several contact edges at distinct points, all carrying the same triple. The ambiguity is recognised topologically, with no geometry consulted: extraction knows which edges it emitted from a polygon-of-contact, and a triple whose crossings involve two or more distinct contact edges is refined by the canonical edge pair, which keeps the pack's crossings distinct. Both decisions are scoped to the triple and all its observations, so every face keys a given crossing identically.

Faces with fewer than 32 segments run a quadratic crossing check; larger faces build a segment tree and find crossings in $O(n \log n)$. Each new crossing is classified by three types (VV, VE, EE), the 2D restriction of the five-type discipline of stage one. Topological exactness of the three types ensures point and segment uniqueness: coincident endpoints --- points landing on the same integer-grid cell --- collapse into a single point as VV, and overlapping segments collapse into one canonical segment. The records on any single face are typically a handful, occasionally a few dozen --- bounded by the local cut geometry --- so $n$ is small. What scales in real workloads is the number of cut faces, parallelised across threads. The strategy is deliberate: keep the heavy algorithmic step inside a small per-face set, and ride throughput on face-level parallelism rather than per-face triangulation.

Identity stays globally consistent through a canonicalisation step --- stage two's geometric merge level --- that sorts every edge instance by canonical endpoint pair and groups all per-face copies of the same intersection edge into a single offset block. This is the reverse map from canonical edges to containing faces, built by structure alone. Each canonical intersection edge is then processed once, in parallel across blocks, to absorb the crossings naming it: crossings are sorted by parametric position $t$ along the segment, the segment is split into sub-segments at the sorted positions, and two crossings landing at the same $t$ merge into a single point by the same VV mechanism. The split rebuilds the shared buffer in place; every face's slice then sees the new sub-segments by lookup. Point identities are remapped in a single parallel pass over the buffer, so the next phase sees a globally consistent segment graph on every face. Cross-face consistency follows from identity, not from coordinate comparison: any point created on any face propagates to every face containing its canonical edge.

\paragraph*{Region extraction.}
With the segment graph in place, regions of the cut face follow from a path classification. Segments form paths in the graph. A path is a \emph{crossing} when both endpoints sit on the cut base loop. It is \emph{non-crossing} when its terminus is an internal junction of degree three or more --- a condition that arises only at points where three or more faces meet, or at non-manifold edge junctions. The majority of faces produce only crossing paths.

Region extraction dispatches on this distinction. When every path is a crossing, regions are extracted in linear time by dividing the cut base loop at each crossing into sub-loops. The trivial common case --- a single crossing edge --- yields two sub-loops with no predicate work at all. When non-crossing paths are present, an exact planar walk runs on the full edge set: half-edges at each vertex are placed in radial order by an exact \texttt{orient2d} comparator, and the regions are walked.

Closed paths and dangling cut paths are patched into the regions that contain them by an exact point-in-region test. The output of stage two is the set of sub-loops belonging to each cut face.

\begin{figure*}[t]
  \centering
  \input{diagrams/mel_components_to_wedges.tex}
  \caption{
    Left: the reduced graph $\mathcal{R}$ holds MEL components
    $c_1, \ldots, c_4$ joined by a relation $r$ --- the equivalence class
    of non-manifold edges along a polyline that share the incident-component
    set $\{c_1, c_2, c_3, c_4\}$. Right: each edge in $r$ realises as
    a wedge ring with cyclic order $\rho_e = (c_1, c_2, c_4, c_3)$. The
    graph is exact; the cyclic order is a geometric observation ---
    exact within the build (Section~\ref{sec:graphs}), perturbable once
    the arrangement is read from materialised coordinates. A majority vote
    within $r$ (Section~\ref{sec:aggregation}) recovers the canonical $\rho_e$.
  }
  \label{fig:reduced_graph}
\end{figure*}
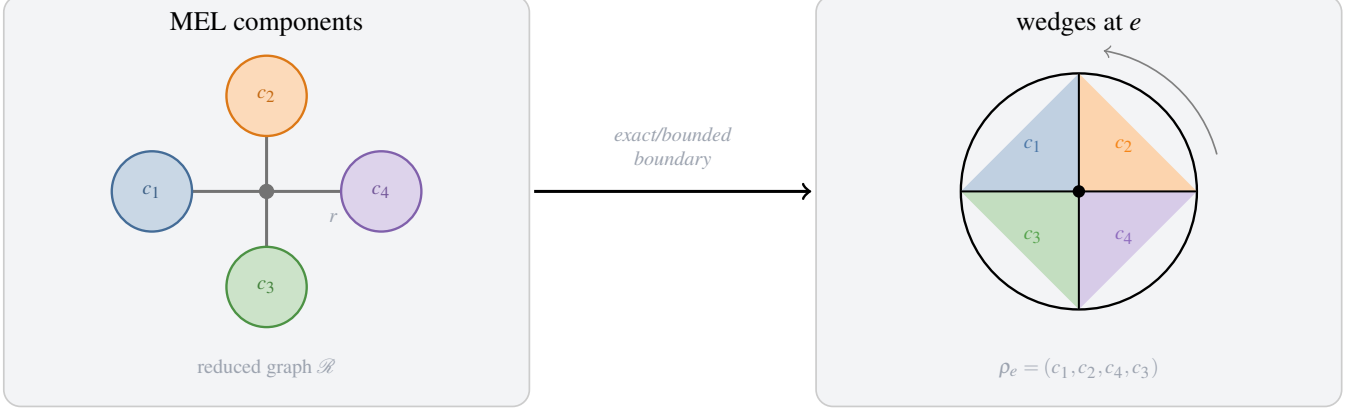
\subsection{Implicit graph structures}
\label{sec:graphs}

Every stage of the pipeline is a graph problem on graphs that are never explicitly constructed; the next three subsections define the implicit graph that the domain partition and classification operate on (Section~\ref{sec:aggregation}).

Two implementation primitives appear throughout: connected-component labelling and a dense equivalence-class mapping that replaces pointer-based union-find. Below we use ``union-find'' generically for the latter; their low-level implementations are not material to the CSG algorithm.\footnote{Available as \texttt{tf::label\_connected\_components} and \texttt{tf::make\_dense\_equivalence\_class\_map} in the trueform implementation.}

\subsubsection*{The reduced graph: components as nodes}

The domain partition and classification (Section~\ref{sec:aggregation}) operate on connected components of the manifold-edge link (MEL components) of the full arrangement. Running an MEL flood-fill directly on the whole arrangement would work but throw away the structure already built --- uncut faces inherit the input mesh's manifold-edge link, and cut sub-loops carry the per-face cut graph from stage two of the arrangement (Section~\ref{sec:arrangement}). We compute on the two sets separately and join.

\textbf{Non-intersected MEL components} flood-fill through the input manifold-edge link with intersected faces masked out. \textbf{Intersected MEL components} flood-fill through the per-face cut graph, in which coincident sub-loops from coplanar contacts are collapsed to a single face. The collapse means coplanar overlaps end up in one intersected component automatically. A union-find then joins the two sets across the cut/uncut interface: each cut sub-loop edge on the original face boundary contributes a pair with its uncut neighbour through the input manifold-edge link, and the dense equivalence-class map collapses the pairs. The result is the MEL components of the full arrangement. Components are not instantiated as explicit data structures: we keep two label arrays --- one over original faces, one over cut sub-loops --- each entry holding its MEL-component id, queried by lookup.

The \textbf{reduced graph} has these MEL components as nodes. Edges between nodes are the non-manifold edges shared by their faces in the arrangement --- the relations of the next subsection. Within each component the winding is consistent: coherent by construction where no coplanar pack was collapsed, and where one was --- collapsing it keeps a single loop with arbitrary winding --- the component is re-oriented to the area-weighted majority of its faces, the orientation vote of Section~\ref{sec:aggregation}. The per-face side map (stored-orientation side, reversed side) is then globally coherent.

\subsubsection*{Component relations and the radial permutation}

Around the directed axis of a non-manifold edge $e$, the incident faces have a well-defined cyclic order; listing them in this order yields a cyclic sequence $\rho_e$ of MEL-component labels, computed by exact sign predicates on the integer kernel (Section~\ref{sec:predicates}). The choice of axis direction is arbitrary, and reversing it reverses $\rho_e$. The non-manifold edges of an arrangement form \textbf{polylines} under shared-vertex adjacency --- maximal chains whose internal vertices have non-manifold-edge degree two. The \textbf{canonical permutation} of $e$ is the unique linear realisation of $\rho_e$: (i) inherit the axis direction from $e$'s polyline traversal, so every edge of one polyline reads its $\rho_e$ from the same side; (ii) start at the smallest label. The first step fixes the axis, the second the rotation.

Every angular comparison is an exact sign test on the \emph{original} input planes alone. The intersection edge lies in every incident original plane, so each incident face's wedge direction is exactly perpendicular to the fan's carrier line; comparing two wedges reduces to signs of degree-two products of original plane coefficients, and no constructed coordinate enters the comparison. Two wedges tie precisely when their original planes coincide --- exactly the coplanarity the contact classification has already decided (Section~\ref{sec:predicates}) --- so materialisation can neither create nor destroy an ordering: within the build, $\rho_e$ is exact. The per-apex perturbations of Figure~\ref{fig:observed_wedge} --- a near-coplanar pair swapping, a near-axial apex flipping --- arise only when the order is read back from materialised geometry, as when an arrangement is consumed as a mesh (\S\ref{sec:implementation}). The materialised endpoints are consulted in exactly one place: a single sign aligning the ring's direction with the stored edge. A wrong sign mirrors the ring; cyclic adjacency is preserved, the canonical-permutation equivalence below is order-blind, and the per-polyline alignment keeps the choice consistent, so the merging vote (Section~\ref{sec:aggregation}) absorbs it.

Two canonical permutations are \textbf{equivalent} when they sort to the same sequence --- deliberately order-blind: equivalence only groups observations of the same relation; the merging vote settles their order. A fan containing a degenerate incident face --- one whose original vertices admit no plane --- cannot be ordered and is flagged invalid whole: it contributes no observation to the merging vote (Section~\ref{sec:aggregation}), and the components it touches participate through the relation's other non-manifold edges. The same guard rejects fans whose member planes do not all contain one line, as when snapping merges two distinct intersection edges.

A \textbf{relation between components} is an equivalence class of non-manifold edges sharing both a polyline and a set of incident components. Within a relation, multiple canonical radial permutations may be observed; the merging vote of Section~\ref{sec:aggregation} aggregates these disagreeing observations into a single intended permutation. L\'evy organises the radial structure as a \textbf{radial polyline} \cite{Levy2025} and runs one radial sort per chain. This is correct on closed-manifold input, where a polyline carries one relation by construction. Non-manifold flaps --- open MEL components attached partway along a polyline --- raise the incident-component set on the edges they meet without breaking the chain: the polyline still walks as one structure but now carries two relations, one per distinct incident-component set. Our framing handles both cases uniformly: every (polyline, component-set) pair is a relation with its own vote. Even when the chain carries a single relation, the vote remains a distribution over canonical radial permutations rather than one sort, so an edge whose reading is disturbed --- a materialised re-reading, a mirrored direction sign, an inconsistently wound input sheet --- is absorbed by the majority.

\textbf{Computing the equivalence classes.} Non-manifold edges are grouped into polylines by a single union-find on shared endpoints, with edge axes aligned to the polyline's traversal direction so that canonical radial permutations are well-defined within the polyline. For each non-manifold edge we store two contiguous blocks: the incident face IDs in canonical-permutation order, and the corresponding MEL-component labels in sorted form. These form two flat arrays of total incident-face length, indexed by offsets over the non-manifold edges. An argsort over the key (polyline id, sorted incident-component set) then groups non-manifold edges into contiguous relations; a polyline carrying multiple distinct component sets contributes one relation per set.

In applications like geomodelling, an arrangement mixes closed primitives (volumes, intrusions) with partial cuts, fault surfaces, and probe geometry. Open patches sometimes form valid domain separators (a fault that fully cuts the model) and sometimes are intrusion artefacts: a half-inserted patch sits inside an otherwise closed body without splitting it, leaving a non-manifold fin in the domain's boundary. The application chooses the policy.

\begin{figure}[t]
\centering
\begin{tikzpicture}
  \clip[rounded corners=6pt]
    (0,0) rectangle (\columnwidth,0.745\columnwidth);
  \fill[tfsoft] (0,0) rectangle (\columnwidth,0.745\columnwidth);
  \node[anchor=south west, inner sep=0] at (0,0)
    {\includegraphics[width=\columnwidth]{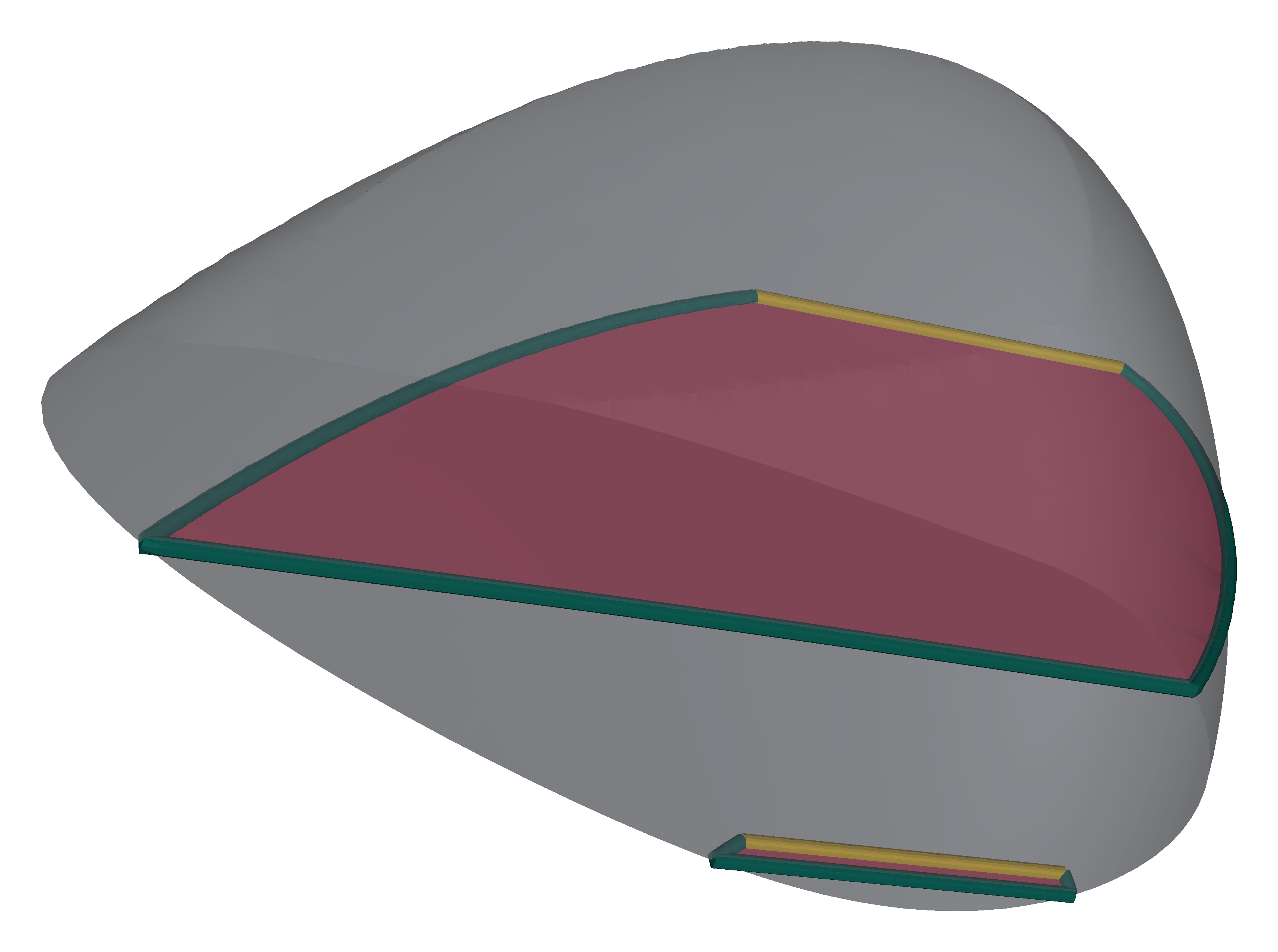}};
\end{tikzpicture}
\caption{A non-manifold domain with fins (Mode 2). A single body, shown translucent, wraps two open patches --- half-inserted intrusion surfaces --- as fins, each part of the body's boundary. Teal traces where a fin meets the body: a non-manifold edge whose two sides are fused into one domain ($d_0 = d_1$); amber traces the fin's free open edge. Mode 1 (\texttt{ignore\_open\_fragments}) drops the fins, recovering a closed manifold of equal volume. Rendered from the geological arrangement of Figure~\ref{fig:geological}.}
\label{fig:fin}
\end{figure}

An open MEL component, one carrying a boundary edge, does not separate two volumes: its two sides are the same region, joined around the boundary, so we always fuse them into one domain ($d_0 = d_1$) via the merge pairs of Section~\ref{sec:aggregation}. A fin never splits the region it sits in. The modes differ only in whether the fused component is then kept or discarded. \textbf{Mode 2}, the default, keeps it: the intrusion's interior stays a single non-manifold domain with the fin as part of its boundary (Figure~\ref{fig:fin}). \textbf{Mode 1} discards it: the open component's faces are dropped from the radial votes and all open components collapse into one outer domain, so an intrusion with a half-inserted patch returns to a single closed manifold domain with the fin gone.\footnote{Selectable in the trueform implementation via the \texttt{tf::domain\_config} flag set: \texttt{tf::domain\_config::ignore\_open\_fragments} for Mode 1, omitted for Mode 2 (the default).}

\subsubsection*{Equivalence to the Weiler 3-map}

The Weiler 3-map~\cite{Weiler1985,Levy2025}, in the language of combinatorial maps~\cite{Lienhardt1988}, stores three per-dart links over a duplicated triangle set: $\sigma_1$ within a triangle, $\alpha_2$ across an edge within the same sheet, $\alpha_3$ across the same edge in the twin sheet. We carry the same content without storage: $\sigma_1$ in each face's vertex order, $\alpha_2$ in the manifold-edge link and cut graph (Sections~\ref{sec:predicates},~\ref{sec:arrangement}), $\alpha_3$ as the cyclic adjacency at each non-manifold edge derived from its canonical permutation. The orbits map one-to-one onto our structures: $\langle\sigma_1, \alpha_2\rangle$, L\'evy's \emph{shell}, is our MEL component; $\langle\alpha_2, \alpha_3\rangle$, L\'evy's \emph{bundle}, is our wedge ring around a non-manifold edge --- wedge rings sharing a polyline and incident-component set form a relation; and $\langle\sigma_1, \alpha_2, \alpha_3\rangle$, L\'evy's \emph{connected component}, is our connected component, generated by our merge pairs (\S\ref{sec:classification}).

\subsection{Domain partition}
\label{sec:aggregation}

Each relation is observed once per non-manifold edge, and the observations can disagree: within the build only through the aligned direction sign or input sheets inconsistently wound (Section~\ref{sec:graphs}); read back from a materialised arrangement, through every rounded coordinate --- and any materialisation downstream only widens the spread. Prior pipelines handle it implicitly --- L\'evy by raising precision until it vanishes, per-edge methods by trusting one reading~\cite{Levy2025,BohmRunge2025}; we make the per-unit majority the explicit rule --- topological aggregation --- which Appendix~\ref{app:bayesian} casts as maximum-a-posteriori estimation. The pipeline of Section~\ref{sec:graphs} groups non-manifold edges into relations by (polyline, incident-component set), and within each relation collects the canonical radial permutations observed on its edges (Figure~\ref{fig:reduced_graph}). Each relation votes its canonical radial permutation (Figure~\ref{fig:relation_majority_vote}) and emits merge pairs over \textbf{fragment-sides}, each a $(\text{component id}, \text{side bit})$ pair whose side bit selects a component's stored-orientation side or its reverse. A dense equivalence-class map consumes the merge buffer, assigning one \textbf{domain id} per fragment-side. The final per-face per-side labels follow by lookup in one parallel pass: $\text{labels}[f][s] = \text{domain\_of}[2c + s]$ for face $f$ in MEL component $c$. The whole pipeline operates on the reduced graph (buffers proportional to the number of MEL components), not on per-face data.

\subsubsection*{The relation-merge vote}

\textbf{Distribution.} Offsets over the argsort of Section~\ref{sec:graphs} demarcate each relation as a parallel slice. Within each slice we sort by canonical radial permutation and count runs of identical permutation. The run lengths form the distribution of canonical radial permutations within the relation.

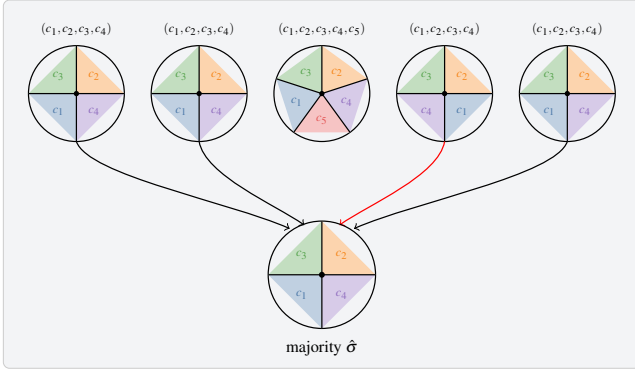
\begin{figure}[t]
\centering
\input{diagrams/relation_majority_vote.tex}
\caption{An intersection polyline along which five MEL components meet, bounding a domain. Four of its five non-manifold edges are incident to $c_1, \ldots, c_4$ alone and form a relation $r_1$ with canonical permutation $\rho_1 = (c_1, c_2, c_3, c_4)$; the remaining edge is incident to a fifth component $c_5$ contributed by a non-manifold fin from a half-inserted open operand, and forms a separate relation $r_2$ with $\rho_2 = (c_1, c_2, c_3, c_4, c_5)$. Each relation votes its own canonical permutation. One $\rho_1$ observation is flipped by a perturbation (red arrow); the majority $\hat\rho$ recovers the intended one. Under Mode 1, the fin is masked out, $c_5$ disappears, and all five edges fall into the single relation $r_1$.}
\label{fig:relation_majority_vote}
\end{figure}

\textbf{Vote.} The relation's canonical permutation is the longest run --- the majority. One linear walk per slice tracks the longest run seen so far and its representative non-manifold edge (Algorithm~\ref{alg:vote}). Figure~\ref{fig:relation_majority_vote} illustrates the vote on a polyline whose edges fall into two relations. The reduction extends to weights with no structural change. Each observation's weight is the minimum area of the faces incident to its non-manifold edge. Near-degenerate slivers (Figure~\ref{fig:observed_wedge}, $T_4$ flip) contribute less.

\begin{algorithm}[t]
\caption{Per-relation canonical-permutation vote}
\label{alg:vote}
\begin{algorithmic}[1]
\Require \texttt{edge\_order}, argsorted by ((polyline id, incident-component set), canonical radial permutation)
\Require \texttt{class\_offsets}, slice boundaries per relation
\Ensure \texttt{rel\_perm}[$r$], voted canonical radial permutation per relation $r$
\For{each class slice $[b, e)$ in \texttt{class\_offsets} \textbf{in parallel}}
  \State $\textit{longest}, \textit{longest\_rep} \gets 1,\, \texttt{edge\_order}[b]$
  \State $\textit{current}, \textit{current\_rep} \gets 1,\, \texttt{edge\_order}[b]$
  \For{$i \in [b+1, e)$}
    \If{$\textrm{perm}(\texttt{edge\_order}[i]) = \textrm{perm}(\texttt{edge\_order}[i-1])$}
      \State $\textit{current} \gets \textit{current} + 1$
    \Else
      \If{$\textit{current} > \textit{longest}$}
        \State $\textit{longest}, \textit{longest\_rep} \gets \textit{current}, \textit{current\_rep}$
      \EndIf
      \State $\textit{current}, \textit{current\_rep} \gets 1,\, \texttt{edge\_order}[i]$
    \EndIf
  \EndFor
  \If{$\textit{current} > \textit{longest}$}
    \State $\textit{longest}, \textit{longest\_rep} \gets \textit{current}, \textit{current\_rep}$
  \EndIf
  \State $\texttt{rel\_perm}[r] \gets \textrm{perm}(\textit{longest\_rep})$
\EndFor
\end{algorithmic}
\end{algorithm}

\textbf{Merge pairs.} The voted permutation's representative edge emits merge pairs once for the entire relation: in a consistent configuration every edge in the relation would emit identical pairs, so the representative suffices. At every pair of cyclic-adjacent positions in the voted permutation, the two fragment-sides bounding the wedge between them are joined into one pair. Mode-2 MEL components carrying boundary edges (Section~\ref{sec:graphs}) contribute one extra pair $(c, 0) \leftrightarrow (c, 1)$.

For a single connected component this partition is final. Geometrically disjoint connected components, and the operand inclusion a boolean selects on, are taken up in classification (Section~\ref{sec:classification}).

\subsection{Classification}
\label{sec:classification}

The domain partition (Section~\ref{sec:aggregation}) tells each face-side which region it bounds. CSG asks a different question of the same partition: not which region a face bounds, but whether a boolean expression selects it. The answer is one \textbf{operand-inclusion bitvector} per domain, $b(d) \in \{0,1\}^N$, with bit $i$ set iff domain $d$ lies inside operand $i$. Crossing a surface toggles membership in exactly the operands that surface carries, and the boolean reads the result. Inclusion bitvectors as labels go back to L\'evy~\cite{Levy2025}, who seeds them with a ray per connected component; we read the seeds off the votes of Section~\ref{sec:aggregation} instead, leaving a single geometric test for the cross-component nesting below, a closed connected component nested inside another with no surface contact.

\textbf{The inclusion potential.} Crossing one MEL component $c$ between its two incident domains toggles membership in the operands whose surface $c$ carries: $b(d') = b(d) \oplus B(c)$, where $B(c)$ is the OR of $c$'s own operand bit and any coplanar duplicate folded into it (Section~\ref{sec:graphs}). Inclusion is therefore a potential on the domain-adjacency graph: known increments $B(c)$ on every edge, fixed once anchors pin the absolute values.

\textbf{Local seeds carry the aggregation.} At each wedge the interior side of a component's loop lies inside that component's operand, so the interior-side domain $\text{domain\_of}[2c+1]$ takes the bit directly; a coplanar duplicate sets the side its winding selects. The wedge that places this bit is read off the voted radial permutation of Section~\ref{sec:aggregation} (Figure~\ref{fig:relation_majority_vote}). Because that permutation is the relation's majority vote over its (possibly disagreeing) per-edge observations, the seed already carries the aggregated answer: the uncertainty is absorbed where the seed is set, not repaired afterward. This fixes every domain reachable by surface contact, the whole partition for a single connected component, with no ray and no per-component seeding pass.

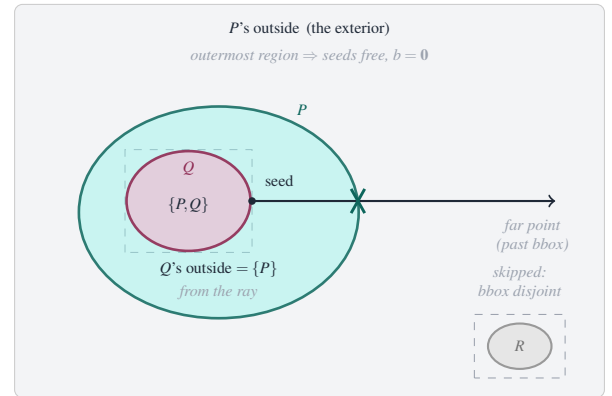
\begin{figure}[t]
  \centering
  \input{diagrams/cross_component_nesting.tex}
  \caption{Cross-component nesting. $Q$ sits inside $P$ without touching it, so they form two separate connected components. $P$'s outside is the outermost region, inside nothing, and seeds for free ($b = \mathbf{0}$, no ray). $Q$'s outside is the gap between the two surfaces, which may lie inside other operands; one ray from $Q$ crosses $P$ once, so the gap is inside $P$. The flood then makes $Q$'s interior $\{P,Q\}$. A form like $R$, whose bounding box misses the connected component, is skipped without a ray.}
  \label{fig:cross_component_nesting}
\end{figure}

\textbf{Anchoring across connected components.} The wedge seeds reach only within a \textbf{connected component}, the equivalence class of MEL components joined by shared non-manifold edges; they say nothing about how one connected component sits inside another with no surface contact between them (L\'evy's ``internal-boundary''~\cite{Levy2025}; Figure~\ref{fig:cross_component_nesting}). Per-domain signed volumes follow by scatter from per-component integer-exact contributions, and each connected component's \textbf{outer-environment domain} is its most-negative-volume incident domain. The connected component whose outer-environment is the most negative of all cannot be contained in anything, so that domain anchors at $b = \mathbf{0}$, inside nothing, with no ray. Every other connected component's outer-environment may lie inside some operands. We cast an SoS-perturbed segment from any vertex of the connected component to a fixed far point past the global bounding box; SoS here decides no topology, only steering the ray clear of grazing contacts without changing the crossing parity. The cast runs only against operands whose bounding box overlaps the connected component's: an operand that cannot enclose it contributes no crossings and is skipped before any ray. The parity of the crossings against the rest, per operand, fixes the outer-environment's bitvector. Open patches not declared sheets (Section~\ref{sec:sheets}) are transparent to the cast: the front and back of such an open component are fused into one domain (Section~\ref{sec:aggregation}, $d_0 = d_1$), so a ray crossing a fin flips no parity, with no geometric test needed. The cast crosses every enclosing domain in one pass, so an outer connected component with several interior domains, such as a cube split by an internal plane, is resolved in that pass.

\textbf{The flood.} A single connected component is already fully seeded; with disjoint connected components a multi-source XOR-BFS carries each ray-anchored outer-environment inward, every unvisited domain taking $b(d') = b(d) \oplus B(c)$ from a fixed neighbour across component $c$. One pass over the domain adjacencies leaves every domain's inclusion bitvector exact.

\subsection{Open operands: sheets}
\label{sec:sheets}

\begin{figure*}[t]
  \centering
  \begin{tikzpicture}
    \newdimen\sheetw
    \sheetw=0.32\textwidth
    \newcommand{\sheetplate}[2]{%
      \begin{scope}[shift={(#1,0)}]
        \clip[rounded corners=6pt] (0,0) rectangle (\sheetw,\sheetw);
        \node[anchor=south west, inner sep=0] at (0,0)
          {\includegraphics[width=\sheetw]{#2}};
      \end{scope}}
    \sheetplate{0pt}{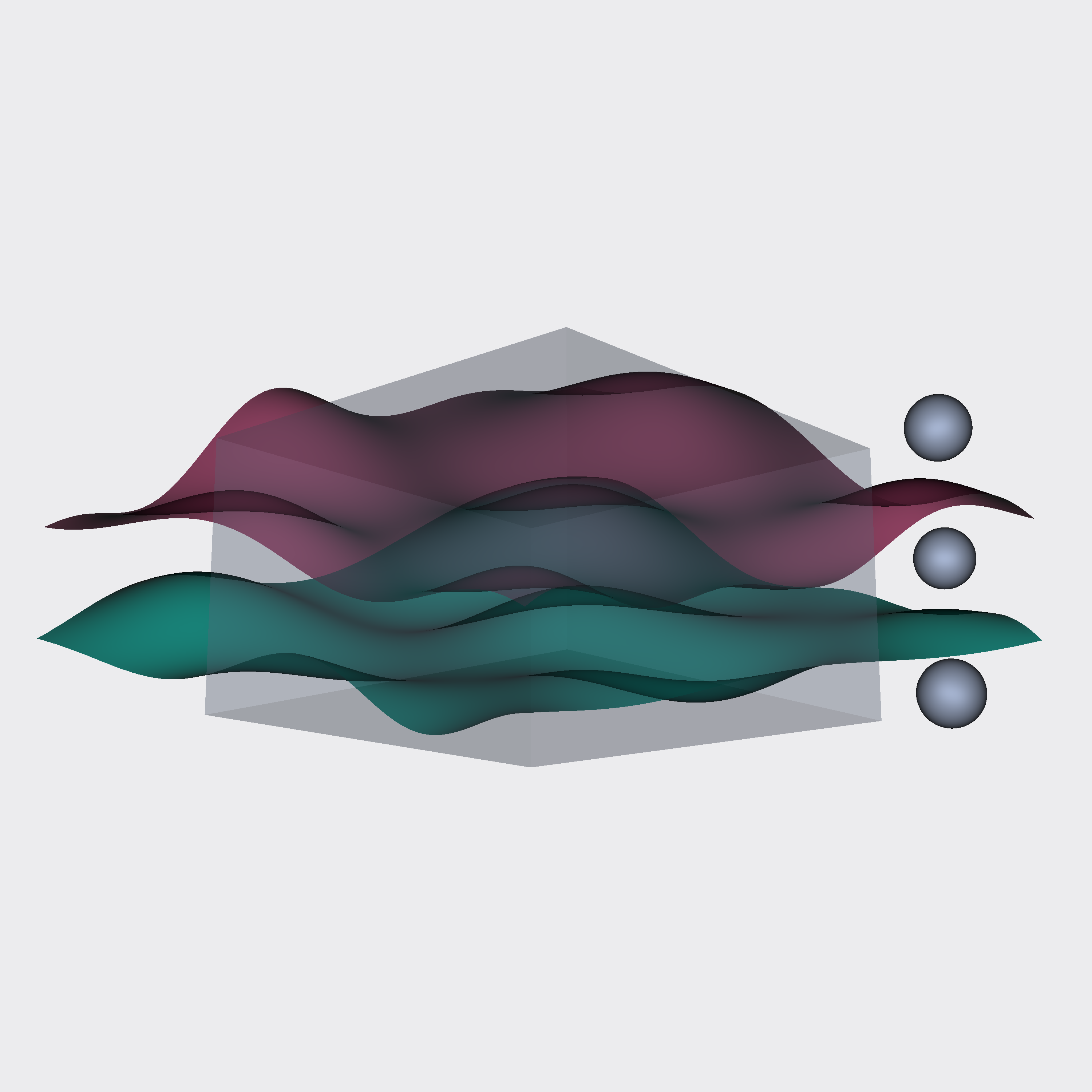}
    \begin{scope}[shift={(0.34\textwidth,0)}]
      \input{diagrams/sheets_expressions.tex}
    \end{scope}
    \sheetplate{0.68\textwidth}{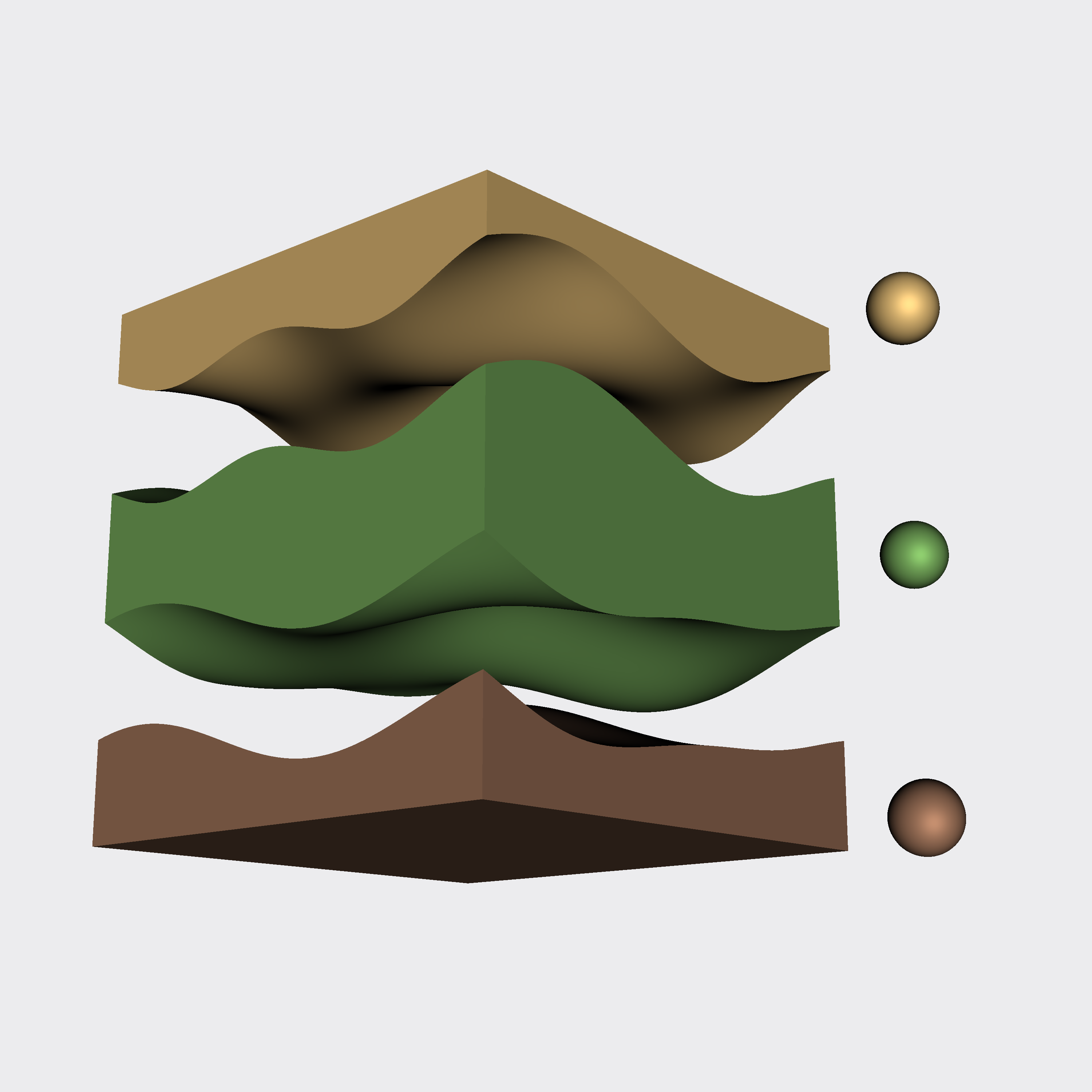}
  \end{tikzpicture}
  \caption{\textbf{Open operands: sheets.} A box and three disconnected spheres (volumes, $M$) with two undulating horizon surfaces declared as sheets ($T$, $B$). \emph{Left}: the inputs. \emph{Middle}: the strata as ordinary boolean expressions --- a sheet's operand bit reads ``behind its normal'', so intersection and difference cut. \emph{Right}: the three strata, exploded: each closed, capped by the horizons it meets; the spheres never touch a sheet and take their stratum from the sign of the generalized winding number.}
  \label{fig:sheets}
\end{figure*}

A volume operand's bit means \emph{inside}. An open surface has no inside, but it has sides: an operand declared a \textbf{sheet} is an oriented separator whose bit means \emph{behind its normal}. Nothing downstream changes --- domains carry the same bitvectors and expressions evaluate unchanged --- so intersection against a sheet keeps what lies behind it, difference keeps what lies in front, and the result is capped by the sheet faces it crosses (Figure~\ref{fig:sheets}). A clean closed mesh declared a sheet behaves identically to the volume it bounds: behind its outward normals \emph{is} its inside, so the two readings coincide and the semantics strictly generalise.

Three local rules implement this. A sheet is exempt from the open-patch fusion: where an ordinary open component's two sides merge into one domain (Section~\ref{sec:aggregation}), a sheet's sides stay distinct, so its fragments always separate their neighbourhood. Its bit is anchored, per region of the sheet, to the side behind its normal. And a connected component the sheet never touches --- a floater above or below a horizon --- takes its side from the sign of the sheet's generalized winding number~\cite{JacobsonKavanSorkine2013} evaluated at the component: the crossing-parity cast of Section~\ref{sec:classification} is meaningless against a surface a path can walk around, but the sign of the summed solid angle is not. The generality costs nothing measurable: on the $1000$ corpus pairs (\S\ref{sec:results-scaling}) run with one operand declared a sheet, outputs are identical and build times agree within $2\%$ (untabulated).

\subsection{Selection and output}
\label{sec:selection}

The inclusion bitvectors are the per-domain truth the arrangement was built to produce. A boolean expression reads them; the result is the boundary between selected and unselected domains.

\textbf{Evaluating the expression.} The expression is an arbitrary tree of \texttt{merge} (union), \texttt{intersection}, \texttt{difference}, and \texttt{complement} over operand leaves, of any arity. It selects each domain by one evaluation against that domain's bitvector, $\text{in}(d) = E(b(d))$. Compiling the tree collapses every chain of like-kind leaves into a single per-word mask: a union or intersection of $k$ operands becomes one masked compare per occupied word of the bitvector, not $k$ leaf tests, so cost scales with the number of occupied words rather than with operand count. Each expression is then a single linear pass over the domains.

\textbf{Output.} A face lies on the result boundary when its two sides fall in different selection classes, and it is emitted oriented from the selected side outward. Selected uncut original faces pass through unchanged; selected cut sub-faces are Delaunay-triangulated. The output coordinates materialise to IEEE double here.

\textbf{Volumetric regions.} If only the regions are wanted, not a boolean over them, the same partition yields them directly: reading the cross-component nesting as a merge rather than a seed collapses each inner connected component's outer-environment into its container, and the domains become volumetric regions. This skips the inclusion flood and the expression entirely. One construction serves both reads.

\section{Implementation}
\label{sec:implementation}

The pipeline is part of \emph{trueform}~\cite{trueform2025}, a header-only C\texttt{++}17 geometry library. \texttt{tf::make\_csg\_graph} builds the arrangement once into a reusable graph that holds the domain partition and the per-domain inclusion bitvectors (Sections~\ref{sec:aggregation},~\ref{sec:classification}); each boolean expression is then one \texttt{tf::make\_csg\_mesh} call --- a per-domain evaluation and a boundary extraction (Section~\ref{sec:selection}) --- or one \texttt{tf::make\_csg\_domains} call for its volumetric domains, so a family of such queries over the same operands shares a single arrangement. Trueform runs in one of two precision modes, chosen from the input coordinate type (Table~\ref{tab:precision}): \texttt{float} input uses the int32 kernel, with exact predicates in $128$-bit arithmetic, and \texttt{double} the int64 kernel, exact in $256$-bit. The wider kernel is forced on float input with \texttt{tf::make\_csg\_graph<tf::exact::int64>}. The output coordinate type is the input's by default and is set per query: \texttt{tf::make\_csg\_mesh<double>} materialises to double, while an integer output returns the kernel's integer-grid coordinates directly, with no floating-point round-trip.

\begin{table}[t]
\centering
\caption{The two precision modes, chosen from the input coordinate type. Predicates are exact in the kernel's integer arithmetic; the output coordinate type is set per query, and an integer output returns the grid coordinates directly.}
\label{tab:precision}
\begin{tabular*}{\columnwidth}{@{\extracolsep{\fill}}lll@{}}
\toprule
 & single & double \\
\midrule
input            & \texttt{float} & \texttt{double} \\
exact predicates & $128$-bit      & $256$-bit \\
output (default) & \texttt{float} & \texttt{double} \\
\bottomrule
\end{tabular*}
\end{table}

The pipeline is also exposed as separable stages. \texttt{tf::make\_mesh\_arrangements} materialises the arrangement: the co-refined, intersection-tagged mesh on floating-point coordinates, or the intersection curves alone. \texttt{tf::make\_domain\_labels} then partitions any such arrangement into per-face per-side domains, and \texttt{tf::split\_into\_domains} emits the per-domain meshes. This labeller still evaluates exact predicates internally. The difference is that materialisation has already happened upstream, so the predicates read materialised coordinates rather than building from exact ones. It also takes a bare arrangement with no provenance, so it cannot reach back to the original planes of Section~\ref{sec:graphs} and relies on the relation vote alone. The vote already absorbs disagreement (Section~\ref{sec:aggregation}), so the partition stays correct on such input. This is what makes the float entry point viable, and lets a caller who already holds an arrangement recover its domains without adopting our arithmetic. The library exposes the pipeline through idiomatic bindings in two further languages: in Python meshes are NumPy arrays, and in TypeScript operations are asynchronous calls over a WebAssembly runtime that runs in the browser and in Node.js. The browser runtime backs \emph{Lunar}\footnote{\texttt{lunar.polydera.com}}, an in-browser mesh editor that surfaces booleans, arrangements, and the rest of trueform as interactive UI, with all computation running client-side in the WebAssembly build.

\textbf{Examples.} Both the C\texttt{++} and Python distributions ship interactive examples built on VTK, runnable on the reader's own meshes: one operand is dragged through another while the boolean is re-evaluated each frame and a rolling average reports the per-operation time. The C\texttt{++} examples are in \texttt{vtk/examples/} (among them \texttt{boolean.cpp}) and the Python ones in \texttt{python/examples/vtk/} (\texttt{boolean\_difference.py}) --- the desktop counterpart of the browser demo of Section~\ref{sec:results-performance}.

\section{Results}
\label{sec:results}

We evaluate three properties in turn: the robustness of the relation vote that underwrites correctness (Section~\ref{sec:results-robustness}), performance as the operand count grows and the build/extract split that amortises a family of booleans (Section~\ref{sec:results-performance}), and comparison against prior art (Section~\ref{sec:results-scaling}).

\subsection{Robustness of the relation vote}
\label{sec:results-robustness}

The pipeline's correctness rests on one decision: a relation aggregates the radial observations of all its non-manifold edges into a single voted permutation (Section~\ref{sec:aggregation}), rather than trusting any edge alone. In the intended geometry the radial order is constant along a non-manifold polyline, so each edge is one noisy reading of a single shared order --- an observation we share with L\'evy~\cite{Levy2025}, whose pipeline sorts once per polyline on the strength of it. The radial sort is common to all three approaches; they differ in how they treat the uncertainty in it. B\"ohm and Runge~\cite{BohmRunge2025} take one reading and assume it exact: correct when it is, as on the given meshes their domain extraction targets, with nothing to fall back on when it is not. L\'evy eliminates the uncertainty --- exact homogeneous-coordinate constructions make every reading agree, so one sort is provably right --- but needs exact provenance to redo the arithmetic, which a materialised or composed arrangement no longer carries, and pays the precision-escalation cost. We do both, each where it is cheap: within the build the sort is decided on the original input planes and is exact without exact constructions (Section~\ref{sec:graphs}); beyond the build --- an arrangement already materialised, a composed pipeline, input sheets inconsistently wound --- the majority over the polyline's many readings recovers the shared order with no exact provenance to lean on. The vote is the per-edge sort generalised --- when the readings are unanimous it returns the permutation a single edge would, and it degrades gracefully only as they disagree. We measure it against per-edge on the materialised reading, where construction alone already disagrees --- from a controlled two-sphere probe through a real geomodelling workflow to the Thingi10K corpus --- and under the error a composition injects.

\begin{figure}[t]
\centering
\begin{tikzpicture}
  % Square render; the transparent top/bottom margins are clipped away by
  % a 0.81:1 panel with the image shifted down to centre the content.
  \clip[rounded corners=6pt]
    (0,0) rectangle (\columnwidth,0.79\columnwidth);
  \fill[tfsoft] (0,0) rectangle (\columnwidth,0.79\columnwidth);
  \node[anchor=south west, inner sep=0] at (0,-0.126\columnwidth)
    {\includegraphics[width=\columnwidth]{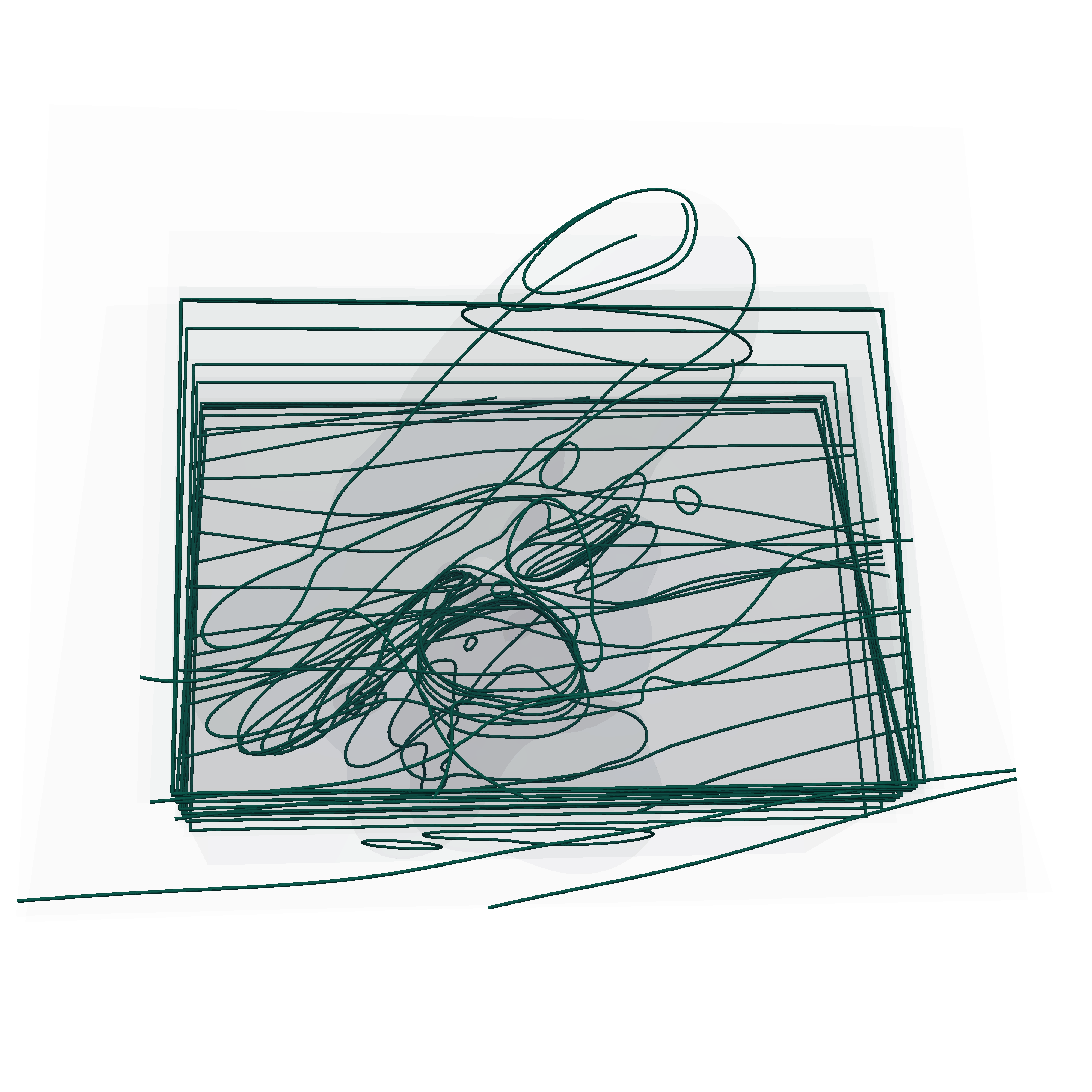}};
\end{tikzpicture}
\caption{The arrangement of eleven geological surfaces --- deposits, dykes, intrusions, erosion, and a bounding box. Its $588$k triangles carry $25$k non-manifold edges, grouped into $283$ polylines (drawn as tubes), and partition into $90$ volumetric domains. The surfaces share the bounding box and graze one another, so the arrangement is dense with coplanar and near-coplanar contacts.}
\label{fig:geological}
\end{figure}

\textbf{What we measure.} A relation is partitioned correctly when its extracted merges match the consistent (intended) partition. The baseline, \emph{per-edge}, is the vote with its aggregation removed --- the same pipeline reading the merges off each edge's own radial sort, the zero-uncertainty case of B\"ohm and Runge~\cite{BohmRunge2025}. It requires \emph{every} edge of the relation to agree; the vote requires only the \emph{majority}. Per-edge therefore mispartitions a relation precisely when its edges' sorts disagree. Within the build the sorts cannot disagree (Section~\ref{sec:graphs}); the measurements that follow read the arrangement the way any consumer of its output must --- from materialised coordinates --- where construction alone causes disagreement even before any perturbation.

\begin{figure}[t]
\centering
\input{diagrams/sphere_offset.tex}
\caption{Two unit UV spheres ($32{\times}32$), one offset by $d$ along the pole axis, arranged as $d$ is swept over $191$ values. Each row reports a reading of the radial order across the sweep at $\varepsilon = 0$: green where the relation is partitioned correctly, red where it is mispartitioned. Reading the order from the materialised, snap-rounded geometry disagrees at $54$ offsets --- per-edge breaks at every one, while the vote over the same readings is unbroken. Ordering from the original planes (bottom row) never disagrees: there is nothing to recover. The failures cluster near the latitude vertex-ring offsets $d = 2\cos(\pi k/32)$ (axis ticks), where the circle grazes near-coplanar contact and its edges round to disagreeing radial orders --- isolated on the cleanest possible geometry.}
\label{fig:sphere_offset}
\end{figure}
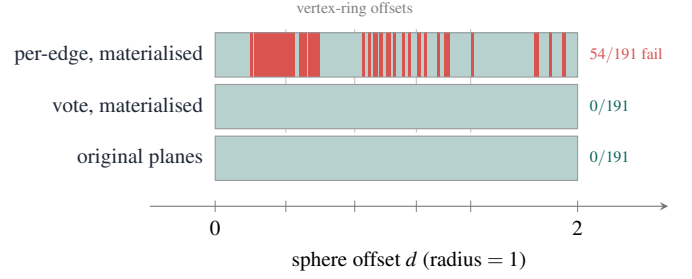

\textbf{A controlled probe.} The mechanism is visible on the simplest geometry there is: two UV spheres of unit radius, one offset by $d$ along the pole axis, arranged as $d$ is swept (Figure~\ref{fig:sphere_offset}). Their intersection is a single circle. At most offsets every reading agrees, but where the circle grazes a latitude vertex ring --- near $d = 2\cos(\pi k/32)$ --- the constructions on different edges of the circle round to different radial orders: reading the order from the materialised, snap-rounded geometry splits the relation's edges at $54$ of $191$ offsets, construction alone, before any perturbation. Per-edge mispartitions the circle at every one of them; the vote, taking the majority, is correct at all $191$. Ordering from the original planes splits none --- the failures are not in the arrangement but in its materialised reading. Nothing here is degenerate by design: two clean, convex, watertight spheres, and the materialised reading still disagrees at more than a quarter of all offsets --- the faces whose radial sort flips are full near-coplanar triangles, not slivers a quality filter could remove.

\textbf{Native failures on real geometry.} The same failure appears unprompted in practice, at $\varepsilon = 0$ with nothing injected. The geological arrangement (Figure~\ref{fig:geological}) --- eleven geomodelling surfaces sharing a bounding box and grazing one another, dense with coplanar contacts --- splits four of its $220$ relations under native materialisation alone, with overwhelming majorities ($9{:}1$ up to $267{:}2$); per-edge misses them, the vote recovers all (Table~\ref{tab:eps0}). On the Thingi10K dataset~\cite{Thingi10K} --- the standard corpus of real, self-intersecting 3D-printing models --- self-arranging each of $59$ models splits $178$ of their $704$ relations, across $25$ of the $59$; per-edge mispartitions every one, the vote none. This is not an artefact of double precision: materialising onto an integer grid instead of to double splits the geological relations the same way --- the disagreement follows materialisation itself, whatever the target representation. Per-edge fails on real data the moment the arrangement is read from its coordinates; the vote does not.

\begin{table}[t]
\centering
\caption{Relations partitioned correctly at $\varepsilon = 0$ --- native materialisation, no added error, double precision. Per-edge mispartitions a relation when any of its edges' sorts disagree; the vote takes the majority. The real geomodelling arrangement splits four of $220$ relations (majorities $9{:}1$ up to $267{:}2$), and the $59$ self-intersecting Thingi10K 3D-printing models~\cite{Thingi10K} split $178$ of $704$ across $25$ models --- per-edge misses them, the vote recovers all. The fifteen standard model/rotated-copy pairs meet transversally and tie. The two-sphere sweep of Figure~\ref{fig:sphere_offset} adds $54$ further materialised-reading failures, the vote none --- and none at all when the order is read from the original planes.}
\label{tab:eps0}
\begin{tabular*}{\columnwidth}{@{\extracolsep{\fill}}lrrcc@{}}
\toprule
configuration & relations & split & per-edge & vote \\
\midrule
geological (real workflow)        & $220$ & $4$   & $216$ & $220$ \\
Thingi10K ($59$ models)           & $704$ & $178$ & $526$ & $704$ \\
$15$ standard model pairs         & $130$ & $0$   & $130$ & $130$ \\
\bottomrule
\end{tabular*}
\end{table}

\begin{figure}[t]
\centering
\input{diagrams/robustness_models.tex}
\caption{Domain-partition correctness under added construction error, in double precision, across the twenty-two model configurations of Section~\ref{sec:results-robustness} (fifteen standard meshes as model/rotated-copy pairs, seven as random-rotation arrangements), $1000$ perturbed trials per $\varepsilon$. Each band is the min--max envelope of correctness over the configurations: per-edge (red) collapses to nothing by $\varepsilon \approx 10^{-3}$ at a model-dependent rate, while the relation vote (green) stays exact through $\varepsilon \approx 10^{-5}$ and correct on at least $78\%$ of relations through $10^{-3}$ --- two to three orders of magnitude more, on every model. The geological model of Figure~\ref{fig:geological} degrades within these same bands. Each added flip emits a wrong merge that per-edge cannot undo; the vote takes the majority.}
\label{fig:robustness}
\end{figure}
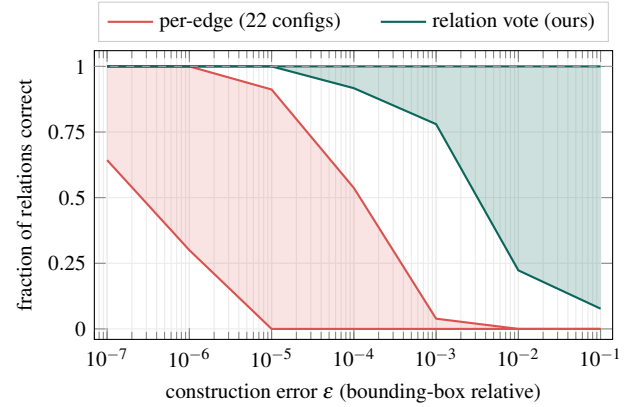

\textbf{Under perturbation.} Native failure is the floor; any composition adds more. An arrangement materialised to floating point, remeshed, and re-extracted accumulates construction error with no exact provenance to recover (Section~\ref{sec:implementation}); once the arrangement exists only as coordinates, the radial order is the one cross-face decision a construction's rounding can reach (Section~\ref{sec:graphs}). We model the round-trip by perturbing the materialised non-manifold-edge endpoints by a uniform box of half-width $\varepsilon$, relative to the bounding box, redrawn over $1000$ trials, and re-running the sort --- single-precision rounding sits near $10^{-7}$ of that scale, double near $10^{-16}$. We sweep $\varepsilon$ over twenty-two configurations spanning fifteen standard meshes (CAD, organic, and scanned), each arranged against a $30^\circ$-rotated copy and, for seven, as dense random-rotation clusters. At $\varepsilon = 0$ the pairs tie: on generic transversal contact the vote invents no false splits; it is never worse than per-edge. Under added error every configuration behaves the same way (Figure~\ref{fig:robustness}): per-edge collapses to nothing by $\varepsilon \approx 10^{-3}$, at a model-dependent rate, while the vote stays correct on $78$ to $100\%$ of relations there --- two to three orders of magnitude more construction error before the partition breaks, on every model. The advantage tracks construction error, not the model.

\textbf{Composition after materialisation.} This margin is what lets operations chain: the materialised entry points of Section~\ref{sec:implementation} hand the labeller an arrangement stripped of its exact history, and the vote alone keeps its partition correct --- the regime L\'evy's exact constructions cannot enter.

\subsection{Performance}
\label{sec:results-performance}

The pipeline builds one arrangement and answers a whole family of booleans from it (Section~\ref{sec:selection}). We measure both halves --- the build, which scales with the input, and the per-expression extraction, which does not --- in two ways. A broad random sample of arbitrary $N$-ary unions isolates how the build scales with operand count. A controlled swarm of Stanford bunnies then exposes the amortisation, where one build answers a whole family of booleans. All timings are on an Apple M4 Max ($12$ performance and $4$ efficiency cores, arm64, $64$\,GB), multi-threaded over TBB. We report the minimum over repeated runs of each deterministic case. The minimum strips operating-system scheduling noise, which only ever inflates a run.

\begin{figure}[t]
  \centering
  \input{diagrams/nary_scaling.tex}
  \caption{\textbf{Operand count, not triangle count, drives the build.} Build time of $1989$ random $N$-ary unions of Thingi10K meshes against total input triangles, one point per union, coloured by operand count $N$ (full geometry-to-build time, M4 Max). The sampling generalises the EMBER protocol~\cite{Trettner2022} to $N$ operands, drawing operand count and triangle budget independently to fill the plane. Total triangles alone is a loose predictor. At any vertical slice the points stratify by $N$, the low-$N$ configurations tracing the geometry floor along the bottom; the build is sub-linear in operand count.}
  \label{fig:nary-scaling}
\end{figure}
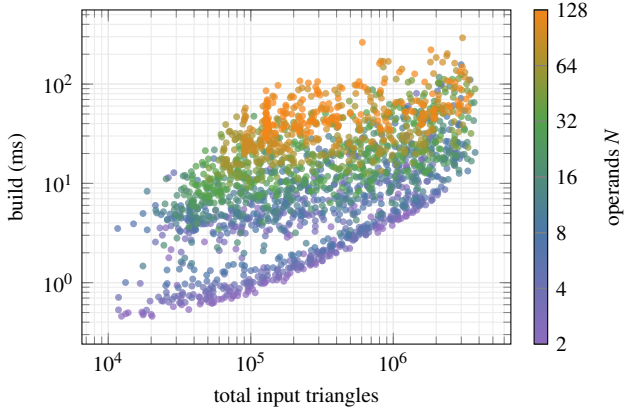

\textbf{Scaling with operand count.} How does the build scale as operands multiply, apart from raw geometry? To answer without bias, we generalise the EMBER protocol~\cite{Trettner2022} to $N$ operands. EMBER benchmarks a boolean by drawing random Thingi10K meshes, normalising each to unit size, and placing them with overlapping bounding boxes. We do the same, but with $N$ operands rather than two. Crucially, each configuration draws its operand count and its total triangle budget \emph{independently}, sampling operands from a size-diverse pool of $369$ meshes ($1$\,k--$1$\,M triangles each). Decoupling count from size fills the (operands, total-triangles) plane, where equal-size operands would trace only a single diagonal. Figure~\ref{fig:nary-scaling} plots the build time of $1989$ such unions against total input triangles, each point coloured by its operand count $N$. Triangle count alone is a weak predictor: it accounts for only $27\%$ of the variance in build time. Operand count is what orders the cloud. At every fixed total the points fan out and stratify by $N$, the $N{=}2$ configurations tracing a clean geometry floor along the bottom. The dependence on operand count is \emph{sub-linear}: at a fixed $\approx\!1$\,M-triangle budget, raising the operand count from $2$ to $128$ --- a $64$-fold increase --- lifts the build only about $10$-fold. The per-domain build cost stays flat to mildly falling, from $\approx\!380$ to $\approx\!115$\,$\mu$s per domain across $N{=}2\to128$. The machinery does not degrade with arity. The build grows only because there are more domains, each of which stays cheap.

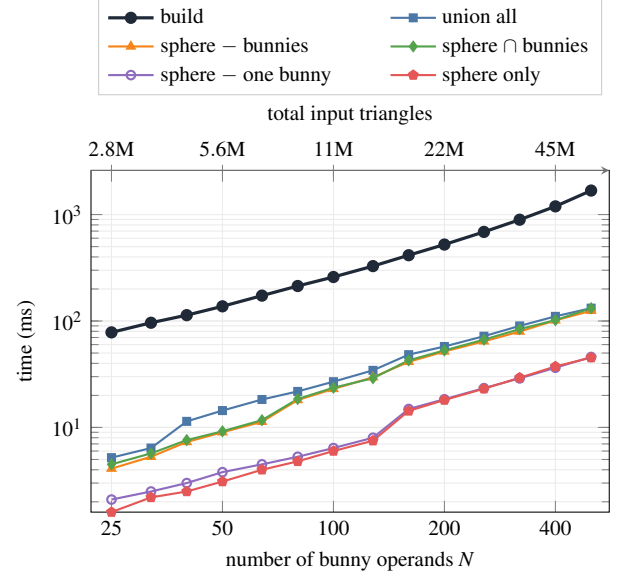
\begin{figure}[t]
  \centering
  \input{diagrams/performance.tex}
  \caption{Build versus extract on the Fibonacci bunny swarm, $N$ copies of the Stanford bunny unioned with a sphere ($N{+}1$ operands), as $N$ grows. The build (one arrangement over all operands) grows sub-quadratically with input size and is the larger cost. The five extraction curves --- \texttt{union\_all}, \texttt{diff\_all}, \texttt{isect\_all}, \texttt{diff\_one}, and the \texttt{sphere\_only} construction floor --- are each the best of five runs of the same extraction; they sit far below the build, in proportion to their output size. The gap between build and extract is the amortisation: one build serves a whole family of booleans. At $N{=}500$ ($56$\,M input triangles) the build is $1.68$\,s and the five extractions $45$--$132$\,ms.}
  \label{fig:performance}
\end{figure}

\begin{figure}[t]
  \centering
  \input{diagrams/stages.tex}
  \caption{Where the build time goes: the arrangement build of Figure~\ref{fig:performance} decomposed across the five algorithm stages, normalised per $N$, the absolute build total printed above each bar (best of five; totals agree with Figure~\ref{fig:performance} to within run-to-run variation). Pairwise intersection dominates throughout ($52$--$59\%$). The arrangement, graph, and partition stages are scale-stable, and the relation-merge vote of Section~\ref{sec:results-robustness} costs under $0.3$\,ms at every $N$. Classification --- per-domain signed-volume seeding --- is the one growing share, from $7\%$ at $N{=}25$ to $23\%$ at $N{=}500$ as the domains grow $190 \to 2404$, the second cost after intersection at the largest swarm.}
  \label{fig:stages}
\end{figure}
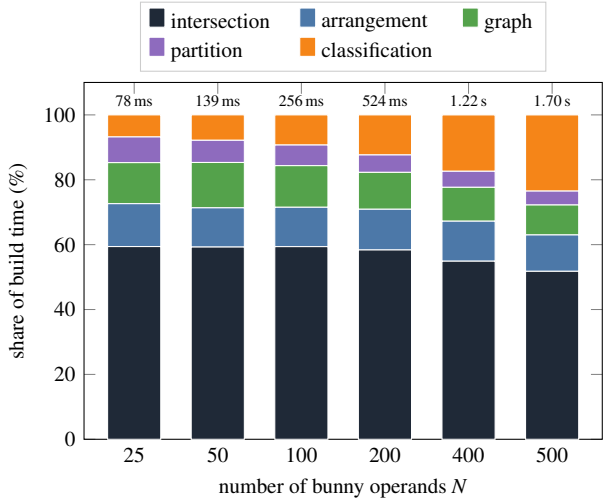

\textbf{Amortising a family of booleans.} The broad sample varies the operands. The swarm instead fixes them and varies the query, exposing the build/extract split. We place $N$ scaled Stanford bunnies on a Fibonacci lattice over a sphere (the swarm of Figure~\ref{fig:teaser}, right), taking the sphere as operand $0$. This gives $N{+}1$ operands and $\approx\!112\text{k}\,N$ input triangles, with $N$ ranging over fourteen values from $25$ to $500$ ($2.8$\,M to $56$\,M triangles). One arrangement is built over all operands, and a fixed family of five extractions is read from it (Table~\ref{tab:extractions}). The build scales with the input, from $78$\,ms at $N{=}25$ to $1.68$\,s at $N{=}500$, at $33$ to $44$ million triangles per second. The AABB-tree broadphase holds the in-principle-quadratic pairwise intersection sub-quadratic in practice. Pairwise intersection dominates the build throughout ($52$--$59\%$), and only classification grows with the domain count, from $7\%$ to $23\%$ (Figure~\ref{fig:stages}). Each extraction then reads the per-domain inclusion bitvectors and walks the boundary between selected and unselected domains. Its cost follows the output size, not the operand count. It ranges from $2$ to $45$\,ms for the \emph{sphere only} surface up to $5$ to $132$\,ms for the full \emph{union}, $13$ to $37\times$ below the build at every $N$ (Figure~\ref{fig:performance}). A set of $k$ expressions therefore costs $\text{build} + k\,\text{extract}$ rather than $k\,(\text{build} + \text{extract})$, the arrangement and its bitvectors built once and reused. Peak memory is linear in the input, $0.21$ to $1.3$\,GB, also paid once and shared across extractions.

\begin{table}[t]
\centering
\caption{A family of booleans read from one arrangement, on the $N{=}500$ bunny swarm ($56$\,M input triangles, $2404$ domains). The build is paid once ($1.68$\,s); each expression is then a per-domain bit test and a boundary extraction, $13$ to $37\times$ cheaper, regardless of its arity. Operands are $O_0$ (the sphere) and $O_1,\dots,O_N$ (the bunnies).}
\label{tab:extractions}
\small
\begin{tabular*}{\columnwidth}{@{\extracolsep{\fill}}llrrr@{}}
\toprule
extraction & expression & out.\ tris & time & vs build \\
\midrule
union all             & $\bigcup_i O_i$                   & $29.1$\,M & $132$\,ms & $13\times$ \\
sphere $-$ bunnies    & $O_0 \setminus \bigcup_{i>0} O_i$ & $20.9$\,M & $125$\,ms & $13\times$ \\
sphere $\cap$ bunnies & $O_0 \cap \bigcup_{i>0} O_i$      & $21.0$\,M & $132$\,ms & $13\times$ \\
sphere $-$ one bunny  & $O_0 \setminus O_1$               & $0.83$\,M & $46$\,ms  & $37\times$ \\
sphere only           & $O_0$                             & $0.77$\,M & $45$\,ms  & $37\times$ \\
\bottomrule
\end{tabular*}
\end{table}

\textbf{Volumetric domains.} The same build answers a different read entirely --- volumetric regions rather than a boolean (Section~\ref{sec:selection}). The cellular fracture of Figure~\ref{fig:teaser} (left) --- a Stanford bunny against $60$ axis-aligned cutting planes, $112$k input triangles --- is arranged and partitioned into its $8665$ bounded domains in $59$\,ms (arrangement $25$\,ms, cleanup $6$\,ms, domain labelling $28$\,ms); the $3459$ cells interior to the bunny are then read straight from the partition, with no boolean and no per-cell work. This figure is produced by a repository example \texttt{csg\_fracture.cpp} \cite{trueform2025}.

\textbf{Interactive in the browser.} The WebAssembly build runs the same pipeline client-side in the browser. A live demo\footnote{\texttt{trueform.polydera.com/live-examples/boolean}} differences a Stanford dragon against a sphere, $\approx\!550$k triangles, re-evaluating the boolean every frame as the sphere is dragged through the dragon. Each frame rebuilds the arrangement and re-extracts the difference in $12$\,ms on an M4 Max, on desktop and mobile alike.

\subsection{Comparison with prior art}
\label{sec:results-scaling}

We measure performance against representatives spanning the prior approaches: Geogram~\cite{Geogram,Levy2025} (exact predicates and constructions), EMBER~\cite{Trettner2022} (exact constructions via octree-embedded BSP, specialised for iterated CSG; implementation~\cite{Solidean}), CGAL~\cite{CGAL2023} corefinement~\cite{CGALCorefinement} (exact predicates, inexact constructions), the mesh-arrangement booleans of Cherchi et al.~\cite{Cherchi2022} (exact via indirect predicates; implementation~\cite{CherchiImpl}), MeshLib~\cite{MeshLib} (Simulation of Simplicity for degeneracy handling; proprietary core), and Manifold~\cite{Manifold} (deterministic floating point, degeneracies resolved by symbolic perturbation on exact equality); for the in-browser comparison we run our own and Manifold's WebAssembly builds and add three-bvh-csg~\cite{ThreeBvhCsg}, the CSG evaluator for three.js. We compare on three fronts: a single pairwise boolean (Section~\ref{sec:results-pairwise}), $N$-ary and iterated booleans (Section~\ref{sec:results-nary}), and the same operations run client-side in the browser (Section~\ref{sec:results-browser}).

\textbf{Scope.} The baselines all compute booleans only between \emph{closed, manifold} solids: each operand must bound a volume. The comparison is therefore confined to that case --- the booleans every library supports. The remainder of our pipeline --- domain extraction against open surfaces, arrangements of triangle soup, $N$-mesh cuts, embedded curves on non-manifold input --- has no counterpart to compare against and is demonstrated in Sections~\ref{sec:results-robustness}--\ref{sec:results-performance}; we restrict the head-to-head to the operations the prior art is built for, and report them on the same machine and inputs for all libraries.

\textbf{Protocol.} Each method receives, for every operand, a vertex-coordinate array (float triples) and a triangle-index array (integer triples) --- the form its own import entry point accepts --- and is timed until it returns the result as that same pair of arrays. Everything between is inside the timer: import, the construction of whatever spatial and topological structures the algorithm uses, the boolean, classification, and the emission of the result's coordinate and index arrays --- the result first exists only inside the library's own representation, and emitting it to arrays is part of the measured work. Only file I/O --- parsing the OBJ into the input arrays --- is outside, and no method amortises: each rebuilds every structure on every operation, ours included. The one relaxation is the $N$-ary and iterated comparison (Section~\ref{sec:results-nary}), where a method that supports it may keep its result inside its own representation across the chain of operations and emit the arrays only once, at the end. We report the minimum of $K$ repetitions per operation. All runs are on an Apple M4 Max with multithreading enabled for every implementation; the drivers are compiled with \texttt{clang -O3 -mcpu=native}. Library versions: our method is trueform~$0.9.8$~\cite{trueform2025}; Geogram~$1.10$, CGAL~$6.1$ (the \texttt{EPICK} kernel --- exact predicates, double constructions --- the fast configuration used for corefinement), Cherchi et al.\ (reference implementation~\cite{CherchiImpl}), MeshLib~$3.1$, Manifold~(\texttt{manifold3d})~$3.5$, EMBER (Solidean Community~\cite{Solidean}, preview build \texttt{2026-04-07-be517c}). Each baseline is built and run as shipped on the host above.

\textbf{Rationale.} A mesh-boolean library is staged: it ingests the input, builds the internal representation its algorithm needs --- spatial acceleration, topological connectivity, an exact-arithmetic context --- runs the boolean, and emits an output mesh. ``The boolean'' can name the whole path or only the middle, with the structures already in hand, and prior comparisons rarely say which. The difference is a factor, not a rounding: for our method the spatial and topological structures alone are $20$--$40\%$ of a single pairwise boolean ($28\%$ at the geometric mean, untabulated), and they are reusable --- built once, they serve a whole sequence of operations over moving geometry. The output is the same ambiguity at the other end: every method holds its result in an internal representation before emitting it --- ours an implicit arrangement graph from which the surface is extracted --- and charging or not charging that emission again moves the number. Anchoring the timed boundary at the data --- arrays in, arrays out --- removes both: there is no internal stage to include or omit, and every method is measured over the identical, complete task a user performs. Several libraries could do less if we let them --- Manifold builds its collider at construction, MeshLib can cache half-edge connectivity, EMBER can pre-build its exact operands, and we can reuse our structures --- but the head-to-head denies that to all uniformly; the $N$-ary relaxation is the one place where staying inside the representation reflects the operation itself, not a measurement choice.

\textbf{Validity.} Every output is graded by one oracle on the emitted floating-point arrays: the result must be a watertight solid --- closed and consistently oriented --- whose signed volume and surface area match the boolean's reference values to floating-point tolerance. Every method computes the same solid, so the grade turns on whether each returns it as a closed, watertight mesh. ``Valid'' in the tables below means the output meets this bar, and the same oracle grades every comparison.

\textbf{Corpus.} Following the EMBER protocol~\cite{Trettner2022}, we draw random pairs of solid, manifold, non-self-intersecting Thingi10K~\cite{Thingi10K} meshes, normalise each to unit extent, apply an independent random rotation, and offset the second so the bounding boxes overlap. Each pair is assigned a random operation --- union, intersection, or difference --- that every method computes. We sample at industrial resolution --- $100\mathrm{k}$ to $1\mathrm{M}$ triangles per operand --- the scale at which mesh booleans are a real cost in production pipelines. Pairs are baked to indexed OBJ so every driver reads byte-identical input.

\subsubsection{Pairwise booleans}
\label{sec:results-pairwise}

\begin{figure}[t]
  \centering
  \input{diagrams/comparison_pairwise.tex}
  \caption{\textbf{Pairwise booleans against prior art.} Per-operation wall-clock time (log scale) over 1000 random Thingi10K pairs at $100\mathrm{k}$--$1\mathrm{M}$ triangles, one boolean per pair, full geometry-to-output timing under the protocol above (best of $K$, M4 Max). Each violin is the distribution over the corpus; the bar and dot mark the geometric mean, the label its slowdown over our method. We are fastest on every one of the $1000$ pairs: $5.5\times$, $7.6\times$, $10.4\times$, $29\times$, $32\times$, and $100\times$ at the geometric mean over MeshLib, Manifold, EMBER, Cherchi et al., CGAL, and Geogram respectively.}
  \label{fig:comparison-pairwise}
\end{figure}
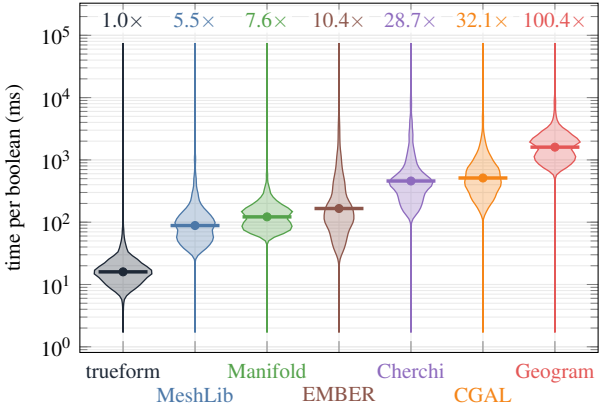

Figure~\ref{fig:comparison-pairwise} reports the per-boolean wall-clock distribution over the corpus~\cite{trueform2025}. Our method is fastest on every one of the $1000$ pairs. At the geometric mean it runs $5.5\times$ faster than MeshLib, $7.6\times$ faster than Manifold, $10.4\times$ faster than EMBER, $29\times$ faster than Cherchi et al., $32\times$ faster than CGAL, and $100\times$ faster than Geogram.

\begin{table}[t]
\centering
\caption{Pairwise boolean over the $1000$-pair corpus of Section~\ref{sec:results-scaling} ($100$k--$1$M triangles per operand), one boolean per pair, on an Apple M4 Max. Median wall-clock, geometric-mean slowdown relative to our method, and valid outputs, graded by the oracle of Section~\ref{sec:results-scaling}; CGAL's single timeout is reported in the text.}
\label{tab:pairwise}
\small
\input{diagrams/comparison_pairwise_table.tex}
\end{table}

Every method computes the same solid: the signed volumes and areas agree to floating-point tolerance on every pair, and each returns it as a watertight mesh (Table~\ref{tab:pairwise}). Validity does not separate the methods here --- only MeshLib misses, on a single pair, and CGAL on the one pair it fails to finish within the cap (below). The comparison turns on speed.

Completion is not guaranteed either. CGAL's corefinement can blow up super-polynomially on near-degenerate contacts, its exact predicates falling repeatedly to expensive exact evaluation: on one corpus pair --- Thingi10K models $844209$ and $518083$, $0.86$\,M input triangles --- CGAL ran for over five hours ($315$ minutes of CPU on a single core) without producing a result before we terminated it, while our method returned the same result, watertight, in $20$\,ms. We therefore cap each operation at $120$\,s and exclude from a method's distribution any pair it fails to finish within it; CGAL timed out on this one pair, and every other method on none.

\subsubsection{N-ary and iterated booleans}
\label{sec:results-nary}

A single $N$-ary boolean measures how a method scales in the number of operands. For our method it is one arrangement build: all $N$ operands are arranged together and the result is read from the resulting domain inclusion (Section~\ref{sec:results-performance}), so its cost follows the sub-linear scaling reported there. A library with no native $N$-ary operator chains $N{-}1$ pairwise booleans, the result accumulating one operand at a time; the native-$N$-ary engines --- Manifold, Cherchi, and Geogram --- arrange all operands at once, as we do. EMBER sits between: it accumulates pairwise, but keeps each intermediate inside its own representation across the chain --- the relaxation the protocol grants (Section~\ref{sec:results-scaling}) --- so it re-imports no growing mesh and tracks the native engines rather than climbing with the chain.

We draw $40$ random $N$-operand sets for each $N \in \{4, 16, 64\}$ from the Thingi10K pool~\cite{trueform2025}, each operand randomly rotated and placed in an overlapping cluster, and time the single boolean of each set under the protocol of Section~\ref{sec:results-scaling} (full geometry-to-output, best of $K$, $120$\,s cap, multithreading enabled). CGAL, like MeshLib, has no native $N$-ary operator and accumulates pairwise; at this scale it is too slow to complete the sweep within the cap, so it is omitted here and its pairwise cost reported in Section~\ref{sec:results-pairwise}.

\begin{figure}[t]
  \centering
  \input{diagrams/comparison_nary.tex}
  \caption{\textbf{$N$-ary booleans against prior art.} Median operation time relative to our method against operand count $N$ (log scale on both axes), over $40$ random Thingi10K $N$-operand sets per $N$, full geometry-to-output timing under the protocol of Section~\ref{sec:results-scaling} (best of $K$, M4 Max). EMBER is the closest baseline to us, accumulating pairwise yet keeping its intermediates internal so it does not grow with the chain; Manifold and Cherchi arrange all operands at once and hold a fixed factor too, while MeshLib, also accumulating but carrying a materialised mesh, grows with the chain. CGAL accumulates pairwise as MeshLib does, but is too slow to complete the sweep within the cap and is excluded.}
  \label{fig:comparison-nary}
\end{figure}
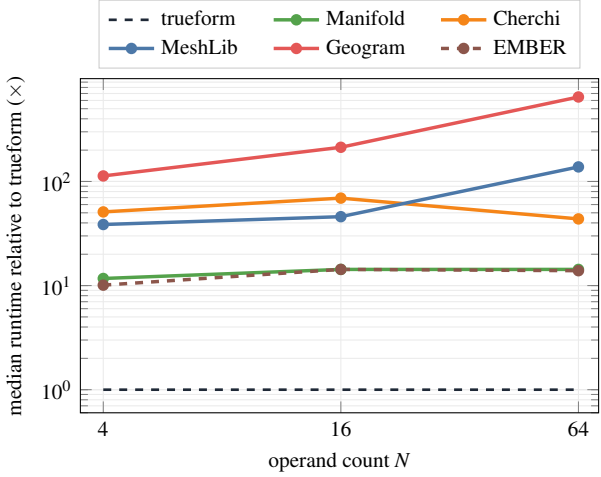

Figure~\ref{fig:comparison-nary} reports each method's median operation time relative to ours as $N$ grows; Table~\ref{tab:nary} gives the absolute medians (printed to the millisecond; quoted factors use the unrounded values) and validity. Three methods stay near a fixed factor across the range. EMBER is the closest to us, $10$ to $14\times$: it accumulates pairwise but keeps its intermediates internal, so it does not grow with the chain. Of the native-$N$-ary engines, Manifold and Cherchi hold their factor too, at $12$ to $14\times$ and $44$ to $69\times$. The accumulator that carries a materialised mesh climbs with the chain: MeshLib runs $39\times$ slower than ours at $N{=}4$ and $138\times$ at $N{=}64$, the mesh it re-imports growing at every step. Geogram, though it arranges all operands at once, scales worst in absolute terms, from $113\times$ at $N{=}4$ to $648\times$ at $N{=}64$, as its exact arithmetic grows with the combined complexity. All methods agree on signed volume and area on every set, so each computes the same solid, and every method returns a watertight solid on every set at every $N$ --- save MeshLib, on two sets at $N{=}16$ and three at $N{=}64$ (Table~\ref{tab:nary}). As in the pairwise case (Section~\ref{sec:results-pairwise}), validity does not separate the methods; the comparison is one of speed.

\begin{table}[t]
\centering
\caption{$N$-ary boolean: median wall-clock and validity per operand count $N$, over $40$ random Thingi10K $N$-operand sets per $N$. Validity counts, out of the $40$ per $N$, the sets the oracle of Section~\ref{sec:results-scaling} grades valid; a set that times out or errors counts invalid. CGAL is excluded: it accumulates pairwise, like MeshLib, but is too slow to complete the sweep within the cap.}
\label{tab:nary}
\small
\input{diagrams/comparison_nary_table.tex}
\end{table}

\subsubsection{In the browser}
\label{sec:results-browser}

The same pipeline runs client-side in the WebAssembly build (Section~\ref{sec:results-performance}). We measure it against the two booleans deployed for the browser: Manifold~\cite{Manifold}, whose WebAssembly build is a common in-browser solid-modelling kernel, and three-bvh-csg~\cite{ThreeBvhCsg}, the BVH-based CSG evaluator for three.js. Each is built and run as shipped. We run the pairwise corpus of Section~\ref{sec:results-scaling} ($100$k--$1$M triangles per operand), one boolean per pair, in Chrome~$148$ on the same Apple M4 Max, and grade every output with the oracle of Section~\ref{sec:results-scaling}.

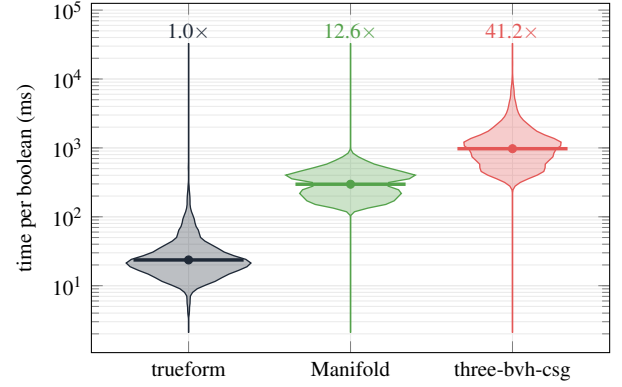
\begin{figure}[t]
  \centering
  \input{diagrams/comparison_browser.tex}
  \caption{\textbf{In-browser pairwise booleans.} Per-operation wall-clock time (log scale) for the WebAssembly/JavaScript build of each library, over the corpus of Section~\ref{sec:results-scaling} ($100$k--$1$M triangles per operand), one boolean per pair, in Chrome~$148$ on an Apple M4 Max. Each violin is the distribution over the corpus; the bar and dot mark the geometric mean, the label its slowdown over our method. Our method and Manifold produce a valid (watertight solid) result on all $1000$ pairs; three-bvh-csg on only $22$.}
  \label{fig:comparison-browser}
\end{figure}

Figure~\ref{fig:comparison-browser} and Table~\ref{tab:browser} report the per-operation wall-clock and validity. Our method is fastest; at the geometric mean it runs $12.6\times$ faster than Manifold and $41.2\times$ faster than three-bvh-csg. Running in WebAssembly rather than native costs our method about $1.4\times$ ($15.7$\,ms native to $21.9$\,ms in the browser on the same corpus), so the desktop ranking carries over and the exact pipeline stays interactive client-side. In fact, our method is faster in the browser than any baseline runs natively: its $21.9$\,ms median sits below MeshLib's native median ($86$\,ms), the fastest of the prior methods. Validity separates them here. Our method and Manifold return a watertight solid on all $1000$ pairs; three-bvh-csg on only $22$ --- on the other $978$ it returns an open mesh, not a closed solid at all.

\begin{table}[t]
\centering
\caption{In-browser pairwise boolean over the corpus of Section~\ref{sec:results-scaling} ($1000$ pairs, $100$k--$1$M triangles per operand), each library's WebAssembly/JavaScript build as shipped, Chrome~$148$ on an Apple M4 Max. Median wall-clock, geometric-mean slowdown relative to our method, and valid outputs, graded by the oracle of Section~\ref{sec:results-scaling}.}
\label{tab:browser}
\small
\begin{tabular*}{\columnwidth}{@{\extracolsep{\fill}}lrrr@{}}
\toprule
library & median (ms) & geomean $\times$ & valid \\
\midrule
trueform      & $21.9$  & $1.0\times$   & $1000/1000$ \\
Manifold      & $303.4$ & $12.6\times$  & $1000/1000$ \\
three-bvh-csg & $978$   & $41.2\times$  & $22/1000$   \\
\bottomrule
\end{tabular*}
\end{table}

\section{Conclusions}
\label{sec:conclusions}

We have presented a mesh-CSG pipeline organised around a single fact: a CSG result loses exactness the moment it materialises, so a stage meant to compose with the rest of a pipeline can keep only its topology exact. The arrangement is computed locally with integer-exact predicates, every stage a graph problem on graphs it never explicitly constructs; where a decision spans faces, the disagreeing \texttt{orient3d} observations that materialisation perturbs are resolved by topological aggregation --- a weighted-majority vote within their topological unit, the maximum-a-posteriori estimate of the unit's intended outcome. The arrangement and its domain partition are built once, and a boolean of any arity is then a per-domain bit test, so a single build answers a whole family of CSG queries. The \emph{trueform} implementation runs the same exact kernel in C\texttt{++}, Python, and the browser.

The vote is what makes the stage composable. An exact-construction pipeline cannot redo its arithmetic on a materialised arrangement --- the regime in which a CSG step actually composes --- whereas the vote keeps the partition correct from the observations alone: it tolerates two to three orders of magnitude more construction error than per-edge extraction, and recovers the failures construction alone already causes when the order is read from materialised coordinates on clean, watertight input --- failures the build itself no longer makes, its radial order decided on the original planes without exact constructions. Against prior art the pipeline is the fastest method in the comparison --- $5.5\times$ to $100\times$ at the geometric mean, depending on the baseline, and, compiled to WebAssembly, faster in the browser than any baseline runs natively.

\subsection{Limitations and future work}

The kernel operates on convex faces, fan-triangulated online; non-convex inputs, rare in mesh CSG, are triangulated upstream. Coordinates are snapped onto one of two integer grids (\texttt{int32} or \texttt{int64}), chosen from the input type, so inputs beyond the wider grid's range must be rescaled.

Two limitations are intrinsic to the approach. First, the vote resolves the \emph{partition} --- which domain each face bounds --- from observations that have already been materialised; it does not constrain the geometry on the far side of materialisation. Materialising the output can introduce self-intersections the exact arrangement did not have: the partition stays correct as a combinatorial object, but the coordinates carrying it are not guaranteed crossing-free, and a stage that re-arranges the result may find new intersections. We make the topology robust to the crossing; we do not remove the crossing itself. Second, the vote is a majority estimator, and its guarantee is statistical, not worst-case. It recovers the intended order when the corrupted observations are a minority within their unit --- the right model for construction error, which flips a few edges of a relation --- but a configuration can be built to defeat it: a near-coplanar interface engineered so that, once materialised, more than half of a relation's edges round to the wrong side leaves the majority itself corrupted, and the vote returns the wrong order. The method is robust against the error a real pipeline injects, not against an adversary who controls the geometry.

Making materialisation lossless is the natural next step: a snap-rounding of the output that introduces no new intersections, bounding the first limitation above, would let exactness and composability hold together rather than trade off. It is the open problem L\'evy~\cite{Levy2025} and Cherchi~\cite{Cherchi2022} both leave standing, and the one piece our approach does not yet supply.

\section*{Acknowledgements}
We thank Luna Sajovic for her help designing the figures, David Coeurjolly for his early review of the manuscript, and Philip Trettner for providing the Solidean preview binary used in our comparison.

\appendix
\section{Aggregation as maximum-a-posteriori estimation}
\label{app:bayesian}

The aggregation of Sections~\ref{sec:aggregation} and~\ref{sec:classification} is an estimator, and writing it as one makes precise why the majority is the right rule. Per-face exact predicates lift into a globally trusted topology, the reduced graph $\mathcal{R}$ (Section~\ref{sec:graphs}), which is exact and fixed. The geometric relations between its wedges --- the radial order at each non-manifold edge, the nesting between disjoint connected components --- are not: each is an \emph{observation}, a categorical reading of one intended topological outcome.

\paragraph*{Model.} Fix a topological unit with a finite categorical space $\mathcal{C}$ of size $K$: for a component the candidate orientations, for a relation the canonical radial permutations. The unit has one intended category, the latent $c^\star \in \mathcal{C}$. Its $n$ observations $g_1, \dots, g_n \in \mathcal{C}$ are independent categorical reads of an unknown category distribution $\theta \in \Delta^{K-1}$ whose mode is $c^\star$; the noise perturbs the reads, not the intent --- for the radial vote, a materialised re-reading, a mirrored direction sign, or an inconsistently wound input sheet. With $N_k = \lvert\{\, i : g_i = k \,\}\rvert$ the count of category $k$, the likelihood is
\begin{equation}
  P(G \mid \theta) \;=\; \prod_{k=1}^{K} \theta_k^{\,N_k}.
\end{equation}

\paragraph*{Prior and posterior.} A symmetric Dirichlet prior $\theta \sim \mathrm{Dir}(\alpha, \dots, \alpha)$ favours no category,
\begin{equation}
  P(\theta) \;\propto\; \prod_{k=1}^{K} \theta_k^{\,\alpha - 1},
\end{equation}
and conjugacy makes the posterior a Dirichlet whose parameters are the counts plus the prior,
\begin{equation}
  P(\theta \mid G) \;=\; \mathrm{Dir}\!\bigl(N_1 + \alpha, \dots, N_K + \alpha\bigr).
\end{equation}
The intended category is the most probable one, estimated at the posterior mode
\begin{equation}
  \hat\theta_k \;=\; \frac{N_k + \alpha - 1}{\,n + K(\alpha - 1)\,},
  \qquad
  \hat c \;=\; \arg\max_k \hat\theta_k \;=\; \arg\max_k N_k .
\end{equation}
For $\alpha \ge 1$ this expression applies; the normaliser and the constant $\alpha - 1$ are then common to every category, so they leave the argmax untouched: the MAP category is the count argmax, the majority, independent of $\alpha$.

\paragraph*{Weights.} Replacing each unit count by a weighted count $W_k = \sum_{i :\, g_i = k} w_i$ scales the pseudocounts with weight and leaves the form unchanged, $\hat c = \arg\max_k W_k$. Trueform weights each radial observation by the minimum area of the faces incident to its non-manifold edge, so near-degenerate slivers --- the reads most likely to be wrong --- contribute least, and observations from invalid fans carry $w_i = 0$ and drop out.

\paragraph*{Epistemic status.} The Dirichlet model is a subjective prior: it makes the pipeline's uncertainty over a relation's intended permutation precise, and identifies the majority as the coherent estimate under it. It claims nothing about objective chances, and none is needed --- the operative guarantee is definitional: the vote returns the intended permutation exactly when a majority of the relation's observations carry it.

\paragraph*{Scope.} The framing adds no machinery to the algorithms of Sections~\ref{sec:aggregation} and~\ref{sec:classification}: the MAP estimate under a uniform Dirichlet \emph{is} the weighted majority they already compute, run on two nested units --- a component (orientation) and a relation (merging) --- each instantiating $\mathcal{C}$; the domain inherits its inside/outside from the voted wedges that bound it, with no further estimate. Its value is in stating the assumption plainly: the intended geometry is a single latent category per unit, and materialisation's perturbations are independent categorical noise around it. The independence is an idealisation --- adjacent edges of a relation share an endpoint, so neighbouring reads are weakly correlated --- but the correlation is local and does not bias the majority, which aggregates over all the relation's edges. Prior pipelines treat the same observations differently: L\'evy raises precision until they agree~\cite{Levy2025}, B\"ohm and Runge trust each edge in isolation~\cite{BohmRunge2025}; naming the rule as estimation is what lets one vote serve orientation and merging alike, with the domain classification inheriting from them.

\bibliographystyle{eg-alpha-doi}
\bibliography{references}

\end{document}

%% file: diagrams/observed_wedge.tex
% Observed wedge: intended -> three quantization buckets -> two manifestations.
% Used in section sec:observations to motivate set-valued FP uncertainty.

\def\sclI{0.55}
\def\panelw{4.0}
\def\panelh{4.6}
\def\panelgap{0.4}

% Isometric projection from 3D (x,y,z) to 2D screen.
% +x and +y come toward the viewer; +z is up.
\newcommand{\iso}[3]{({((#1) - (#2)) * 0.866 * \sclI}, {((#3) - ((#1) + (#2)) * 0.5) * \sclI})}

% Apex positions:
%   T1 at angle   0 deg, radius 1.3:  (+1.300,  0.000, 0)   cOne
%   T2 at angle 110 deg, radius 1.3:  (-0.445,  1.222, 0)   cTwo
%   T3 at angle 135 deg, radius 1.3:  (-0.919,  0.919, 0)   cThree (close to T2)
%   T4 near-degenerate (tiny):
%       intended apex ( 0.000, -0.200, 0)  cFive
%       flipped  apex ( 0.000,  0.200, 0)  cFive (Panel 4)
% Edge e: (0,0,-1.5) -- (0,0,+1.5)
% Buckets at T2, T3, T4 apices, half-extent 0.30 each.
% T2 and T3 buckets overlap in 3D. T4's bucket crosses the y=0 plane,
% so T4's apex can land on either side of the edge.

\begin{tikzpicture}[font=\sffamily\footnotesize, x=1cm, y=1cm]

  % =====================================================================
  % PANEL 1 -- Intended
  % =====================================================================
  \begin{scope}[shift={(0,0)}]
    \filldraw[fill=black!3, draw=black!20, line width=0.4pt, rounded corners=3pt]
      (-\panelw/2, -\panelh/2) rectangle (\panelw/2, \panelh/2);

    % T4 (back-most by depth) -- drawn first
    \fill[cFive, opacity=0.50]
      \iso{0}{0}{-1.5} -- \iso{0}{0}{1.5} -- \iso{0}{-0.20}{0} -- cycle;
    \draw[cFive, line width=0.7pt, opacity=0.9]
      \iso{0}{0}{-1.5} -- \iso{0}{0}{1.5} -- \iso{0}{-0.20}{0} -- cycle;

    % T3 -- back-upper-left
    \fill[cThree, opacity=0.50]
      \iso{0}{0}{-1.5} -- \iso{0}{0}{1.5} -- \iso{-0.919}{0.919}{0} -- cycle;
    \draw[cThree, line width=0.7pt, opacity=0.9]
      \iso{0}{0}{-1.5} -- \iso{0}{0}{1.5} -- \iso{-0.919}{0.919}{0} -- cycle;

    % T2 -- front-upper-left
    \fill[cTwo, opacity=0.50]
      \iso{0}{0}{-1.5} -- \iso{0}{0}{1.5} -- \iso{-0.445}{1.222}{0} -- cycle;
    \draw[cTwo, line width=0.7pt, opacity=0.9]
      \iso{0}{0}{-1.5} -- \iso{0}{0}{1.5} -- \iso{-0.445}{1.222}{0} -- cycle;

    % T1 -- front-right
    \fill[cOne, opacity=0.50]
      \iso{0}{0}{-1.5} -- \iso{0}{0}{1.5} -- \iso{1.3}{0}{0} -- cycle;
    \draw[cOne, line width=0.7pt, opacity=0.9]
      \iso{0}{0}{-1.5} -- \iso{0}{0}{1.5} -- \iso{1.3}{0}{0} -- cycle;

    % Edge e
    \draw[black!85, line width=1.0pt]
      \iso{0}{0}{-1.5} -- \iso{0}{0}{1.5};
    \fill[black] \iso{0}{0}{-1.5} circle [radius=1.5pt];
    \fill[black] \iso{0}{0}{1.5} circle [radius=1.5pt];

    % Apex dots
    \fill[cOne]   \iso{1.3}{0}{0}        circle [radius=1.7pt];
    \fill[cTwo]   \iso{-0.445}{1.222}{0} circle [radius=1.7pt];
    \fill[cThree] \iso{-0.919}{0.919}{0} circle [radius=1.7pt];
    \fill[cFive]  \iso{0}{-0.20}{0}      circle [radius=1.7pt];

    % Labels
    \node[text=cOne,   font=\sffamily\small\bfseries, anchor=west]
      at \iso{1.55}{0}{0}        {$T_1$};
    \node[text=cTwo,   font=\sffamily\small\bfseries, anchor=east]
      at \iso{-0.633}{1.738}{0}  {$T_2$};
    \node[text=cThree, font=\sffamily\small\bfseries, anchor=east]
      at \iso{-1.308}{1.308}{0}  {$T_3$};
    \node[text=cFive,  font=\sffamily\small\bfseries, anchor=west]
      at \iso{0.7}{-0.6}{0}      {$T_4$};

    % Edge label
    \node[font=\sffamily\small, anchor=south]
      at \iso{0}{0}{1.7} {$e$};

    % Panel header
    \node[anchor=north, font=\sffamily\small\itshape] at (0, \panelh/2 - 0.25) {intended};
  \end{scope}

  % =====================================================================
  % PANEL 2 -- Three buckets: T2 & T3 overlap; T4 straddles edge plane
  % =====================================================================
  \begin{scope}[shift={(\panelw + \panelgap, 0)}]
    \filldraw[fill=black!3, draw=black!20, line width=0.4pt, rounded corners=3pt]
      (-\panelw/2, -\panelh/2) rectangle (\panelw/2, \panelh/2);

    % T4 (back-most)
    \fill[cFive, opacity=0.50]
      \iso{0}{0}{-1.5} -- \iso{0}{0}{1.5} -- \iso{0}{-0.20}{0} -- cycle;
    \draw[cFive, line width=0.7pt, opacity=0.9]
      \iso{0}{0}{-1.5} -- \iso{0}{0}{1.5} -- \iso{0}{-0.20}{0} -- cycle;

    % T4's bucket: center (0, -0.20, 0), half-extent 0.30
    % Spans x: [-0.30, 0.30], y: [-0.50, 0.10], z: [-0.30, 0.30].
    % Crosses y=0 by 0.10.
    % Visible faces in iso: top (z=+0.30), +x face (x=+0.30), +y face (y=+0.10).
    \filldraw[fill=cFive, fill opacity=0.30, draw=cFive, line width=0.45pt]
         \iso{-0.30}{-0.50}{0.30}
      -- \iso{0.30}{-0.50}{0.30}
      -- \iso{0.30}{0.10}{0.30}
      -- \iso{-0.30}{0.10}{0.30}
      -- cycle;
    \filldraw[fill=cFive, fill opacity=0.22, draw=cFive, line width=0.45pt]
         \iso{0.30}{-0.50}{0.30}
      -- \iso{0.30}{0.10}{0.30}
      -- \iso{0.30}{0.10}{-0.30}
      -- \iso{0.30}{-0.50}{-0.30}
      -- cycle;
    \filldraw[fill=cFive, fill opacity=0.22, draw=cFive, line width=0.45pt]
         \iso{-0.30}{0.10}{0.30}
      -- \iso{0.30}{0.10}{0.30}
      -- \iso{0.30}{0.10}{-0.30}
      -- \iso{-0.30}{0.10}{-0.30}
      -- cycle;

    % T3 -- back-upper-left
    \fill[cThree, opacity=0.50]
      \iso{0}{0}{-1.5} -- \iso{0}{0}{1.5} -- \iso{-0.919}{0.919}{0} -- cycle;
    \draw[cThree, line width=0.7pt, opacity=0.9]
      \iso{0}{0}{-1.5} -- \iso{0}{0}{1.5} -- \iso{-0.919}{0.919}{0} -- cycle;

    % T3's bucket: center (-0.919, 0.919, 0), half-extent 0.30
    \filldraw[fill=cThree, fill opacity=0.30, draw=cThree, line width=0.45pt]
         \iso{-1.219}{0.619}{0.30}
      -- \iso{-0.619}{0.619}{0.30}
      -- \iso{-0.619}{1.219}{0.30}
      -- \iso{-1.219}{1.219}{0.30}
      -- cycle;
    \filldraw[fill=cThree, fill opacity=0.22, draw=cThree, line width=0.45pt]
         \iso{-0.619}{0.619}{0.30}
      -- \iso{-0.619}{1.219}{0.30}
      -- \iso{-0.619}{1.219}{-0.30}
      -- \iso{-0.619}{0.619}{-0.30}
      -- cycle;
    \filldraw[fill=cThree, fill opacity=0.22, draw=cThree, line width=0.45pt]
         \iso{-1.219}{1.219}{0.30}
      -- \iso{-0.619}{1.219}{0.30}
      -- \iso{-0.619}{1.219}{-0.30}
      -- \iso{-1.219}{1.219}{-0.30}
      -- cycle;

    % T2 -- front-upper-left
    \fill[cTwo, opacity=0.50]
      \iso{0}{0}{-1.5} -- \iso{0}{0}{1.5} -- \iso{-0.445}{1.222}{0} -- cycle;
    \draw[cTwo, line width=0.7pt, opacity=0.9]
      \iso{0}{0}{-1.5} -- \iso{0}{0}{1.5} -- \iso{-0.445}{1.222}{0} -- cycle;

    % T2's bucket: center (-0.445, 1.222, 0), half-extent 0.30
    \filldraw[fill=cTwo, fill opacity=0.30, draw=cTwo, line width=0.45pt]
         \iso{-0.745}{0.922}{0.30}
      -- \iso{-0.145}{0.922}{0.30}
      -- \iso{-0.145}{1.522}{0.30}
      -- \iso{-0.745}{1.522}{0.30}
      -- cycle;
    \filldraw[fill=cTwo, fill opacity=0.22, draw=cTwo, line width=0.45pt]
         \iso{-0.145}{0.922}{0.30}
      -- \iso{-0.145}{1.522}{0.30}
      -- \iso{-0.145}{1.522}{-0.30}
      -- \iso{-0.145}{0.922}{-0.30}
      -- cycle;
    \filldraw[fill=cTwo, fill opacity=0.22, draw=cTwo, line width=0.45pt]
         \iso{-0.745}{1.522}{0.30}
      -- \iso{-0.145}{1.522}{0.30}
      -- \iso{-0.145}{1.522}{-0.30}
      -- \iso{-0.745}{1.522}{-0.30}
      -- cycle;

    % T1 -- front-right
    \fill[cOne, opacity=0.50]
      \iso{0}{0}{-1.5} -- \iso{0}{0}{1.5} -- \iso{1.3}{0}{0} -- cycle;
    \draw[cOne, line width=0.7pt, opacity=0.9]
      \iso{0}{0}{-1.5} -- \iso{0}{0}{1.5} -- \iso{1.3}{0}{0} -- cycle;

    % Edge e
    \draw[black!85, line width=1.0pt]
      \iso{0}{0}{-1.5} -- \iso{0}{0}{1.5};
    \fill[black] \iso{0}{0}{-1.5} circle [radius=1.5pt];
    \fill[black] \iso{0}{0}{1.5} circle [radius=1.5pt];

    % Apex dots
    \fill[cOne]   \iso{1.3}{0}{0}        circle [radius=1.7pt];
    \fill[cTwo]   \iso{-0.445}{1.222}{0} circle [radius=1.7pt];
    \fill[cThree] \iso{-0.919}{0.919}{0} circle [radius=1.7pt];
    \fill[cFive]  \iso{0}{-0.20}{0}      circle [radius=1.7pt];

    % Labels
    \node[text=cOne,   font=\sffamily\small\bfseries, anchor=west]
      at \iso{1.55}{0}{0}        {$T_1$};
    \node[text=cTwo,   font=\sffamily\small\bfseries, anchor=east]
      at \iso{-0.633}{1.738}{0}  {$T_2$};
    \node[text=cThree, font=\sffamily\small\bfseries, anchor=east]
      at \iso{-1.308}{1.308}{0}  {$T_3$};
    \node[text=cFive,  font=\sffamily\small\bfseries, anchor=west]
      at \iso{0.7}{-0.6}{0}      {$T_4$};

    % Edge label
    \node[font=\sffamily\small, anchor=south]
      at \iso{0}{0}{1.7} {$e$};

    % Panel header
    \node[anchor=north, font=\sffamily\small\itshape] at (0, \panelh/2 - 0.25) {buckets};
  \end{scope}

  % =====================================================================
  % PANEL 3 -- T2/T3 swap
  % T2 at T3's intended slot (back, angle 135); T3 at T2's (front, angle 110)
  % T1 and T4 unchanged.
  % =====================================================================
  \begin{scope}[shift={(2*\panelw + 2*\panelgap, 0)}]
    \filldraw[fill=black!3, draw=black!20, line width=0.4pt, rounded corners=3pt]
      (-\panelw/2, -\panelh/2) rectangle (\panelw/2, \panelh/2);

    % T4 (unchanged)
    \fill[cFive, opacity=0.50]
      \iso{0}{0}{-1.5} -- \iso{0}{0}{1.5} -- \iso{0}{-0.20}{0} -- cycle;
    \draw[cFive, line width=0.7pt, opacity=0.9]
      \iso{0}{0}{-1.5} -- \iso{0}{0}{1.5} -- \iso{0}{-0.20}{0} -- cycle;

    % T2 (swapped to T3's slot)
    \fill[cTwo, opacity=0.50]
      \iso{0}{0}{-1.5} -- \iso{0}{0}{1.5} -- \iso{-0.919}{0.919}{0} -- cycle;
    \draw[cTwo, line width=0.7pt, opacity=0.9]
      \iso{0}{0}{-1.5} -- \iso{0}{0}{1.5} -- \iso{-0.919}{0.919}{0} -- cycle;

    % T3 (swapped to T2's slot)
    \fill[cThree, opacity=0.50]
      \iso{0}{0}{-1.5} -- \iso{0}{0}{1.5} -- \iso{-0.445}{1.222}{0} -- cycle;
    \draw[cThree, line width=0.7pt, opacity=0.9]
      \iso{0}{0}{-1.5} -- \iso{0}{0}{1.5} -- \iso{-0.445}{1.222}{0} -- cycle;

    % T1 (unchanged)
    \fill[cOne, opacity=0.50]
      \iso{0}{0}{-1.5} -- \iso{0}{0}{1.5} -- \iso{1.3}{0}{0} -- cycle;
    \draw[cOne, line width=0.7pt, opacity=0.9]
      \iso{0}{0}{-1.5} -- \iso{0}{0}{1.5} -- \iso{1.3}{0}{0} -- cycle;

    % Edge e
    \draw[black!85, line width=1.0pt]
      \iso{0}{0}{-1.5} -- \iso{0}{0}{1.5};
    \fill[black] \iso{0}{0}{-1.5} circle [radius=1.5pt];
    \fill[black] \iso{0}{0}{1.5} circle [radius=1.5pt];

    % Apex dots (T2 and T3 at swapped positions)
    \fill[cTwo]   \iso{-0.919}{0.919}{0} circle [radius=1.7pt];
    \fill[cThree] \iso{-0.445}{1.222}{0} circle [radius=1.7pt];
    \fill[cOne]   \iso{1.3}{0}{0}        circle [radius=1.7pt];
    \fill[cFive]  \iso{0}{-0.20}{0}      circle [radius=1.7pt];

    % Labels (T2 and T3 labels follow their triangles to swapped positions)
    \node[text=cOne,   font=\sffamily\small\bfseries, anchor=west]
      at \iso{1.55}{0}{0}        {$T_1$};
    \node[text=cThree, font=\sffamily\small\bfseries, anchor=east]
      at \iso{-0.633}{1.738}{0}  {$T_3$};
    \node[text=cTwo,   font=\sffamily\small\bfseries, anchor=east]
      at \iso{-1.308}{1.308}{0}  {$T_2$};
    \node[text=cFive,  font=\sffamily\small\bfseries, anchor=west]
      at \iso{0.7}{-0.6}{0}      {$T_4$};

    % Edge label
    \node[font=\sffamily\small, anchor=south]
      at \iso{0}{0}{1.7} {$e$};

    % Panel header
    \node[anchor=north, font=\sffamily\small\itshape] at (0, \panelh/2 - 0.25) {$T_2, T_3$ swap};
  \end{scope}

  % =====================================================================
  % PANEL 4 -- T4 flip
  % T4 perturbed across the edge plane: apex (0, -0.20, 0) -> (0, +0.20, 0)
  % T1, T2, T3 unchanged.
  % =====================================================================
  \begin{scope}[shift={(3*\panelw + 3*\panelgap, 0)}]
    \filldraw[fill=black!3, draw=black!20, line width=0.4pt, rounded corners=3pt]
      (-\panelw/2, -\panelh/2) rectangle (\panelw/2, \panelh/2);

    % T3 (unchanged)
    \fill[cThree, opacity=0.50]
      \iso{0}{0}{-1.5} -- \iso{0}{0}{1.5} -- \iso{-0.919}{0.919}{0} -- cycle;
    \draw[cThree, line width=0.7pt, opacity=0.9]
      \iso{0}{0}{-1.5} -- \iso{0}{0}{1.5} -- \iso{-0.919}{0.919}{0} -- cycle;

    % T2 (unchanged)
    \fill[cTwo, opacity=0.50]
      \iso{0}{0}{-1.5} -- \iso{0}{0}{1.5} -- \iso{-0.445}{1.222}{0} -- cycle;
    \draw[cTwo, line width=0.7pt, opacity=0.9]
      \iso{0}{0}{-1.5} -- \iso{0}{0}{1.5} -- \iso{-0.445}{1.222}{0} -- cycle;

    % T4 FLIPPED to y > 0 side -- apex now at (0, +0.20, 0)
    \fill[cFive, opacity=0.50]
      \iso{0}{0}{-1.5} -- \iso{0}{0}{1.5} -- \iso{0}{0.20}{0} -- cycle;
    \draw[cFive, line width=0.7pt, opacity=0.9]
      \iso{0}{0}{-1.5} -- \iso{0}{0}{1.5} -- \iso{0}{0.20}{0} -- cycle;

    % T1 (unchanged)
    \fill[cOne, opacity=0.50]
      \iso{0}{0}{-1.5} -- \iso{0}{0}{1.5} -- \iso{1.3}{0}{0} -- cycle;
    \draw[cOne, line width=0.7pt, opacity=0.9]
      \iso{0}{0}{-1.5} -- \iso{0}{0}{1.5} -- \iso{1.3}{0}{0} -- cycle;

    % Edge e
    \draw[black!85, line width=1.0pt]
      \iso{0}{0}{-1.5} -- \iso{0}{0}{1.5};
    \fill[black] \iso{0}{0}{-1.5} circle [radius=1.5pt];
    \fill[black] \iso{0}{0}{1.5} circle [radius=1.5pt];

    % Apex dots (T4 at flipped position)
    \fill[cOne]   \iso{1.3}{0}{0}        circle [radius=1.7pt];
    \fill[cTwo]   \iso{-0.445}{1.222}{0} circle [radius=1.7pt];
    \fill[cThree] \iso{-0.919}{0.919}{0} circle [radius=1.7pt];
    \fill[cFive]  \iso{0}{0.20}{0}       circle [radius=1.7pt];

    % Labels (T4 label placed near the flipped apex)
    \node[text=cOne,   font=\sffamily\small\bfseries, anchor=west]
      at \iso{1.55}{0}{0}        {$T_1$};
    \node[text=cTwo,   font=\sffamily\small\bfseries, anchor=east]
      at \iso{-0.633}{1.738}{0}  {$T_2$};
    \node[text=cThree, font=\sffamily\small\bfseries, anchor=east]
      at \iso{-1.308}{1.308}{0}  {$T_3$};
    \node[text=cFive,  font=\sffamily\small\bfseries, anchor=west]
      at \iso{0.3}{0.3}{0}       {$T_4$};

    % Edge label
    \node[font=\sffamily\small, anchor=south]
      at \iso{0}{0}{1.7} {$e$};

    % Panel header
    \node[anchor=north, font=\sffamily\small\itshape] at (0, \panelh/2 - 0.25) {$T_4$ flip};
  \end{scope}

\end{tikzpicture}

%% file: diagrams/box_rectangle_split.tex
\def\sclI{4.15}
\def\panelw{7.4}
\def\panelh{5.4}

\definecolor{tfd1face}{HTML}{97F4EB}
\definecolor{tfd1edge}{HTML}{0D665C}
\definecolor{tfd2face}{HTML}{008777}
\definecolor{tfd2edge}{HTML}{063A33}
\definecolor{tfrface}{HTML}{C8467A}
\definecolor{tfredge}{HTML}{5A1F36}
\definecolor{tfvedot}{HTML}{5A1F36}
\definecolor{tfbg}{HTML}{ECECEE}

\newcommand{\proj}[3]{%
  ({((#1) * 0.7880107536 + (#2) * 0.6156614753) * \sclI},
   {((#1) * (-0.1901783216) + (#2) * 0.2433065857 + (#3) * 0.9510565163) * \sclI - 0.35})%
}

\begin{tikzpicture}[font=\sffamily\footnotesize, x=1cm, y=1cm]

  \filldraw[
    fill=tfbg,
    draw=black!20,
    line width=0.4pt,
    rounded corners=3pt
  ]
    (-\panelw/2, -\panelh/2 - 0.35) rectangle (\panelw/2, \panelh/2 - 0.35);

  % --------------------------------------------------------------------
  % D1 outer faces
  % --------------------------------------------------------------------

  \filldraw[
    fill=tfd1face,
    draw=tfd1edge,
    fill opacity=0.40,
    draw opacity=0.75,
    line width=0.45pt
  ]
    \proj{-0.5}{-0.5}{-0.3}
    -- \proj{0}{-0.5}{-0.3}
    -- \proj{0}{0.5}{-0.3}
    -- \proj{-0.5}{0.5}{-0.3}
    -- cycle;

  \filldraw[
    fill=tfd1face,
    draw=tfd1edge,
    fill opacity=0.40,
    draw opacity=0.75,
    line width=0.45pt
  ]
    \proj{-0.5}{-0.5}{0.3}
    -- \proj{0}{-0.5}{0.3}
    -- \proj{0}{0.5}{0.3}
    -- \proj{-0.5}{0.5}{0.3}
    -- cycle;

  \filldraw[
    fill=tfd1face,
    draw=tfd1edge,
    fill opacity=0.40,
    draw opacity=0.75,
    line width=0.45pt
  ]
    \proj{-0.5}{-0.5}{-0.3}
    -- \proj{0}{-0.5}{-0.3}
    -- \proj{0}{-0.5}{0.3}
    -- \proj{-0.5}{-0.5}{0.3}
    -- cycle;

  \filldraw[
    fill=tfd1face,
    draw=tfd1edge,
    fill opacity=0.40,
    draw opacity=0.75,
    line width=0.45pt
  ]
    \proj{0}{0.5}{-0.3}
    -- \proj{-0.5}{0.5}{-0.3}
    -- \proj{-0.5}{0.5}{0.3}
    -- \proj{0}{0.5}{0.3}
    -- cycle;

  \filldraw[
    fill=tfd1face,
    draw=tfd1edge,
    fill opacity=0.40,
    draw opacity=0.75,
    line width=0.45pt
  ]
    \proj{-0.5}{-0.5}{-0.3}
    -- \proj{-0.5}{0.5}{-0.3}
    -- \proj{-0.5}{0.5}{0.3}
    -- \proj{-0.5}{-0.5}{0.3}
    -- cycle;

  % --------------------------------------------------------------------
  % D2 outer faces
  % --------------------------------------------------------------------

  \filldraw[
    fill=tfd2face,
    draw=tfd2edge,
    fill opacity=0.40,
    draw opacity=0.75,
    line width=0.45pt
  ]
    \proj{0}{-0.5}{-0.3}
    -- \proj{0.5}{-0.5}{-0.3}
    -- \proj{0.5}{0.5}{-0.3}
    -- \proj{0}{0.5}{-0.3}
    -- cycle;

  \filldraw[
    fill=tfd2face,
    draw=tfd2edge,
    fill opacity=0.40,
    draw opacity=0.75,
    line width=0.45pt
  ]
    \proj{0}{-0.5}{0.3}
    -- \proj{0.5}{-0.5}{0.3}
    -- \proj{0.5}{0.5}{0.3}
    -- \proj{0}{0.5}{0.3}
    -- cycle;

  \filldraw[
    fill=tfd2face,
    draw=tfd2edge,
    fill opacity=0.40,
    draw opacity=0.75,
    line width=0.45pt
  ]
    \proj{0}{-0.5}{-0.3}
    -- \proj{0.5}{-0.5}{-0.3}
    -- \proj{0.5}{-0.5}{0.3}
    -- \proj{0}{-0.5}{0.3}
    -- cycle;

  \filldraw[
    fill=tfd2face,
    draw=tfd2edge,
    fill opacity=0.40,
    draw opacity=0.75,
    line width=0.45pt
  ]
    \proj{0.5}{0.5}{-0.3}
    -- \proj{0}{0.5}{-0.3}
    -- \proj{0}{0.5}{0.3}
    -- \proj{0.5}{0.5}{0.3}
    -- cycle;

  \filldraw[
    fill=tfd2face,
    draw=tfd2edge,
    fill opacity=0.40,
    draw opacity=0.75,
    line width=0.45pt
  ]
    \proj{0.5}{-0.5}{-0.3}
    -- \proj{0.5}{0.5}{-0.3}
    -- \proj{0.5}{0.5}{0.3}
    -- \proj{0.5}{-0.5}{0.3}
    -- cycle;

  % --------------------------------------------------------------------
  % Cutting rectangle R
  % --------------------------------------------------------------------

  \filldraw[
    fill=tfrface,
    draw=tfredge,
    fill opacity=0.70,
    draw opacity=0.95,
    line width=0.9pt
  ]
    \proj{0}{-0.5}{-0.3}
    -- \proj{0}{0.5}{-0.3}
    -- \proj{0}{0.5}{0.3}
    -- \proj{0}{-0.5}{0.3}
    -- cycle;

  % --------------------------------------------------------------------
  % VE intersection points
  % --------------------------------------------------------------------

  \foreach \x/\y/\z in {
    0/-0.5/-0.3,
    0/0.5/-0.3,
    0/0.5/0.3,
    0/-0.5/0.3
  }{
    \filldraw[
      fill=tfvedot,
      draw=white,
      line width=0.45pt
    ] \proj{\x}{\y}{\z} circle [radius=3.2pt];
  }

  % --------------------------------------------------------------------
  % Legend above drawing
  % --------------------------------------------------------------------

  \node[
    anchor=south,
    fill=tfbg,
    draw=black!18,
    rounded corners=1pt,
    inner xsep=5pt,
    inner ysep=4pt,
    font=\footnotesize,
    minimum width=\panelw cm
  ] at (0, 2.55)
  {
    \begin{tabular}{@{}l@{\hspace{0.4em}}l@{\hspace{1.6cm}}l@{\hspace{0.4em}}l@{}}
      \tikz{\draw[fill=tfd1face, draw=tfd1edge, fill opacity=0.40] (0,0) rectangle (0.28,0.16);}
      &
      Sub-domain $D_1$
      &
      \tikz{\draw[fill=tfd2face, draw=tfd2edge, fill opacity=0.40] (0,0) rectangle (0.28,0.16);}
      &
      Sub-domain $D_2$
      \\[2pt]
      \tikz{\draw[fill=tfrface, draw=tfredge, fill opacity=0.70] (0,0) rectangle (0.28,0.16);}
      &
      Cutting rectangle $R$
      &
      \tikz{\filldraw[fill=tfvedot, draw=white, line width=0.4pt] (0.14,0.08) circle (0.10);}
      &
      VE intersection
    \end{tabular}
  };

\end{tikzpicture}

%% file: diagrams/algorithm_pipeline.tex
% Whole-algorithm overview. Two phases, grouped: BUILD (one pass over the
% operand set, six stages) feeds EXTRACT (one boolean expression -> one
% mesh, repeated). Each stage box carries a schematic glyph, the structure
% it produces, and its section. Single-word titles + wrapped descriptors
% keep every label inside its box.
\begin{tikzpicture}[font=\sffamily\footnotesize, x=1cm, y=1cm,
                    >=stealth, line join=round]

  \def\bw{2.4}    % build-box width
  \def\bh{2.95}   % build-box height
  \def\dx{2.92}   % centre-to-centre spacing
  \def\cx{7.3}    % row centre (2.5*\dx)
  \def\tw{2.05}   % text wrap width inside a box

  % ---- stage-box macro: #1 x-centre  #2 title  #3 produces  #4 sec
  \newcommand{\stage}[4]{
    \begin{scope}[shift={(#1,0)}]
      \filldraw[fill=white, draw=black!25, line width=0.5pt, rounded corners=3pt]
        (-\bw/2,-\bh/2) rectangle (\bw/2,\bh/2);
      \node[anchor=north, text=tfink, font=\sffamily\small\bfseries]
        at (0,\bh/2-0.10) {#2};
      \node[anchor=base, text=black!60, font=\sffamily\scriptsize,
            align=center, text width=\tw cm] at (0,-\bh/2+0.46) {#3};
      \node[anchor=south, text=black!42, font=\sffamily\scriptsize\itshape]
        at (0,-\bh/2+0.07) {#4};
    \end{scope}}

  % =====================================================================
  % BUILD container (drawn first, behind the boxes) + header tab
  % =====================================================================
  \def\ctop{2.18}
  \def\cbot{-1.78}
  % equal side margins: boxes span [-1.2, 15.8]; 0.35 padding each side
  \filldraw[fill=tfteal!4, draw=tfteal!45, line width=0.6pt, rounded corners=5pt]
    (-1.55,\cbot) rectangle (16.15,\ctop);
  \node[anchor=west, text=tfteal!75!black, font=\sffamily\footnotesize\bfseries]
    at (-1.40,\ctop-0.30) {BUILD};
  \node[anchor=west, text=black!50, font=\sffamily\footnotesize]
    at (-0.42,\ctop-0.30) {once per operand set};

  % forward arrows between the six stages
  \foreach \i in {0,...,4} {
    \draw[->, black!45, line width=0.9pt]
      ({\i*\dx+\bw/2},0) -- ({(\i+1)*\dx-\bw/2},0);
  }

  % the six build stages
  \stage{0}{Operands}{$N$ tagged meshes}{input}
  \stage{\dx}{Intersection}{pairwise records}{\S\ref{sec:predicates}--\ref{sec:arrangement}}
  \stage{2*\dx}{Arrangement}{per-face 2D cuts}{\S\ref{sec:arrangement}}
  \stage{3*\dx}{Graph}{reduced graph}{\S\ref{sec:graphs}}
  \stage{4*\dx}{Partition}{voted domains}{\S\ref{sec:aggregation}}
  \stage{5*\dx}{Classification}{inclusion bits}{\S\ref{sec:classification}}

  % =====================================================================
  % GLYPHS  (centred a little above box middle, clear of the title)
  % =====================================================================
  \def\gy{0.28}

  % 1. Operands -- two overlapping blobs
  \begin{scope}[shift={(0,\gy)}]
    \fill[tfteal,opacity=0.20] (-0.28,0) circle [radius=0.40];
    \draw[tfteal,opacity=0.75,line width=0.6pt] (-0.28,0) circle [radius=0.40];
    \fill[tfteal,opacity=0.20] (0.28,0) circle [radius=0.40];
    \draw[tfteal,opacity=0.75,line width=0.6pt] (0.28,0) circle [radius=0.40];
  \end{scope}

  % 2. Intersection -- two crossing faces, rose segment + dots
  \begin{scope}[shift={(\dx,\gy)}]
    \fill[tfteal,opacity=0.12] (-0.52,-0.28) -- (0.52,-0.28) -- (0.28,0.40) -- (-0.76,0.40) -- cycle;
    \draw[tfteal,opacity=0.70,line width=0.5pt] (-0.52,-0.28) -- (0.52,-0.28) -- (0.28,0.40) -- (-0.76,0.40) -- cycle;
    \fill[tfteal,opacity=0.12] (-0.28,-0.40) -- (0.76,-0.40) -- (0.52,0.28) -- (-0.52,0.28) -- cycle;
    \draw[tfteal,opacity=0.70,line width=0.5pt] (-0.28,-0.40) -- (0.76,-0.40) -- (0.52,0.28) -- (-0.52,0.28) -- cycle;
    \draw[tfrose,line width=1pt] (-0.50,0.0) -- (0.50,0.0);
    \fill[tfrose] (-0.50,0.0) circle [radius=1.5pt];
    \fill[tfrose] (0.50,0.0) circle [radius=1.5pt];
  \end{scope}

  % 3. Arrangement -- one face cut by 3 segments, 2 crossings
  \begin{scope}[shift={(2*\dx,\gy)}]
    \fill[tfteal,opacity=0.07] (-0.58,-0.46) rectangle (0.58,0.46);
    \draw[tfteal,opacity=0.55,line width=0.5pt] (-0.58,-0.46) rectangle (0.58,0.46);
    \draw[tfrose,line width=0.8pt] (-0.20,-0.46) -- (-0.20,0.46);
    \draw[tfrose,line width=0.8pt] ( 0.20,-0.46) -- ( 0.20,0.46);
    \draw[tfrose,line width=0.7pt] (-0.58,0.11) -- (0.58,0.11);
    \fill[tfrose] (-0.20,0.11) circle [radius=1.5pt];
    \fill[tfrose] ( 0.20,0.11) circle [radius=1.5pt];
  \end{scope}

  % 4. Graph -- 4 nodes, edges
  \begin{scope}[shift={(3*\dx,\gy)}]
    \coordinate (n1) at (-0.42,0.28);
    \coordinate (n2) at (0.42,0.34);
    \coordinate (n3) at (-0.28,-0.32);
    \coordinate (n4) at (0.38,-0.28);
    \draw[black!45,line width=0.7pt] (n1)--(n2) (n1)--(n3) (n3)--(n4) (n2)--(n4) (n1)--(n4);
    \foreach \n in {n1,n2,n3,n4} \fill[tfteal,opacity=0.85] (\n) circle [radius=2.3pt];
  \end{scope}

  % 5. Partition -- region split into 3 colours
  \begin{scope}[shift={(4*\dx,\gy)}]
    \fill[cOne,opacity=0.55]  (-0.56,-0.42) rectangle (-0.06,0.42);
    \fill[cThree,opacity=0.55] (-0.06,-0.42) rectangle (0.28,0.42);
    \fill[cFive,opacity=0.55]  (0.28,-0.42) rectangle (0.56,0.42);
    \draw[black!30,line width=0.5pt] (-0.56,-0.42) rectangle (0.56,0.42);
    \draw[black!30,line width=0.5pt] (-0.06,-0.42)--(-0.06,0.42);
    \draw[black!30,line width=0.5pt] (0.28,-0.42)--(0.28,0.42);
  \end{scope}

  % 6. Classification -- domains carrying bitvectors
  \begin{scope}[shift={(5*\dx,\gy)}]
    \node[font=\sffamily\scriptsize\ttfamily, text=cOne]  at (0,0.30) {1 0 1};
    \node[font=\sffamily\scriptsize\ttfamily, text=cThree] at (0,0.0)  {1 1 0};
    \node[font=\sffamily\scriptsize\ttfamily, text=cFive]  at (0,-0.30) {0 0 1};
  \end{scope}

  % =====================================================================
  % connector  BUILD -> EXTRACT
  % =====================================================================
  \draw[->, tfteal!60!black, line width=1.3pt]
    (\cx,\cbot) -- (\cx,-2.55);

  % =====================================================================
  % EXTRACT box (centred) + symmetric fan to three booleans
  % =====================================================================
  \def\ecy{-3.55}
  \def\ebw{3.4}
  \def\ebh{2.0}
  \begin{scope}[shift={(\cx,\ecy)}]
    \filldraw[fill=tfsoft, draw=black!30, line width=0.6pt, rounded corners=4pt]
      (-\ebw/2,-\ebh/2) rectangle (\ebw/2,\ebh/2);
    \node[anchor=north, text=tfink, font=\sffamily\small\bfseries]
      at (0,\ebh/2-0.10) {Extraction};
    \node[anchor=south, text=black!60, font=\sffamily\scriptsize] at (0,-\ebh/2+0.34)
      {boolean expression $\to$ mesh};
    \node[anchor=south, text=black!42, font=\sffamily\scriptsize\itshape]
      at (0,-\ebh/2+0.08) {\S\ref{sec:selection}};
    % glyph: select-from-domains
    \fill[cOne,opacity=0.5]  (-0.62,0.05) rectangle (-0.16,0.52);
    \fill[cThree,opacity=0.5](-0.16,0.05) rectangle (0.18,0.52);
    \fill[cFive,opacity=0.5] (0.18,0.05) rectangle (0.46,0.52);
    \draw[tfink,line width=0.9pt] (-0.62,0.05) rectangle (0.46,0.52);
  \end{scope}

  % EXTRACT header tab (left, mirrors BUILD)
  \node[anchor=east, text=black!55, font=\sffamily\footnotesize\bfseries]
    at (\cx-\ebw/2-0.45,\ecy+0.45) {EXTRACT};
  \node[anchor=east, text=black!50, font=\sffamily\footnotesize]
    at (\cx-\ebw/2-0.45,\ecy+0.08) {per boolean};
  \node[anchor=east, text=black!50, font=\sffamily\footnotesize]
    at (\cx-\ebw/2-0.45,\ecy-0.29) {expression};

  % symmetric fan to three boolean outputs -- orthogonal bus (stub, bus, drops)
  \def\fy{-5.75}
  \def\busy{-4.98}
  \draw[black!45, line width=0.9pt] (\cx,{\ecy-\ebh/2}) -- (\cx,\busy);
  \draw[black!45, line width=0.9pt] (\cx-3.1,\busy) -- (\cx+3.1,\busy);
  \foreach \dxk/\sym/\lbl in {-3.1/\cup/union, 0/-/difference, 3.1/\cap/intersection} {
    \draw[->, black!45, line width=0.9pt]
      ({\cx+\dxk},\busy) -- ({\cx+\dxk},{\fy+0.34});
    \node[draw=black!30, fill=white, rounded corners=3pt, inner sep=3pt,
          font=\sffamily\footnotesize, text=tfink]
      at ({\cx+\dxk},\fy) {$\sym$\;\lbl};
  }

\end{tikzpicture}

%% file: diagrams/arrangement_pipeline.tex
\def\sclI{0.52}
\def\sclF{0.72}
\def\panelw{4.0}
\def\panelh{3.8}
\def\panelgap{0.45}

\newcommand{\iso}[3]{({((#1) - (#2)) * 0.866 * \sclI}, {((#3) - ((#1) + (#2)) * 0.5) * \sclI})}

\begin{tikzpicture}[font=\sffamily\footnotesize, x=1cm, y=1cm]

  % =====================================================================
  % PANEL 1 -- 3D configuration
  % =====================================================================
  \begin{scope}[shift={(0,0)}]
    \filldraw[fill=black!3, draw=black!20, line width=0.4pt, rounded corners=3pt]
      (-\panelw/2, -\panelh/2) rectangle (\panelw/2, \panelh/2);

    % Quad D (back, dimmest)
    \fill[tfteal, opacity=0.10]
      \iso{-2}{0.3}{-1.3} -- \iso{2}{0.3}{-1.3} -- \iso{2}{0.3}{1.3} -- \iso{-2}{0.3}{1.3} -- cycle;
    \draw[tfteal, line width=0.5pt, opacity=0.55]
      \iso{-2}{0.3}{-1.3} -- \iso{2}{0.3}{-1.3} -- \iso{2}{0.3}{1.3} -- \iso{-2}{0.3}{1.3} -- cycle;

    % Quad B
    \fill[tfteal, opacity=0.13]
      \iso{-0.8}{-2}{-1.3} -- \iso{-0.8}{2}{-1.3} -- \iso{-0.8}{2}{1.3} -- \iso{-0.8}{-2}{1.3} -- cycle;
    \draw[tfteal, line width=0.5pt, opacity=0.65]
      \iso{-0.8}{-2}{-1.3} -- \iso{-0.8}{2}{-1.3} -- \iso{-0.8}{2}{1.3} -- \iso{-0.8}{-2}{1.3} -- cycle;

    % Quad C
    \fill[tfteal, opacity=0.13]
      \iso{0.8}{-2}{-1.3} -- \iso{0.8}{2}{-1.3} -- \iso{0.8}{2}{1.3} -- \iso{0.8}{-2}{1.3} -- cycle;
    \draw[tfteal, line width=0.5pt, opacity=0.65]
      \iso{0.8}{-2}{-1.3} -- \iso{0.8}{2}{-1.3} -- \iso{0.8}{2}{1.3} -- \iso{0.8}{-2}{1.3} -- cycle;

    % Quad A (front, most opaque)
    \fill[tfteal, opacity=0.20]
      \iso{-2}{-2}{0} -- \iso{-2}{2}{0} -- \iso{2}{2}{0} -- \iso{2}{-2}{0} -- cycle;
    \draw[tfteal, line width=0.7pt, opacity=0.85]
      \iso{-2}{-2}{0} -- \iso{-2}{2}{0} -- \iso{2}{2}{0} -- \iso{2}{-2}{0} -- cycle;

    % Intersection edges on A
    \draw[tfrose, line width=1pt, opacity=0.95] \iso{-0.8}{-2}{0} -- \iso{-0.8}{2}{0};
    \draw[tfrose, line width=1pt, opacity=0.95] \iso{0.8}{-2}{0}  -- \iso{0.8}{2}{0};
    \draw[tfrose, line width=1pt, opacity=0.85] \iso{-2}{0.3}{0}  -- \iso{2}{0.3}{0};

    % Boundary vertices
    \fill[tfteal] \iso{-0.8}{-2}{0} circle [radius=1.4pt];
    \fill[tfteal] \iso{-0.8}{2}{0}  circle [radius=1.4pt];
    \fill[tfteal] \iso{0.8}{-2}{0}  circle [radius=1.4pt];
    \fill[tfteal] \iso{0.8}{2}{0}   circle [radius=1.4pt];
    \fill[tfteal] \iso{-2}{0.3}{0}  circle [radius=1.4pt];
    \fill[tfteal] \iso{2}{0.3}{0}   circle [radius=1.4pt];

    % Quad labels — anchored at the front/top corners of each quad
    \node[text=black!75, font=\sffamily\small\bfseries, anchor=south east]
      at \iso{2.1}{-2}{0.3} {$A$};
    \node[text=black!55, font=\sffamily\small\bfseries\itshape, anchor=north west]
      at \iso{-0.8}{-2}{1.3} {$B$};
    \node[text=black!55, font=\sffamily\small\bfseries\itshape, anchor=south west]
      at \iso{0.8}{-2}{1.3} {$C$};
    \node[text=black!55, font=\sffamily\small\bfseries\itshape, anchor=south east]
      at \iso{-2}{0.3}{1.3} {$D$};

  \end{scope}

  % =====================================================================
  % PANEL 2 -- edges on face A
  % =====================================================================
  \begin{scope}[shift={(\panelw + \panelgap, 0)}]
    \filldraw[fill=black!3, draw=black!20, line width=0.4pt, rounded corners=3pt]
      (-\panelw/2, -\panelh/2) rectangle (\panelw/2, \panelh/2);

    % Face A as a square (scaled to sclF)
    \fill[tfteal, opacity=0.08] (-2*\sclF, -2*\sclF) rectangle (2*\sclF, 2*\sclF);
    \draw[tfteal, line width=0.7pt, opacity=0.75] (-2*\sclF, -2*\sclF) rectangle (2*\sclF, 2*\sclF);

    % 3 edges
    \draw[tfrose, line width=1pt, opacity=0.95] (-0.8*\sclF, -2*\sclF) -- (-0.8*\sclF, 2*\sclF);
    \draw[tfrose, line width=1pt, opacity=0.95] ( 0.8*\sclF, -2*\sclF) -- ( 0.8*\sclF, 2*\sclF);
    \draw[tfrose, line width=1pt, opacity=0.85] (-2*\sclF,  0.3*\sclF) -- (2*\sclF,  0.3*\sclF);

    % 6 boundary vertices
    \fill[tfteal] (-0.8*\sclF, -2*\sclF)  circle [radius=1.6pt];
    \fill[tfteal] (-0.8*\sclF,  2*\sclF)  circle [radius=1.6pt];
    \fill[tfteal] ( 0.8*\sclF, -2*\sclF)  circle [radius=1.6pt];
    \fill[tfteal] ( 0.8*\sclF,  2*\sclF)  circle [radius=1.6pt];
    \fill[tfteal] (-2*\sclF,    0.3*\sclF) circle [radius=1.6pt];
    \fill[tfteal] ( 2*\sclF,    0.3*\sclF) circle [radius=1.6pt];

    % Edge labels (canonical edge ids, named by endpoints)
    \node[text=black!70, font=\sffamily\scriptsize, anchor=east]
      at (-0.8*\sclF - 0.12, \sclF) {$e_0$};
    \node[text=black!70, font=\sffamily\scriptsize, anchor=west]
      at ( 0.8*\sclF + 0.12, \sclF) {$e_1$};
    \node[text=black!70, font=\sffamily\scriptsize, anchor=south]
      at (0, 0.3*\sclF + 0.08) {$e_2$};
  \end{scope}

  % =====================================================================
  % PANEL 3 -- intersection graph (crossings named by face triple)
  % =====================================================================
  \begin{scope}[shift={(2*\panelw + 2*\panelgap, 0)}]
    \filldraw[fill=black!3, draw=black!20, line width=0.4pt, rounded corners=3pt]
      (-\panelw/2, -\panelh/2) rectangle (\panelw/2, \panelh/2);

    % Faded face boundary
    \fill[tfteal, opacity=0.04] (-2*\sclF, -2*\sclF) rectangle (2*\sclF, 2*\sclF);
    \draw[tfteal, line width=0.4pt, opacity=0.30] (-2*\sclF, -2*\sclF) rectangle (2*\sclF, 2*\sclF);

    % B: 2 segments (split at y=0.3)
    \draw[tfrose, line width=0.9pt, opacity=0.70] (-0.8*\sclF, -2*\sclF)  -- (-0.8*\sclF, 0.3*\sclF);
    \draw[tfrose, line width=0.9pt, opacity=0.70] (-0.8*\sclF,  0.3*\sclF) -- (-0.8*\sclF, 2*\sclF);

    % C: 2 segments
    \draw[tfrose, line width=0.9pt, opacity=0.70] ( 0.8*\sclF, -2*\sclF)  -- ( 0.8*\sclF, 0.3*\sclF);
    \draw[tfrose, line width=0.9pt, opacity=0.70] ( 0.8*\sclF,  0.3*\sclF) -- ( 0.8*\sclF, 2*\sclF);

    % D: 3 segments (split at x=-0.8 and x=0.8)
    \draw[tfrose, line width=0.9pt, opacity=0.60] (-2*\sclF,    0.3*\sclF) -- (-0.8*\sclF, 0.3*\sclF);
    \draw[tfrose, line width=0.9pt, opacity=0.60] (-0.8*\sclF,  0.3*\sclF) -- ( 0.8*\sclF, 0.3*\sclF);
    \draw[tfrose, line width=0.9pt, opacity=0.60] ( 0.8*\sclF,  0.3*\sclF) -- ( 2*\sclF,   0.3*\sclF);

    % Endpoints (faded teal)
    \fill[tfteal, opacity=0.50] (-0.8*\sclF, -2*\sclF)  circle [radius=1.1pt];
    \fill[tfteal, opacity=0.50] (-0.8*\sclF,  2*\sclF)  circle [radius=1.1pt];
    \fill[tfteal, opacity=0.50] ( 0.8*\sclF, -2*\sclF)  circle [radius=1.1pt];
    \fill[tfteal, opacity=0.50] ( 0.8*\sclF,  2*\sclF)  circle [radius=1.1pt];
    \fill[tfteal, opacity=0.50] (-2*\sclF,    0.3*\sclF) circle [radius=1.1pt];
    \fill[tfteal, opacity=0.50] ( 2*\sclF,    0.3*\sclF) circle [radius=1.1pt];

    % Crossings (rose, prominent)
    \fill[tfrose] (-0.8*\sclF, 0.3*\sclF) circle [radius=2pt];
    \fill[tfrose] ( 0.8*\sclF, 0.3*\sclF) circle [radius=2pt];

    % Crossing labels (face triples, near each crossing)
    \node[text=black!75, font=\sffamily\scriptsize\bfseries, anchor=south east]
      at (-0.8*\sclF + 0.06, 0.3*\sclF + 0.05) {$\{A, B, D\}$};
    \node[text=black!75, font=\sffamily\scriptsize\bfseries, anchor=south west]
      at ( 0.8*\sclF - 0.06, 0.3*\sclF + 0.05) {$\{A, C, D\}$};

  \end{scope}

  % =====================================================================
  % PANEL 4 -- the coplanar exception (B coplanar with A)
  % =====================================================================
  \begin{scope}[shift={(3*\panelw + 3*\panelgap, 0)}]
    \filldraw[fill=black!3, draw=black!20, line width=0.4pt, rounded corners=3pt]
      (-\panelw/2, -\panelh/2) rectangle (\panelw/2, \panelh/2);

    % Face A
    \fill[tfteal, opacity=0.08] (-2*\sclF, -2*\sclF) rectangle (2*\sclF, 2*\sclF);
    \draw[tfteal, line width=0.7pt, opacity=0.75] (-2*\sclF, -2*\sclF) rectangle (2*\sclF, 2*\sclF);

    % B coplanar with A: its footprint on A, bounded by the contact edges
    \fill[tfteal, opacity=0.12] (-0.8*\sclF, -2*\sclF) rectangle (0.8*\sclF, 2*\sclF);
    \node[text=black!55, font=\sffamily\small\bfseries\itshape, anchor=south]
      at (0, 2*\sclF + 0.05) {$B$};

    % Contact edges (boundary of the polygon-of-contact, one face pair)
    \draw[tfrose, line width=1pt, opacity=0.95] (-0.8*\sclF, -2*\sclF) -- (-0.8*\sclF, 2*\sclF);
    \draw[tfrose, line width=1pt, opacity=0.95] ( 0.8*\sclF, -2*\sclF) -- ( 0.8*\sclF, 2*\sclF);

    % Transversal edge (A, D)
    \draw[tfrose, line width=1pt, opacity=0.85] (-2*\sclF,  0.3*\sclF) -- (2*\sclF,  0.3*\sclF);

    % Boundary vertices
    \fill[tfteal] (-0.8*\sclF, -2*\sclF)  circle [radius=1.6pt];
    \fill[tfteal] (-0.8*\sclF,  2*\sclF)  circle [radius=1.6pt];
    \fill[tfteal] ( 0.8*\sclF, -2*\sclF)  circle [radius=1.6pt];
    \fill[tfteal] ( 0.8*\sclF,  2*\sclF)  circle [radius=1.6pt];
    \fill[tfteal] (-2*\sclF,    0.3*\sclF) circle [radius=1.6pt];
    \fill[tfteal] ( 2*\sclF,    0.3*\sclF) circle [radius=1.6pt];

    % Edge labels
    \node[text=black!70, font=\sffamily\scriptsize, anchor=east]
      at (-0.8*\sclF - 0.12, -\sclF) {$e_0$};
    \node[text=black!70, font=\sffamily\scriptsize, anchor=west]
      at ( 0.8*\sclF + 0.12, -\sclF) {$e_1$};
    \node[text=black!70, font=\sffamily\scriptsize, anchor=north]
      at (0, 0.3*\sclF - 0.10) {$e_2$};

    % Crossings (rose, prominent)
    \fill[tfrose] (-0.8*\sclF, 0.3*\sclF) circle [radius=2pt];
    \fill[tfrose] ( 0.8*\sclF, 0.3*\sclF) circle [radius=2pt];

    % Crossing labels (refined by canonical edge pair)
    \node[text=black!75, font=\sffamily\scriptsize\bfseries, anchor=south east]
      at (-0.8*\sclF - 0.04, 0.3*\sclF + 0.05) {$(e_0, e_2)$};
    \node[text=black!75, font=\sffamily\scriptsize\bfseries, anchor=south west]
      at ( 0.8*\sclF + 0.04, 0.3*\sclF + 0.05) {$(e_1, e_2)$};

  \end{scope}

\end{tikzpicture}

%% file: diagrams/mel_components_to_wedges.tex
\definecolor{cOne}{HTML}{4C78A8}
\definecolor{cTwo}{HTML}{F58518}
\definecolor{cThree}{HTML}{54A24B}
\definecolor{cFour}{HTML}{8E6BBE}
\definecolor{tfgrey}{HTML}{9CA3AF}
\definecolor{tfsoft}{HTML}{F3F4F6}

\providecommand{\melnode}[4]{%
  \fill[#3!30] (#1,#2) circle (0.42);
  \draw[#3!90!black, line width=0.7pt] (#1,#2) circle (0.42);
  \node[text=#3!80!black, font=\scriptsize] at (#1,#2) {$#4$};
}

\resizebox{\textwidth}{!}{%
\begin{tikzpicture}[font=\rmfamily\footnotesize]

  % --------------------------------------------------------------------------
  % Panels
  % --------------------------------------------------------------------------
  \fill[tfsoft, rounded corners=5pt] (-7.00,-1.70) rectangle (-1.50,2.60);
  \fill[tfsoft, rounded corners=5pt] ( 1.50,-1.70) rectangle ( 7.00,2.60);
  \draw[black!20, line width=0.5pt, rounded corners=5pt] (-7.00,-1.70) rectangle (-1.50,2.60);
  \draw[black!20, line width=0.5pt, rounded corners=5pt] ( 1.50,-1.70) rectangle ( 7.00,2.60);

  \node[font=\footnotesize] at (-4.25,2.30)
    {MEL components};

  \node[font=\footnotesize] at (4.25,2.30)
    {wedges at $e$};

  % --------------------------------------------------------------------------
  % LEFT: MEL components + a relation r between them.
  % --------------------------------------------------------------------------
  \coordinate (rcenter) at (-4.25,0.55);
  \draw[black!55, line width=0.9pt] (-5.45,0.55) -- (rcenter);
  \draw[black!55, line width=0.9pt] (rcenter) -- (-4.25,1.55);
  \draw[black!55, line width=0.9pt] (rcenter) -- (-3.05,0.55);
  \draw[black!55, line width=0.9pt] (rcenter) -- (-4.25,-0.45);
  \fill[black!55] (rcenter) circle (0.08);
  \node[font=\scriptsize, text=tfgrey] at (-3.55,0.28) {$r$};

  \melnode{-5.45}{ 0.55}{cOne}{c_1}
  \melnode{-4.25}{ 1.55}{cTwo}{c_2}
  \melnode{-3.05}{ 0.55}{cFour}{c_4}
  \melnode{-4.25}{-0.45}{cThree}{c_3}

  \node[font=\scriptsize, align=center, text=tfgrey] at (-4.25,-1.30)
    {reduced graph $\mathcal{R}$};

  % --------------------------------------------------------------------------
  % RIGHT: wedges at the non-manifold edge -- one cross-section, four
  % face-sectors coloured by their owning MEL component.
  % --------------------------------------------------------------------------
  \begin{scope}[shift={(4.25,0.55)}, scale=1.30]
    \draw[black, line width=0.7pt] (0,0) circle (0.95);
    \fill[cTwo!35]   (0,0) -- (90:0.95) -- (0:0.95) -- cycle;
    \fill[cFour!35]  (0,0) -- (0:0.95) -- (-90:0.95) -- cycle;
    \fill[cThree!35] (0,0) -- (-90:0.95) -- (180:0.95) -- cycle;
    \fill[cOne!35]   (0,0) -- (180:0.95) -- (90:0.95) -- cycle;

    \node[text=cTwo,   font=\scriptsize] at (45:0.52)   {$c_2$};
    \node[text=cFour,  font=\scriptsize] at (-45:0.52)  {$c_4$};
    \node[text=cThree, font=\scriptsize] at (-135:0.52) {$c_3$};
    \node[text=cOne,   font=\scriptsize] at (135:0.52)  {$c_1$};

    \draw[black, line width=0.6pt] (0,0) -- (90:0.95);
    \draw[black, line width=0.6pt] (0,0) -- (0:0.95);
    \draw[black, line width=0.6pt] (0,0) -- (-90:0.95);
    \draw[black, line width=0.6pt] (0,0) -- (180:0.95);
    \fill[black] (0,0) circle (0.05);

    % Cyclic order hint (small arc arrow around the rim).
    \draw[->, black!50, line width=0.5pt]
      (15:1.15) arc[start angle=15, end angle=80, radius=1.15];
  \end{scope}

  \node[font=\scriptsize, align=center, text=tfgrey] at (4.25,-1.30)
    {$\rho_e = (c_1, c_2, c_4, c_3)$};

  % --------------------------------------------------------------------------
  % Bridge between the two boxes.
  % --------------------------------------------------------------------------
  \draw[->, line width=0.9pt, black]
    (-1.45,0.55) -- (1.45,0.55);

  \node[font=\scriptsize\itshape, align=center, text=tfgrey] at (0,1.00)
    {exact/bounded\\boundary};

\end{tikzpicture}%
}

%% file: diagrams/relation_majority_vote.tex
\definecolor{cOne}{HTML}{4C78A8}
\definecolor{cTwo}{HTML}{F58518}
\definecolor{cThree}{HTML}{54A24B}
\definecolor{cFour}{HTML}{8E6BBE}
\definecolor{cFive}{HTML}{E45756}
\definecolor{tfgrey}{HTML}{9CA3AF}
\definecolor{tfsoft}{HTML}{F3F4F6}

\newcommand{\wedgeobs}[6]{%
  % #1 x, #2 y, #3 scale, #4 label, #5 swapped flag, #6 subtitle
  \begin{scope}[shift={(#1,#2)}, scale=#3]
    \draw[black, line width=0.6pt] (0,0) circle (0.82);
    \ifnum#5=0
      \fill[cTwo!35]   (0,0) -- (90:0.82) -- (0:0.82) -- cycle;
      \fill[cFour!35]  (0,0) -- (0:0.82) -- (-90:0.82) -- cycle;
      \fill[cOne!35]   (0,0) -- (-90:0.82) -- (180:0.82) -- cycle;
      \fill[cThree!35] (0,0) -- (180:0.82) -- (90:0.82) -- cycle;

      \node[text=cTwo,   font=\scriptsize] at (45:0.43) {$c_2$};
      \node[text=cFour,  font=\scriptsize] at (-45:0.43) {$c_4$};
      \node[text=cOne,   font=\scriptsize] at (-135:0.43) {$c_1$};
      \node[text=cThree, font=\scriptsize] at (135:0.43) {$c_3$};
    \else
      \fill[cTwo!35]   (0,0) -- (90:0.82) -- (0:0.82) -- cycle;
      \fill[cOne!35]   (0,0) -- (0:0.82) -- (-90:0.82) -- cycle;
      \fill[cFour!35]  (0,0) -- (-90:0.82) -- (180:0.82) -- cycle;
      \fill[cThree!35] (0,0) -- (180:0.82) -- (90:0.82) -- cycle;

      \node[text=cTwo,   font=\scriptsize] at (45:0.43) {$c_2$};
      \node[text=cOne,   font=\scriptsize] at (-45:0.43) {$c_1$};
      \node[text=cFour,  font=\scriptsize] at (-135:0.43) {$c_4$};
      \node[text=cThree, font=\scriptsize] at (135:0.43) {$c_3$};
    \fi

    \draw[black, line width=0.6pt] (0,0) -- (90:0.82);
    \draw[black, line width=0.6pt] (0,0) -- (0:0.82);
    \draw[black, line width=0.6pt] (0,0) -- (-90:0.82);
    \draw[black, line width=0.6pt] (0,0) -- (180:0.82);
    \fill[black] (0,0) circle (0.045);

    \node[font=\footnotesize] at (0,-1.12) {#4};
    \node[font=\scriptsize\itshape, align=center, text=tfgrey] at (0,-1.45) {#6};
  \end{scope}
}

\resizebox{\columnwidth}{!}{%
\begin{tikzpicture}[font=\rmfamily\footnotesize]

  % Panel background.
  \filldraw[fill=tfsoft, draw=black!20, line width=0.4pt, rounded corners=3pt]
    (-5.45,-2.75) rectangle (5.45,3.55);

  % Top row: four observations of equivalence class (c_1, c_2, c_3, c_4).
  \wedgeobs{-4.20}{1.95}{1.0}{}{0}{}
  \wedgeobs{-2.10}{1.95}{1.0}{}{0}{}
  \wedgeobs{ 2.10}{1.95}{1.0}{}{1}{}
  \wedgeobs{ 4.20}{1.95}{1.0}{}{0}{}

  \node[font=\scriptsize] at (-4.20,3.05) {$(c_1, c_2, c_3, c_4)$};
  \node[font=\scriptsize] at (-2.10,3.05) {$(c_1, c_2, c_3, c_4)$};
  \node[font=\scriptsize] at ( 2.10,3.05) {$(c_1, c_2, c_3, c_4)$};
  \node[font=\scriptsize] at ( 4.20,3.05) {$(c_1, c_2, c_3, c_4)$};

  % Middle wedge: different equivalence class with 5 components.
  % Permutation = majority (c_2, c_4, c_1, c_3) with c_5 inserted between c_4 and c_1.
  \begin{scope}[shift={(0,1.95)}, scale=1.0]
    \draw[black, line width=0.6pt] (0,0) circle (0.82);
    \fill[cTwo!35]   (0,0) -- (90:0.82)   -- (18:0.82)   -- cycle;
    \fill[cFour!35]  (0,0) -- (18:0.82)   -- (-54:0.82)  -- cycle;
    \fill[cFive!35]  (0,0) -- (-54:0.82)  -- (-126:0.82) -- cycle;
    \fill[cOne!35]   (0,0) -- (-126:0.82) -- (162:0.82)  -- cycle;
    \fill[cThree!35] (0,0) -- (162:0.82)  -- (90:0.82)   -- cycle;

    \node[text=cTwo,   font=\scriptsize] at (54:0.45)   {$c_2$};
    \node[text=cFour,  font=\scriptsize] at (-18:0.45)  {$c_4$};
    \node[text=cFive,  font=\scriptsize] at (-90:0.45)  {$c_5$};
    \node[text=cOne,   font=\scriptsize] at (-162:0.45) {$c_1$};
    \node[text=cThree, font=\scriptsize] at (126:0.45)  {$c_3$};

    \draw[black, line width=0.6pt] (0,0) -- (90:0.82);
    \draw[black, line width=0.6pt] (0,0) -- (18:0.82);
    \draw[black, line width=0.6pt] (0,0) -- (-54:0.82);
    \draw[black, line width=0.6pt] (0,0) -- (-126:0.82);
    \draw[black, line width=0.6pt] (0,0) -- (162:0.82);
    \fill[black] (0,0) circle (0.045);
  \end{scope}

  \node[font=\scriptsize] at (0,3.05) {$(c_1, c_2, c_3, c_4, c_5)$};

  % Vote convergence arrows.
  \draw[->, line width=0.6pt, black]
    (-4.20,1.13) .. controls (-4.00,0.55) and (-1.60,0.10) .. (-0.55,-0.36);
  \draw[->, line width=0.6pt, black]
    (-2.10,1.13) .. controls (-2.00,0.55) and (-0.80,0.05) .. (-0.30,-0.30);
  \draw[->, line width=0.6pt, red]
    ( 2.10,1.13) .. controls ( 2.00,0.55) and ( 0.80,0.05) .. ( 0.30,-0.30);
  \draw[->, line width=0.6pt, black]
    ( 4.20,1.13) .. controls ( 4.00,0.55) and ( 1.60,0.10) .. ( 0.55,-0.36);

  \wedgeobs{0}{-1.15}{1.12}{majority $\hat{\sigma}$}{0}{}

\end{tikzpicture}%
}

%% file: diagrams/cross_component_nesting.tex
\def\panelw{7.8}
\def\panelh{5.2}

\definecolor{aFace}{HTML}{97F4EB}
\definecolor{aEdge}{HTML}{0D665C}
\definecolor{bFace}{HTML}{F3B6CE}
\definecolor{bEdge}{HTML}{8E2B53}
\definecolor{tfink}{HTML}{1F2937}
\definecolor{tfgrey}{HTML}{9CA3AF}
\definecolor{tfbg}{HTML}{F3F4F6}

\begin{tikzpicture}[font=\sffamily\footnotesize, x=1cm, y=1cm]

  % --- panel ----------------------------------------------------------------
  \filldraw[fill=tfbg, draw=black!20, line width=0.4pt, rounded corners=3pt]
    (-\panelw/2, -\panelh/2) rectangle (\panelw/2, \panelh/2);

  % --- bundle A's outer-environment is the exterior: the free anchor --------
  \node[text=tfink, font=\scriptsize] at (0,2.28)
    {$P$'s outside \,(the exterior)};
  \node[text=tfgrey, font=\scriptsize\itshape] at (0,1.92)
    {outermost region $\Rightarrow$ seeds free, $b = \mathbf{0}$};

  % --- bundle bounding box (B); A obviously overlaps it ---------------------
  \draw[tfgrey, dashed, line width=0.4pt] (-2.44,-0.68) rectangle (-0.76,0.68);

  % --- domains: fill A (D1), then B (D2) ------------------------------------
  \fill[aFace, fill opacity=0.45] (-1.2,-0.15) ellipse (1.85 and 1.40);
  \fill[bFace, fill opacity=0.60] (-1.6,0.00) ellipse (0.82 and 0.66);
  \draw[aEdge, line width=1.0pt, draw opacity=0.85] (-1.2,-0.15) ellipse (1.85 and 1.40);
  \draw[bEdge, line width=1.0pt, draw opacity=0.90] (-1.6,0.00) ellipse (0.82 and 0.66);
  \node[text=aEdge, font=\scriptsize] at (-0.10,1.20) {$P$};
  \node[text=bEdge, font=\scriptsize] at (-1.60,0.44) {$Q$};

  % --- component Q's outer-environment (set by the ray) and the flood result ---
  \node[text=tfink, font=\scriptsize] at (-1.60,-0.05) {$\{P,Q\}$};
  \node[text=tfink, font=\scriptsize] at (-1.20,-0.90) {$Q$'s outside $= \{P\}$};
  \node[text=tfgrey, font=\scriptsize\itshape] at (-1.20,-1.20) {from the ray};

  % --- the ray: from a vertex of B, crossing A once, to a far point ---------
  \coordinate (seed) at (-0.76,0.00);
  \coordinate (xA)   at (0.64,0.00);
  \coordinate (far)  at (3.25,0.00);
  \draw[tfink, line width=0.8pt, ->] (seed) -- (far);
  \fill[tfink] (seed) circle (0.05);
  \node[text=tfink, font=\scriptsize, anchor=south west] at (-0.70,0.08) {seed};
  \draw[aEdge, line width=1.1pt] ($(xA)+(-0.09,-0.16)$) -- ($(xA)+(0.09,0.16)$);
  \draw[aEdge, line width=1.1pt] ($(xA)+(0.09,-0.16)$) -- ($(xA)+(-0.09,0.16)$);
  \node[text=tfgrey, font=\scriptsize\itshape, align=center] at (2.95,-0.46)
    {far point\\(past bbox)};

  % --- a form whose bbox is disjoint from the bundle's: skipped, no ray -----
  \node[text=tfgrey, font=\scriptsize\itshape, align=center] at (2.78,-1.08)
    {skipped:\\bbox disjoint};
  \begin{scope}[shift={(2.78,-1.92)}]
    \draw[tfgrey, dashed, line width=0.4pt] (-0.60,-0.42) rectangle (0.60,0.42);
    \fill[black!10] (0,0) ellipse (0.42 and 0.30);
    \draw[black!40, line width=0.8pt] (0,0) ellipse (0.42 and 0.30);
    \node[text=black!55, font=\scriptsize] at (0,0) {$R$};
  \end{scope}

\end{tikzpicture}

%% file: diagrams/sheets_expressions.tex
% Middle panel of the sheets figure: the declaration and the three
% stratum expressions, colour-keyed to the rendered slabs. A tikz
% fragment drawn into the parent picture's current scope; the parent
% defines \sheetw (the panel side length).
\definecolor{strSand}{rgb}{0.72, 0.57, 0.30}
\definecolor{strMoss}{rgb}{0.33, 0.52, 0.24}
\definecolor{strClay}{rgb}{0.55, 0.36, 0.26}
\filldraw[fill=black!3, draw=black!20, line width=0.4pt, rounded corners=6pt]
  (0,0) rectangle (\sheetw,\sheetw);

\node[anchor=west, align=left, text=black!75, font=\sffamily\small]
  at (0.08\sheetw, 0.82\sheetw) {
  $M$: box $\cup$ spheres \hfill (volumes)\\[2pt]
  $T$, $B$: horizons \hfill (\textbf{sheets})
};

\draw[black!20, line width=0.4pt]
  (0.08\sheetw, 0.70\sheetw) -- (0.92\sheetw, 0.70\sheetw);

\node[anchor=west, text=strSand, font=\sffamily\small\bfseries]
  at (0.08\sheetw, 0.585\sheetw)
  {upper $\;=\; M - T$};
\node[anchor=west, text=strMoss, font=\sffamily\small\bfseries]
  at (0.08\sheetw, 0.465\sheetw)
  {between $\;=\; (M \wedge T) - B$};
\node[anchor=west, text=strClay, font=\sffamily\small\bfseries]
  at (0.08\sheetw, 0.345\sheetw)
  {lower $\;=\; M \wedge B$};

\draw[black!20, line width=0.4pt]
  (0.08\sheetw, 0.23\sheetw) -- (0.92\sheetw, 0.23\sheetw);

\node[anchor=west, align=left, text=black!55, font=\sffamily\footnotesize]
  at (0.08\sheetw, 0.125\sheetw) {
  for a sheet, $\wedge$ selects behind its\\
  normal (below), $-$ in front (above)
};

%% file: diagrams/sphere_offset.tex
% Two UV spheres (32x32), one offset by d; offset swept (sphere_offset.csv).
% Three-row strip: per-edge on the materialised reading (red bands = the
% relation's edges disagree), the vote on the same reading (majority correct
% everywhere), and the original-planes ordering (no disagreement to recover).
% Generated by experimentation/gen_sphere_offset_tex.py -- do not hand-edit bands.
\definecolor{cVote}{HTML}{0D665C}
\definecolor{cEdge}{HTML}{D9534F}
\begin{tikzpicture}
\begin{axis}[
  width=\columnwidth, height=4.1cm,
  xmin=-0.36, xmax=2.5, ymin=0.2, ymax=2.78,
  axis y line=none,
  axis x line=bottom, x axis line style={black!55, shorten >=0pt},
  xlabel={sphere offset $d$ (radius $=1$)},
  xtick={0,0.39,0.77,1.11,1.41,2}, xticklabels={0,,,,,2},
  enlarge x limits=false,
  tick label style={font=\footnotesize}, label style={font=\footnotesize},
  clip=false,
]
% vertex-ring offsets (faint guides through all rows)
\draw[black!25, dashed, very thin] (axis cs:0.39,0.55) -- (axis cs:0.39,2.65);
\draw[black!25, dashed, very thin] (axis cs:0.77,0.55) -- (axis cs:0.77,2.65);
\draw[black!25, dashed, very thin] (axis cs:1.11,0.55) -- (axis cs:1.11,2.65);
\draw[black!25, dashed, very thin] (axis cs:1.41,0.55) -- (axis cs:1.41,2.65);
% per-edge row: green base + red failure bands
\fill[cVote!30] (axis cs:0,1.95) rectangle (axis cs:2,2.55);
\fill[cEdge] (axis cs:0.194,1.95) rectangle (axis cs:0.207,2.55);
\fill[cEdge] (axis cs:0.213,1.95) rectangle (axis cs:0.227,2.55);
\fill[cEdge] (axis cs:0.224,1.95) rectangle (axis cs:0.237,2.55);
\fill[cEdge] (axis cs:0.233,1.95) rectangle (axis cs:0.246,2.55);
\fill[cEdge] (axis cs:0.243,1.95) rectangle (axis cs:0.257,2.55);
\fill[cEdge] (axis cs:0.254,1.95) rectangle (axis cs:0.267,2.55);
\fill[cEdge] (axis cs:0.264,1.95) rectangle (axis cs:0.277,2.55);
\fill[cEdge] (axis cs:0.274,1.95) rectangle (axis cs:0.287,2.55);
\fill[cEdge] (axis cs:0.283,1.95) rectangle (axis cs:0.296,2.55);
\fill[cEdge] (axis cs:0.293,1.95) rectangle (axis cs:0.306,2.55);
\fill[cEdge] (axis cs:0.303,1.95) rectangle (axis cs:0.317,2.55);
\fill[cEdge] (axis cs:0.314,1.95) rectangle (axis cs:0.327,2.55);
\fill[cEdge] (axis cs:0.324,1.95) rectangle (axis cs:0.337,2.55);
\fill[cEdge] (axis cs:0.334,1.95) rectangle (axis cs:0.347,2.55);
\fill[cEdge] (axis cs:0.343,1.95) rectangle (axis cs:0.356,2.55);
\fill[cEdge] (axis cs:0.353,1.95) rectangle (axis cs:0.366,2.55);
\fill[cEdge] (axis cs:0.363,1.95) rectangle (axis cs:0.377,2.55);
\fill[cEdge] (axis cs:0.373,1.95) rectangle (axis cs:0.387,2.55);
\fill[cEdge] (axis cs:0.394,1.95) rectangle (axis cs:0.407,2.55);
\fill[cEdge] (axis cs:0.403,1.95) rectangle (axis cs:0.416,2.55);
\fill[cEdge] (axis cs:0.413,1.95) rectangle (axis cs:0.426,2.55);
\fill[cEdge] (axis cs:0.423,1.95) rectangle (axis cs:0.436,2.55);
\fill[cEdge] (axis cs:0.463,1.95) rectangle (axis cs:0.476,2.55);
\fill[cEdge] (axis cs:0.473,1.95) rectangle (axis cs:0.486,2.55);
\fill[cEdge] (axis cs:0.483,1.95) rectangle (axis cs:0.496,2.55);
\fill[cEdge] (axis cs:0.493,1.95) rectangle (axis cs:0.506,2.55);
\fill[cEdge] (axis cs:0.514,1.95) rectangle (axis cs:0.526,2.55);
\fill[cEdge] (axis cs:0.524,1.95) rectangle (axis cs:0.536,2.55);
\fill[cEdge] (axis cs:0.534,1.95) rectangle (axis cs:0.546,2.55);
\fill[cEdge] (axis cs:0.554,1.95) rectangle (axis cs:0.567,2.55);
\fill[cEdge] (axis cs:0.564,1.95) rectangle (axis cs:0.576,2.55);
\fill[cEdge] (axis cs:0.814,1.95) rectangle (axis cs:0.826,2.55);
\fill[cEdge] (axis cs:0.844,1.95) rectangle (axis cs:0.856,2.55);
\fill[cEdge] (axis cs:0.874,1.95) rectangle (axis cs:0.886,2.55);
\fill[cEdge] (axis cs:0.884,1.95) rectangle (axis cs:0.896,2.55);
\fill[cEdge] (axis cs:0.904,1.95) rectangle (axis cs:0.916,2.55);
\fill[cEdge] (axis cs:0.914,1.95) rectangle (axis cs:0.926,2.55);
\fill[cEdge] (axis cs:0.944,1.95) rectangle (axis cs:0.956,2.55);
\fill[cEdge] (axis cs:0.954,1.95) rectangle (axis cs:0.966,2.55);
\fill[cEdge] (axis cs:0.984,1.95) rectangle (axis cs:0.996,2.55);
\fill[cEdge] (axis cs:1.034,1.95) rectangle (axis cs:1.046,2.55);
\fill[cEdge] (axis cs:1.064,1.95) rectangle (axis cs:1.077,2.55);
\fill[cEdge] (axis cs:1.114,1.95) rectangle (axis cs:1.127,2.55);
\fill[cEdge] (axis cs:1.123,1.95) rectangle (axis cs:1.136,2.55);
\fill[cEdge] (axis cs:1.153,1.95) rectangle (axis cs:1.166,2.55);
\fill[cEdge] (axis cs:1.224,1.95) rectangle (axis cs:1.236,2.55);
\fill[cEdge] (axis cs:1.264,1.95) rectangle (axis cs:1.276,2.55);
\fill[cEdge] (axis cs:1.284,1.95) rectangle (axis cs:1.296,2.55);
\fill[cEdge] (axis cs:1.413,1.95) rectangle (axis cs:1.426,2.55);
\fill[cEdge] (axis cs:1.764,1.95) rectangle (axis cs:1.776,2.55);
\fill[cEdge] (axis cs:1.774,1.95) rectangle (axis cs:1.786,2.55);
\fill[cEdge] (axis cs:1.844,1.95) rectangle (axis cs:1.857,2.55);
\fill[cEdge] (axis cs:1.913,1.95) rectangle (axis cs:1.926,2.55);
\fill[cEdge] (axis cs:1.923,1.95) rectangle (axis cs:1.936,2.55);
\draw[black!40, line width=0.4pt] (axis cs:0,1.95) rectangle (axis cs:2,2.55);
% unbroken green row
\fill[cVote!30] (axis cs:0,1.25) rectangle (axis cs:2,1.85);
\draw[black!40, line width=0.4pt] (axis cs:0,1.25) rectangle (axis cs:2,1.85);
% unbroken green row
\fill[cVote!30] (axis cs:0,0.55) rectangle (axis cs:2,1.15);
\draw[black!40, line width=0.4pt] (axis cs:0,0.55) rectangle (axis cs:2,1.15);
% row labels + counts
\node[anchor=east, font=\footnotesize, text=tfink] at (axis cs:-0.02,2.25) {per-edge, materialised};
\node[anchor=west, font=\scriptsize, text=cEdge] at (axis cs:2.02,2.25) {$54/191$ fail};
\node[anchor=east, font=\footnotesize, text=tfink] at (axis cs:-0.02,1.55) {vote, materialised};
\node[anchor=west, font=\scriptsize, text=cVote] at (axis cs:2.02,1.55) {$0/191$};
\node[anchor=east, font=\footnotesize, text=tfink] at (axis cs:-0.02,0.85) {original planes};
\node[anchor=west, font=\scriptsize, text=cVote] at (axis cs:2.02,0.85) {$0/191$};
\node[anchor=south, font=\scriptsize, text=black!50] at (axis cs:0.77,2.65) {vertex-ring offsets};
\end{axis}
\end{tikzpicture}

%% file: diagrams/robustness_models.tex
% Vote vs per-edge under added construction error, across twenty-two model
% configurations (fifteen standard meshes as model/rotated-copy pairs +
% seven random-rotation arrangements; csg_wedge_votes, model_perturbation.csv).
% Bands are the min--max envelope of correctness over the configurations.
\definecolor{cVote}{HTML}{0D665C}
\definecolor{cEdge}{HTML}{D9534F}
\begin{tikzpicture}
\begin{axis}[
  width=\columnwidth, height=5.4cm,
  xmode=log, log basis x=10,
  xlabel={construction error $\varepsilon$ (bounding-box relative)},
  ylabel={fraction of relations correct},
  xmin=7e-8, xmax=1.4e-1, ymin=-0.04, ymax=1.06,
  ytick={0,0.25,0.5,0.75,1.0},
  legend cell align=left, legend columns=2,
  grid=both, grid style={black!8},
  tick label style={font=\footnotesize},
  label style={font=\footnotesize},
  legend style={font=\footnotesize, draw=black!20,
                at={(0.5,1.03)}, anchor=south,
                /tikz/every even column/.append style={column sep=1.4em}},
]
% ---- per-edge envelope (red band)
\addplot[name path=pehi, draw=none, forget plot] coordinates {
  (1e-7,1.000)(1e-6,1.000)(1e-5,0.912)(1e-4,0.537)(1e-3,0.039)(1e-2,0)(1e-1,0)
};
\addplot[name path=pelo, draw=none, forget plot] coordinates {
  (1e-7,0.643)(1e-6,0.300)(1e-5,0)(1e-4,0)(1e-3,0)(1e-2,0)(1e-1,0)
};
\addplot[cEdge, opacity=0.16, forget plot] fill between[of=pehi and pelo];
\addplot[cEdge, thick] coordinates {
  (1e-7,0.643)(1e-6,0.300)(1e-5,0)(1e-4,0)(1e-3,0)(1e-2,0)(1e-1,0)
};
\addlegendentry{per-edge (22 configs)}
\addplot[cEdge, thick, forget plot] coordinates {
  (1e-7,1.000)(1e-6,1.000)(1e-5,0.912)(1e-4,0.537)(1e-3,0.039)(1e-2,0)(1e-1,0)
};
% ---- vote envelope (green band)
\addplot[name path=vhi, draw=none, forget plot] coordinates {
  (1e-7,1)(1e-6,1)(1e-5,1)(1e-4,1)(1e-3,1)(1e-2,1)(1e-1,1)
};
\addplot[name path=vlo, draw=none, forget plot] coordinates {
  (1e-7,1)(1e-6,1)(1e-5,1)(1e-4,0.917)(1e-3,0.780)(1e-2,0.223)(1e-1,0.077)
};
\addplot[cVote, opacity=0.20, forget plot] fill between[of=vhi and vlo];
\addplot[cVote, thick] coordinates {
  (1e-7,1)(1e-6,1)(1e-5,1)(1e-4,0.917)(1e-3,0.780)(1e-2,0.223)(1e-1,0.077)
};
\addlegendentry{relation vote (ours)}
\addplot[cVote, thick, forget plot] coordinates {
  (1e-7,1)(1e-6,1)(1e-5,1)(1e-4,1)(1e-3,1)(1e-2,1)(1e-1,1)
};
\draw[black!35, dashed, very thin] (axis cs:7e-8,1.0) -- (axis cs:1.4e-1,1.0);
\end{axis}
\end{tikzpicture}

%% file: diagrams/nary_scaling.tex
% AUTO-GENERATED by experimentation/nary_scaling/make_nary.py --- do not edit.
% Build time vs total input triangles over a broad random N-ary sample
% (Thingi10K operands, mimalloc, -march=native, best-of-K). Each point is one
% N-ary union; colour is operand count N. Geometry alone does not order the
% points --- at fixed total they fan out by N.
% Sequential colormap in the paper palette (cFour->cOne->cThree->cTwo).
\pgfplotsset{colormap={tfseq}{
  rgb255=(142,107,190) rgb255=(76,120,168) rgb255=(84,162,75) rgb255=(245,133,24)
}}
\begin{tikzpicture}
\begin{axis}[
  width=0.86\columnwidth, height=6.0cm,  % leave room for the colorbar
  xmode=log, log basis x=10, ymode=log, log basis y=10,
  xlabel={total input triangles}, ylabel={build (ms)},
  colormap name=tfseq, colorbar,
  colorbar style={width=0.18cm},
  point meta min=1, point meta max=7,
  colorbar style={
    ytick={1,2,3,4,5,6,7}, yticklabels={2,4,8,16,32,64,128},
    ylabel={operands $N$}, ylabel style={font=\footnotesize},
    tick label style={font=\footnotesize},
  },
  grid=both, grid style={black!8},
  tick label style={font=\footnotesize}, label style={font=\footnotesize},
]
\addplot[scatter, only marks, scatter src=explicit,
         mark=*, mark size=1.3pt, draw opacity=0, fill opacity=0.7]
  table[x=tot, y=build, meta=ln] {diagrams/nary_scaling.dat};
\end{axis}
\end{tikzpicture}

%% file: diagrams/performance.tex
% Build vs extract on the Fibonacci bunny swarm (csg_amortize_bench,
% amortize_dense.csv), best-of-5, mimalloc, arm64. Log-log: N (operands)
% vs time (ms). Build is one arrangement over all N+1 operands; the four
% extract curves are fixed booleans read from it.
\begin{tikzpicture}
\begin{axis}[
  width=\columnwidth, height=6.1cm,
  xmode=log, log basis x=2,
  ymode=log, log basis y=10,
  xlabel={number of bunny operands $N$},
  ylabel={time (ms)},
  xmin=22, xmax=560, ymin=1.6, ymax=2600,
  xtick={25,50,100,200,400}, xticklabels={25,50,100,200,400},
  grid=both, grid style={black!8},
  tick label style={font=\footnotesize},
  label style={font=\footnotesize},
  legend cell align=left,
  legend columns=2,
  legend style={font=\footnotesize, draw=black!20,
                at={(0.5,1.22)}, anchor=south,
                /tikz/every even column/.append style={column sep=2.2em}},
]
\addplot[mark=*, color=tfink, very thick, mark size=1.7pt] coordinates {
  (25,78.2)(32,96.3)(40,113.5)(50,137.4)(64,173.2)(80,213.3)(100,259.1)
  (128,328.0)(160,415.2)(200,523.0)(256,687.5)(320,895.1)(400,1194.7)(500,1682.0)
};
\addlegendentry{build}
\addplot[mark=square*, color=cOne, thick, mark size=1.3pt] coordinates {
  (25,5.2)(32,6.4)(40,11.4)(50,14.4)(64,18.3)(80,21.8)(100,26.9)
  (128,34.4)(160,48.3)(200,57.6)(256,72.0)(320,89.9)(400,110.6)(500,132.1)
};
\addlegendentry{union all}
\addplot[mark=triangle*, color=cTwo, thick, mark size=1.6pt] coordinates {
  (25,4.1)(32,5.3)(40,7.3)(50,9.0)(64,11.3)(80,18.1)(100,23.0)
  (128,29.6)(160,41.3)(200,51.7)(256,64.4)(320,79.6)(400,101.4)(500,124.8)
};
\addlegendentry{sphere $-$ bunnies}
\addplot[mark=diamond*, color=cThree, thick, mark size=1.7pt] coordinates {
  (25,4.5)(32,5.7)(40,7.6)(50,9.2)(64,11.7)(80,18.5)(100,23.6)
  (128,29.0)(160,42.7)(200,52.9)(256,66.8)(320,83.9)(400,101.8)(500,131.9)
};
\addlegendentry{sphere $\cap$ bunnies}
\addplot[mark=o, color=cFour, thick, mark size=1.6pt] coordinates {
  (25,2.1)(32,2.5)(40,3.0)(50,3.8)(64,4.5)(80,5.3)(100,6.4)
  (128,8.0)(160,14.9)(200,18.4)(256,23.4)(320,28.9)(400,36.4)(500,45.9)
};
\addlegendentry{sphere $-$ one bunny}
\addplot[mark=pentagon*, color=cFive, thick, mark size=1.7pt] coordinates {
  (25,1.6)(32,2.2)(40,2.5)(50,3.1)(64,4.0)(80,4.8)(100,6.0)
  (128,7.5)(160,14.3)(200,18.1)(256,23.1)(320,29.3)(400,37.4)(500,45.3)
};
\addlegendentry{sphere only}
\end{axis}
% secondary top axis: total input triangles (aligned to the same box)
\begin{axis}[
  width=\columnwidth, height=6.1cm,
  xmode=log, log basis x=2, ymode=log, log basis y=10,
  xmin=22, xmax=560, ymin=1.6, ymax=2600,
  axis x line=top, axis y line=none, ytick=\empty,
  xtick={25,50,100,200,400},
  xticklabels={$2.8$M,$5.6$M,$11$M,$22$M,$45$M},
  xlabel={total input triangles}, xlabel near ticks,
  x axis line style={black!55},
  tick label style={font=\footnotesize}, label style={font=\footnotesize},
]
\end{axis}
\end{tikzpicture}

%% file: diagrams/stages.tex
% Stage breakdown of the arrangement build on the Fibonacci bunny swarm
% (csg_stages_bench, stages.csv), best-of-5. Each bar is one N, normalised
% to 100% of that N's build time; the absolute build total is printed above.
% Five phases map to the algorithm sections:
%   intersection  ibp        -- broadphase + pairwise face records (Sec 3.2/i)
%   arrangement   ig+fc      -- edge graph + per-face planar arrangement (3.2/ii)
%   graph         ag         -- implicit graph structures (3.3)
%   partition     desc+inc   -- domain partition vote (3.4)
%   classification vol+seed+prop -- inclusion bits / domain volumes (3.5)
\begin{tikzpicture}
\begin{axis}[
  ybar stacked,
  bar width=20pt,
  width=\columnwidth, height=6.3cm,
  ymin=0, ymax=110,
  ytick={0,20,40,60,80,100},
  enlarge x limits=0.12,
  symbolic x coords={25,50,100,200,400,500},
  xtick=data,
  xlabel={number of bunny operands $N$},
  ylabel={share of build time (\%)},
  tick label style={font=\footnotesize},
  label style={font=\footnotesize},
  legend cell align=left,
  legend columns=3,
  legend style={font=\footnotesize, draw=black!20,
                at={(0.5,1.03)}, anchor=south,
                /tikz/every even column/.append style={column sep=1.0em}},
]
\addplot[fill=tfink,  draw=white] coordinates {(25,59.41)(50,59.26)(100,59.39)(200,58.36)(400,54.90)(500,51.78)};
\addplot[fill=cOne,   draw=white] coordinates {(25,13.22)(50,12.09)(100,12.13)(200,12.55)(400,12.33)(500,11.21)};
\addplot[fill=cThree, draw=white] coordinates {(25,12.65)(50,13.98)(100,12.83)(200,11.37)(400,10.44)(500, 9.23)};
\addplot[fill=cFour,  draw=white] coordinates {(25, 7.92)(50, 6.84)(100, 6.36)(200, 5.37)(400, 4.95)(500, 4.29)};
\addplot[fill=cTwo,   draw=white] coordinates {(25, 6.80)(50, 7.84)(100, 9.29)(200,12.35)(400,17.37)(500,23.48)};
\legend{intersection, arrangement, graph, partition, classification}
% absolute build totals above each bar
\node[font=\scriptsize, anchor=south] at (axis cs:25,100)  {$78$\,ms};
\node[font=\scriptsize, anchor=south] at (axis cs:50,100)  {$139$\,ms};
\node[font=\scriptsize, anchor=south] at (axis cs:100,100) {$256$\,ms};
\node[font=\scriptsize, anchor=south] at (axis cs:200,100) {$524$\,ms};
\node[font=\scriptsize, anchor=south] at (axis cs:400,100) {$1.22$\,s};
\node[font=\scriptsize, anchor=south] at (axis cs:500,100) {$1.70$\,s};
\end{axis}
\end{tikzpicture}

%% file: diagrams/comparison_pairwise.tex
% AUTO-GENERATED by experimentation/csg_comparison/make_violin.py --- do not edit.
% Pairwise-boolean wall-clock distribution per library (best-of-K, mimalloc,
% Option-A timing, Thingi10K pairs at 100k-1M faces). Re-run the generator to
% refresh from the CSVs. y = time (ms, log); one violin per library.
\begin{tikzpicture}
\begin{axis}[
  width=\columnwidth, height=6.2cm,
  ymode=log, log basis y=10,
  ylabel={time per boolean (ms)},
  xmin=0.4, xmax=7.6,
  xtick={1,2,3,4,5,6,7},
  xticklabels={,,,,,,},
  ymin=0.8111, ymax=3.279e+05,
  grid=both, grid style={black!8},
  clip=false,
  tick label style={font=\footnotesize},
  label style={font=\footnotesize},
]
% --- trueform ---
\addplot[draw=tfink, fill=tfink, fill opacity=0.30, line width=0.5pt] coordinates {
  (1.0000,1.69621)(1.0000,1.94215)(1.0000,2.22376)(1.0000,2.5462)(1.0003,2.91539)(1.0021,3.33811)(1.0047,3.82213)(1.0062,4.37632)(1.0097,5.01088)(1.0185,5.73744)(1.0484,6.56935)(1.1041,7.52189)(1.1743,8.61254)(1.2359,9.86133)(1.2855,11.2912)(1.3348,12.9284)(1.4000,14.803)(1.3957,16.9493)(1.3325,19.4069)(1.2561,22.2209)(1.2000,25.4429)(1.1274,29.132)(1.0702,33.3561)(1.0510,38.1926)(1.0376,43.7304)(1.0252,50.0712)(1.0173,57.3313)(1.0118,65.6442)(1.0096,75.1624)(1.0057,86.0608)(1.0040,98.5393)(1.0027,112.827)(1.0004,129.187)(1.0001,147.919)(1.0011,169.366)(1.0016,193.924)(1.0004,222.042)(1.0000,254.238)(1.0000,291.101)(1.0000,333.31)(1.0000,381.639)(1.0000,436.976)(1.0000,500.336)(1.0000,572.883)(1.0000,655.95)(1.0000,751.06)(1.0000,859.962)(1.0000,984.654)(1.0000,1127.43)(1.0000,1290.9)(1.0000,1478.08)(1.0000,1692.39)(1.0000,1937.78)(1.0000,2218.76)(1.0000,2540.47)(1.0000,2908.83)(1.0000,3330.6)(1.0000,3813.53)(1.0000,4366.48)(1.0000,4999.61)(1.0000,5724.54)(1.0000,6554.58)(1.0000,7504.97)(1.0000,8593.17)(1.0000,9839.15)(1.0000,11265.8)(1.0000,12899.3)(1.0000,14769.7)(1.0000,16911.2)(1.0000,19363.3)(1.0000,22170.9)(1.0000,25385.6)(1.0000,29066.5)(1.0000,33281)(1.0000,38106.7)(1.0000,43632)(1.0000,49958.6)(1.0000,57202.4)(1.0000,65496.6)(1.0000,74993.4)
  (1.0000,74993.4)(1.0000,65496.6)(1.0000,57202.4)(1.0000,49958.6)(1.0000,43632)(1.0000,38106.7)(1.0000,33281)(1.0000,29066.5)(1.0000,25385.6)(1.0000,22170.9)(1.0000,19363.3)(1.0000,16911.2)(1.0000,14769.7)(1.0000,12899.3)(1.0000,11265.8)(1.0000,9839.15)(1.0000,8593.17)(1.0000,7504.97)(1.0000,6554.58)(1.0000,5724.54)(1.0000,4999.61)(1.0000,4366.48)(1.0000,3813.53)(1.0000,3330.6)(1.0000,2908.83)(1.0000,2540.47)(1.0000,2218.76)(1.0000,1937.78)(1.0000,1692.39)(1.0000,1478.08)(1.0000,1290.9)(1.0000,1127.43)(1.0000,984.654)(1.0000,859.962)(1.0000,751.06)(1.0000,655.95)(1.0000,572.883)(1.0000,500.336)(1.0000,436.976)(1.0000,381.639)(1.0000,333.31)(1.0000,291.101)(1.0000,254.238)(0.9996,222.042)(0.9984,193.924)(0.9989,169.366)(0.9999,147.919)(0.9996,129.187)(0.9973,112.827)(0.9960,98.5393)(0.9943,86.0608)(0.9904,75.1624)(0.9882,65.6442)(0.9827,57.3313)(0.9748,50.0712)(0.9624,43.7304)(0.9490,38.1926)(0.9298,33.3561)(0.8726,29.132)(0.8000,25.4429)(0.7439,22.2209)(0.6675,19.4069)(0.6043,16.9493)(0.6000,14.803)(0.6652,12.9284)(0.7145,11.2912)(0.7641,9.86133)(0.8257,8.61254)(0.8959,7.52189)(0.9516,6.56935)(0.9815,5.73744)(0.9903,5.01088)(0.9938,4.37632)(0.9953,3.82213)(0.9979,3.33811)(0.9997,2.91539)(1.0000,2.5462)(1.0000,2.22376)(1.0000,1.94215)(1.0000,1.69621)
} --cycle;
\addplot[tfink, line width=1.3pt] coordinates {(0.660,15.9607)(1.340,15.9607)};
\addplot[only marks, mark=*, mark size=1.5pt, tfink] coordinates {(1,15.9607)};
\node[font=\footnotesize, text=tfink, fill=white, fill opacity=0.85, text opacity=1, rounded corners=1pt, inner sep=1.3pt] at (axis cs:1,168828) {1.0$\times$};
\node[anchor=north, font=\footnotesize, text=tfink, yshift=-1pt] at (axis cs:1,0.811141) {trueform};
% --- MeshLib ---
\addplot[draw=cOne, fill=cOne, fill opacity=0.30, line width=0.5pt] coordinates {
  (2.0000,1.69621)(2.0000,1.94215)(2.0000,2.22376)(2.0000,2.5462)(2.0000,2.91539)(2.0000,3.33811)(2.0000,3.82213)(2.0000,4.37632)(2.0000,5.01088)(2.0000,5.73744)(2.0000,6.56935)(2.0000,7.52189)(2.0000,8.61254)(2.0000,9.86133)(2.0000,11.2912)(2.0000,12.9284)(2.0000,14.803)(2.0000,16.9493)(2.0001,19.4069)(2.0014,22.2209)(2.0099,25.4429)(2.0398,29.132)(2.0965,33.3561)(2.1584,38.1926)(2.2063,43.7304)(2.2413,50.0712)(2.2562,57.3313)(2.2456,65.6442)(2.2404,75.1624)(2.2658,86.0608)(2.2788,98.5393)(2.2557,112.827)(2.2181,129.187)(2.1712,147.919)(2.1229,169.366)(2.0851,193.924)(2.0603,222.042)(2.0489,254.238)(2.0415,291.101)(2.0289,333.31)(2.0201,381.639)(2.0181,436.976)(2.0163,500.336)(2.0109,572.883)(2.0064,655.95)(2.0046,751.06)(2.0053,859.962)(2.0079,984.654)(2.0072,1127.43)(2.0038,1290.9)(2.0011,1478.08)(2.0002,1692.39)(2.0000,1937.78)(2.0000,2218.76)(2.0000,2540.47)(2.0000,2908.83)(2.0000,3330.6)(2.0000,3813.53)(2.0000,4366.48)(2.0000,4999.61)(2.0000,5724.54)(2.0000,6554.58)(2.0000,7504.97)(2.0000,8593.17)(2.0000,9839.15)(2.0000,11265.8)(2.0000,12899.3)(2.0000,14769.7)(2.0000,16911.2)(2.0000,19363.3)(2.0000,22170.9)(2.0000,25385.6)(2.0000,29066.5)(2.0000,33281)(2.0000,38106.7)(2.0000,43632)(2.0000,49958.6)(2.0000,57202.4)(2.0000,65496.6)(2.0000,74993.4)
  (2.0000,74993.4)(2.0000,65496.6)(2.0000,57202.4)(2.0000,49958.6)(2.0000,43632)(2.0000,38106.7)(2.0000,33281)(2.0000,29066.5)(2.0000,25385.6)(2.0000,22170.9)(2.0000,19363.3)(2.0000,16911.2)(2.0000,14769.7)(2.0000,12899.3)(2.0000,11265.8)(2.0000,9839.15)(2.0000,8593.17)(2.0000,7504.97)(2.0000,6554.58)(2.0000,5724.54)(2.0000,4999.61)(2.0000,4366.48)(2.0000,3813.53)(2.0000,3330.6)(2.0000,2908.83)(2.0000,2540.47)(2.0000,2218.76)(2.0000,1937.78)(1.9998,1692.39)(1.9989,1478.08)(1.9962,1290.9)(1.9928,1127.43)(1.9921,984.654)(1.9947,859.962)(1.9954,751.06)(1.9936,655.95)(1.9891,572.883)(1.9837,500.336)(1.9819,436.976)(1.9799,381.639)(1.9711,333.31)(1.9585,291.101)(1.9511,254.238)(1.9397,222.042)(1.9149,193.924)(1.8771,169.366)(1.8288,147.919)(1.7819,129.187)(1.7443,112.827)(1.7212,98.5393)(1.7342,86.0608)(1.7596,75.1624)(1.7544,65.6442)(1.7438,57.3313)(1.7587,50.0712)(1.7937,43.7304)(1.8416,38.1926)(1.9035,33.3561)(1.9602,29.132)(1.9901,25.4429)(1.9986,22.2209)(1.9999,19.4069)(2.0000,16.9493)(2.0000,14.803)(2.0000,12.9284)(2.0000,11.2912)(2.0000,9.86133)(2.0000,8.61254)(2.0000,7.52189)(2.0000,6.56935)(2.0000,5.73744)(2.0000,5.01088)(2.0000,4.37632)(2.0000,3.82213)(2.0000,3.33811)(2.0000,2.91539)(2.0000,2.5462)(2.0000,2.22376)(2.0000,1.94215)(2.0000,1.69621)
} --cycle;
\addplot[cOne, line width=1.3pt] coordinates {(1.660,88.4009)(2.340,88.4009)};
\addplot[only marks, mark=*, mark size=1.5pt, cOne] coordinates {(2,88.4009)};
\node[font=\footnotesize, text=cOne, fill=white, fill opacity=0.85, text opacity=1, rounded corners=1pt, inner sep=1.3pt] at (axis cs:2,168828) {5.5$\times$};
\node[anchor=north, font=\footnotesize, text=cOne, yshift=-10pt] at (axis cs:2,0.811141) {MeshLib};
% --- Manifold ---
\addplot[draw=cThree, fill=cThree, fill opacity=0.30, line width=0.5pt] coordinates {
  (3.0000,1.69621)(3.0000,1.94215)(3.0000,2.22376)(3.0000,2.5462)(3.0000,2.91539)(3.0000,3.33811)(3.0000,3.82213)(3.0000,4.37632)(3.0000,5.01088)(3.0000,5.73744)(3.0000,6.56935)(3.0000,7.52189)(3.0000,8.61254)(3.0000,9.86133)(3.0000,11.2912)(3.0000,12.9284)(3.0000,14.803)(3.0000,16.9493)(3.0000,19.4069)(3.0000,22.2209)(3.0000,25.4429)(3.0000,29.132)(3.0000,33.3561)(3.0001,38.1926)(3.0021,43.7304)(3.0231,50.0712)(3.0999,57.3313)(3.2295,65.6442)(3.3246,75.1624)(3.3477,86.0608)(3.3257,98.5393)(3.2825,112.827)(3.3086,129.187)(3.3532,147.919)(3.2932,169.366)(3.2052,193.924)(3.1347,222.042)(3.0837,254.238)(3.0502,291.101)(3.0386,333.31)(3.0254,381.639)(3.0120,436.976)(3.0077,500.336)(3.0088,572.883)(3.0056,655.95)(3.0034,751.06)(3.0036,859.962)(3.0023,984.654)(3.0015,1127.43)(3.0015,1290.9)(3.0004,1478.08)(3.0000,1692.39)(3.0000,1937.78)(3.0000,2218.76)(3.0000,2540.47)(3.0000,2908.83)(3.0000,3330.6)(3.0000,3813.53)(3.0000,4366.48)(3.0000,4999.61)(3.0000,5724.54)(3.0000,6554.58)(3.0000,7504.97)(3.0000,8593.17)(3.0000,9839.15)(3.0000,11265.8)(3.0000,12899.3)(3.0000,14769.7)(3.0000,16911.2)(3.0000,19363.3)(3.0000,22170.9)(3.0000,25385.6)(3.0000,29066.5)(3.0000,33281)(3.0000,38106.7)(3.0000,43632)(3.0000,49958.6)(3.0000,57202.4)(3.0000,65496.6)(3.0000,74993.4)
  (3.0000,74993.4)(3.0000,65496.6)(3.0000,57202.4)(3.0000,49958.6)(3.0000,43632)(3.0000,38106.7)(3.0000,33281)(3.0000,29066.5)(3.0000,25385.6)(3.0000,22170.9)(3.0000,19363.3)(3.0000,16911.2)(3.0000,14769.7)(3.0000,12899.3)(3.0000,11265.8)(3.0000,9839.15)(3.0000,8593.17)(3.0000,7504.97)(3.0000,6554.58)(3.0000,5724.54)(3.0000,4999.61)(3.0000,4366.48)(3.0000,3813.53)(3.0000,3330.6)(3.0000,2908.83)(3.0000,2540.47)(3.0000,2218.76)(3.0000,1937.78)(3.0000,1692.39)(2.9996,1478.08)(2.9985,1290.9)(2.9985,1127.43)(2.9977,984.654)(2.9964,859.962)(2.9966,751.06)(2.9944,655.95)(2.9912,572.883)(2.9923,500.336)(2.9880,436.976)(2.9746,381.639)(2.9614,333.31)(2.9498,291.101)(2.9163,254.238)(2.8653,222.042)(2.7948,193.924)(2.7068,169.366)(2.6468,147.919)(2.6914,129.187)(2.7175,112.827)(2.6743,98.5393)(2.6523,86.0608)(2.6754,75.1624)(2.7705,65.6442)(2.9001,57.3313)(2.9769,50.0712)(2.9979,43.7304)(2.9999,38.1926)(3.0000,33.3561)(3.0000,29.132)(3.0000,25.4429)(3.0000,22.2209)(3.0000,19.4069)(3.0000,16.9493)(3.0000,14.803)(3.0000,12.9284)(3.0000,11.2912)(3.0000,9.86133)(3.0000,8.61254)(3.0000,7.52189)(3.0000,6.56935)(3.0000,5.73744)(3.0000,5.01088)(3.0000,4.37632)(3.0000,3.82213)(3.0000,3.33811)(3.0000,2.91539)(3.0000,2.5462)(3.0000,2.22376)(3.0000,1.94215)(3.0000,1.69621)
} --cycle;
\addplot[cThree, line width=1.3pt] coordinates {(2.660,121.811)(3.340,121.811)};
\addplot[only marks, mark=*, mark size=1.5pt, cThree] coordinates {(3,121.811)};
\node[font=\footnotesize, text=cThree, fill=white, fill opacity=0.85, text opacity=1, rounded corners=1pt, inner sep=1.3pt] at (axis cs:3,168828) {7.6$\times$};
\node[anchor=north, font=\footnotesize, text=cThree, yshift=-1pt] at (axis cs:3,0.811141) {Manifold};
% --- EMBER ---
\addplot[draw=cSix, fill=cSix, fill opacity=0.30, line width=0.5pt] coordinates {
  (4.0000,1.69621)(4.0000,1.94215)(4.0000,2.22376)(4.0000,2.5462)(4.0000,2.91539)(4.0000,3.33811)(4.0000,3.82213)(4.0000,4.37632)(4.0000,5.01088)(4.0000,5.73744)(4.0000,6.56935)(4.0000,7.52189)(4.0000,8.61254)(4.0000,9.86133)(4.0000,11.2912)(4.0001,12.9284)(4.0003,14.803)(4.0009,16.9493)(4.0025,19.4069)(4.0066,22.2209)(4.0159,25.4429)(4.0326,29.132)(4.0540,33.3561)(4.0741,38.1926)(4.0898,43.7304)(4.1025,50.0712)(4.1141,57.3313)(4.1264,65.6442)(4.1440,75.1624)(4.1692,86.0608)(4.1938,98.5393)(4.2065,112.827)(4.2064,129.187)(4.2000,147.919)(4.1873,169.366)(4.1646,193.924)(4.1354,222.042)(4.1108,254.238)(4.0965,291.101)(4.0896,333.31)(4.0853,381.639)(4.0798,436.976)(4.0715,500.336)(4.0616,572.883)(4.0523,655.95)(4.0446,751.06)(4.0390,859.962)(4.0349,984.654)(4.0307,1127.43)(4.0253,1290.9)(4.0196,1478.08)(4.0155,1692.39)(4.0132,1937.78)(4.0119,2218.76)(4.0111,2540.47)(4.0106,2908.83)(4.0098,3330.6)(4.0084,3813.53)(4.0069,4366.48)(4.0054,4999.61)(4.0040,5724.54)(4.0029,6554.58)(4.0020,7504.97)(4.0012,8593.17)(4.0008,9839.15)(4.0008,11265.8)(4.0009,12899.3)(4.0007,14769.7)(4.0003,16911.2)(4.0001,19363.3)(4.0000,22170.9)(4.0000,25385.6)(4.0000,29066.5)(4.0000,33281)(4.0000,38106.7)(4.0000,43632)(4.0000,49958.6)(4.0000,57202.4)(4.0000,65496.6)(4.0000,74993.4)
  (4.0000,74993.4)(4.0000,65496.6)(4.0000,57202.4)(4.0000,49958.6)(4.0000,43632)(4.0000,38106.7)(4.0000,33281)(4.0000,29066.5)(4.0000,25385.6)(4.0000,22170.9)(3.9999,19363.3)(3.9997,16911.2)(3.9993,14769.7)(3.9991,12899.3)(3.9992,11265.8)(3.9992,9839.15)(3.9988,8593.17)(3.9980,7504.97)(3.9971,6554.58)(3.9960,5724.54)(3.9946,4999.61)(3.9931,4366.48)(3.9916,3813.53)(3.9902,3330.6)(3.9894,2908.83)(3.9889,2540.47)(3.9881,2218.76)(3.9868,1937.78)(3.9845,1692.39)(3.9804,1478.08)(3.9747,1290.9)(3.9693,1127.43)(3.9651,984.654)(3.9610,859.962)(3.9554,751.06)(3.9477,655.95)(3.9384,572.883)(3.9285,500.336)(3.9202,436.976)(3.9147,381.639)(3.9104,333.31)(3.9035,291.101)(3.8892,254.238)(3.8646,222.042)(3.8354,193.924)(3.8127,169.366)(3.8000,147.919)(3.7936,129.187)(3.7935,112.827)(3.8062,98.5393)(3.8308,86.0608)(3.8560,75.1624)(3.8736,65.6442)(3.8859,57.3313)(3.8975,50.0712)(3.9102,43.7304)(3.9259,38.1926)(3.9460,33.3561)(3.9674,29.132)(3.9841,25.4429)(3.9934,22.2209)(3.9975,19.4069)(3.9991,16.9493)(3.9997,14.803)(3.9999,12.9284)(4.0000,11.2912)(4.0000,9.86133)(4.0000,8.61254)(4.0000,7.52189)(4.0000,6.56935)(4.0000,5.73744)(4.0000,5.01088)(4.0000,4.37632)(4.0000,3.82213)(4.0000,3.33811)(4.0000,2.91539)(4.0000,2.5462)(4.0000,2.22376)(4.0000,1.94215)(4.0000,1.69621)
} --cycle;
\addplot[cSix, line width=1.3pt] coordinates {(3.660,165.906)(4.340,165.906)};
\addplot[only marks, mark=*, mark size=1.5pt, cSix] coordinates {(4,165.906)};
\node[font=\footnotesize, text=cSix, fill=white, fill opacity=0.85, text opacity=1, rounded corners=1pt, inner sep=1.3pt] at (axis cs:4,168828) {10.4$\times$};
\node[anchor=north, font=\footnotesize, text=cSix, yshift=-10pt] at (axis cs:4,0.811141) {EMBER};
% --- Cherchi ---
\addplot[draw=cFour, fill=cFour, fill opacity=0.30, line width=0.5pt] coordinates {
  (5.0000,1.69621)(5.0000,1.94215)(5.0000,2.22376)(5.0000,2.5462)(5.0000,2.91539)(5.0000,3.33811)(5.0000,3.82213)(5.0000,4.37632)(5.0000,5.01088)(5.0000,5.73744)(5.0000,6.56935)(5.0000,7.52189)(5.0000,8.61254)(5.0000,9.86133)(5.0000,11.2912)(5.0000,12.9284)(5.0000,14.803)(5.0000,16.9493)(5.0000,19.4069)(5.0000,22.2209)(5.0000,25.4429)(5.0000,29.132)(5.0000,33.3561)(5.0000,38.1926)(5.0000,43.7304)(5.0000,50.0712)(5.0000,57.3313)(5.0001,65.6442)(5.0006,75.1624)(5.0025,86.0608)(5.0088,98.5393)(5.0285,112.827)(5.0660,129.187)(5.1049,147.919)(5.1333,169.366)(5.1527,193.924)(5.1670,222.042)(5.1789,254.238)(5.1959,291.101)(5.2345,333.31)(5.2847,381.639)(5.3039,436.976)(5.2709,500.336)(5.2089,572.883)(5.1517,655.95)(5.1117,751.06)(5.0859,859.962)(5.0721,984.654)(5.0615,1127.43)(5.0508,1290.9)(5.0419,1478.08)(5.0325,1692.39)(5.0255,1937.78)(5.0258,2218.76)(5.0270,2540.47)(5.0237,2908.83)(5.0192,3330.6)(5.0177,3813.53)(5.0155,4366.48)(5.0100,4999.61)(5.0063,5724.54)(5.0062,6554.58)(5.0075,7504.97)(5.0071,8593.17)(5.0060,9839.15)(5.0051,11265.8)(5.0034,12899.3)(5.0021,14769.7)(5.0021,16911.2)(5.0030,19363.3)(5.0035,22170.9)(5.0028,25385.6)(5.0019,29066.5)(5.0015,33281)(5.0012,38106.7)(5.0005,43632)(5.0001,49958.6)(5.0000,57202.4)(5.0000,65496.6)(5.0000,74993.4)
  (5.0000,74993.4)(5.0000,65496.6)(5.0000,57202.4)(4.9999,49958.6)(4.9995,43632)(4.9988,38106.7)(4.9985,33281)(4.9981,29066.5)(4.9972,25385.6)(4.9965,22170.9)(4.9970,19363.3)(4.9979,16911.2)(4.9979,14769.7)(4.9966,12899.3)(4.9949,11265.8)(4.9940,9839.15)(4.9929,8593.17)(4.9925,7504.97)(4.9938,6554.58)(4.9937,5724.54)(4.9900,4999.61)(4.9845,4366.48)(4.9823,3813.53)(4.9808,3330.6)(4.9763,2908.83)(4.9730,2540.47)(4.9742,2218.76)(4.9745,1937.78)(4.9675,1692.39)(4.9581,1478.08)(4.9492,1290.9)(4.9385,1127.43)(4.9279,984.654)(4.9141,859.962)(4.8883,751.06)(4.8483,655.95)(4.7911,572.883)(4.7291,500.336)(4.6961,436.976)(4.7153,381.639)(4.7655,333.31)(4.8041,291.101)(4.8211,254.238)(4.8330,222.042)(4.8473,193.924)(4.8667,169.366)(4.8951,147.919)(4.9340,129.187)(4.9715,112.827)(4.9912,98.5393)(4.9975,86.0608)(4.9994,75.1624)(4.9999,65.6442)(5.0000,57.3313)(5.0000,50.0712)(5.0000,43.7304)(5.0000,38.1926)(5.0000,33.3561)(5.0000,29.132)(5.0000,25.4429)(5.0000,22.2209)(5.0000,19.4069)(5.0000,16.9493)(5.0000,14.803)(5.0000,12.9284)(5.0000,11.2912)(5.0000,9.86133)(5.0000,8.61254)(5.0000,7.52189)(5.0000,6.56935)(5.0000,5.73744)(5.0000,5.01088)(5.0000,4.37632)(5.0000,3.82213)(5.0000,3.33811)(5.0000,2.91539)(5.0000,2.5462)(5.0000,2.22376)(5.0000,1.94215)(5.0000,1.69621)
} --cycle;
\addplot[cFour, line width=1.3pt] coordinates {(4.660,458.666)(5.340,458.666)};
\addplot[only marks, mark=*, mark size=1.5pt, cFour] coordinates {(5,458.666)};
\node[font=\footnotesize, text=cFour, fill=white, fill opacity=0.85, text opacity=1, rounded corners=1pt, inner sep=1.3pt] at (axis cs:5,168828) {28.7$\times$};
\node[anchor=north, font=\footnotesize, text=cFour, yshift=-1pt] at (axis cs:5,0.811141) {Cherchi};
% --- CGAL ---
\addplot[draw=cTwo, fill=cTwo, fill opacity=0.30, line width=0.5pt] coordinates {
  (6.0000,1.69621)(6.0000,1.94215)(6.0000,2.22376)(6.0000,2.5462)(6.0000,2.91539)(6.0000,3.33811)(6.0000,3.82213)(6.0000,4.37632)(6.0000,5.01088)(6.0000,5.73744)(6.0000,6.56935)(6.0000,7.52189)(6.0000,8.61254)(6.0000,9.86133)(6.0000,11.2912)(6.0000,12.9284)(6.0000,14.803)(6.0000,16.9493)(6.0000,19.4069)(6.0000,22.2209)(6.0000,25.4429)(6.0000,29.132)(6.0000,33.3561)(6.0000,38.1926)(6.0000,43.7304)(6.0000,50.0712)(6.0000,57.3313)(6.0000,65.6442)(6.0000,75.1624)(6.0001,86.0608)(6.0012,98.5393)(6.0076,112.827)(6.0245,129.187)(6.0485,147.919)(6.0760,169.366)(6.1096,193.924)(6.1453,222.042)(6.1715,254.238)(6.1902,291.101)(6.2177,333.31)(6.2459,381.639)(6.2537,436.976)(6.2459,500.336)(6.2379,572.883)(6.2325,655.95)(6.2129,751.06)(6.1747,859.962)(6.1403,984.654)(6.1131,1127.43)(6.0855,1290.9)(6.0635,1478.08)(6.0482,1692.39)(6.0360,1937.78)(6.0260,2218.76)(6.0172,2540.47)(6.0101,2908.83)(6.0068,3330.6)(6.0064,3813.53)(6.0071,4366.48)(6.0071,4999.61)(6.0050,5724.54)(6.0025,6554.58)(6.0010,7504.97)(6.0007,8593.17)(6.0011,9839.15)(6.0010,11265.8)(6.0004,12899.3)(6.0001,14769.7)(6.0000,16911.2)(6.0000,19363.3)(6.0000,22170.9)(6.0000,25385.6)(6.0000,29066.5)(6.0000,33281)(6.0000,38106.7)(6.0000,43632)(6.0000,49958.6)(6.0000,57202.4)(6.0000,65496.6)(6.0000,74993.4)
  (6.0000,74993.4)(6.0000,65496.6)(6.0000,57202.4)(6.0000,49958.6)(6.0000,43632)(6.0000,38106.7)(6.0000,33281)(6.0000,29066.5)(6.0000,25385.6)(6.0000,22170.9)(6.0000,19363.3)(6.0000,16911.2)(5.9999,14769.7)(5.9996,12899.3)(5.9990,11265.8)(5.9989,9839.15)(5.9993,8593.17)(5.9990,7504.97)(5.9975,6554.58)(5.9950,5724.54)(5.9929,4999.61)(5.9929,4366.48)(5.9936,3813.53)(5.9932,3330.6)(5.9899,2908.83)(5.9828,2540.47)(5.9740,2218.76)(5.9640,1937.78)(5.9518,1692.39)(5.9365,1478.08)(5.9145,1290.9)(5.8869,1127.43)(5.8597,984.654)(5.8253,859.962)(5.7871,751.06)(5.7675,655.95)(5.7621,572.883)(5.7541,500.336)(5.7463,436.976)(5.7541,381.639)(5.7823,333.31)(5.8098,291.101)(5.8285,254.238)(5.8547,222.042)(5.8904,193.924)(5.9240,169.366)(5.9515,147.919)(5.9755,129.187)(5.9924,112.827)(5.9988,98.5393)(5.9999,86.0608)(6.0000,75.1624)(6.0000,65.6442)(6.0000,57.3313)(6.0000,50.0712)(6.0000,43.7304)(6.0000,38.1926)(6.0000,33.3561)(6.0000,29.132)(6.0000,25.4429)(6.0000,22.2209)(6.0000,19.4069)(6.0000,16.9493)(6.0000,14.803)(6.0000,12.9284)(6.0000,11.2912)(6.0000,9.86133)(6.0000,8.61254)(6.0000,7.52189)(6.0000,6.56935)(6.0000,5.73744)(6.0000,5.01088)(6.0000,4.37632)(6.0000,3.82213)(6.0000,3.33811)(6.0000,2.91539)(6.0000,2.5462)(6.0000,2.22376)(6.0000,1.94215)(6.0000,1.69621)
} --cycle;
\addplot[cTwo, line width=1.3pt] coordinates {(5.660,512.425)(6.340,512.425)};
\addplot[only marks, mark=*, mark size=1.5pt, cTwo] coordinates {(6,512.425)};
\node[font=\footnotesize, text=cTwo, fill=white, fill opacity=0.85, text opacity=1, rounded corners=1pt, inner sep=1.3pt] at (axis cs:6,168828) {32.1$\times$};
\node[anchor=north, font=\footnotesize, text=cTwo, yshift=-10pt] at (axis cs:6,0.811141) {CGAL};
% --- Geogram ---
\addplot[draw=cFive, fill=cFive, fill opacity=0.30, line width=0.5pt] coordinates {
  (7.0000,1.69621)(7.0000,1.94215)(7.0000,2.22376)(7.0000,2.5462)(7.0000,2.91539)(7.0000,3.33811)(7.0000,3.82213)(7.0000,4.37632)(7.0000,5.01088)(7.0000,5.73744)(7.0000,6.56935)(7.0000,7.52189)(7.0000,8.61254)(7.0000,9.86133)(7.0000,11.2912)(7.0000,12.9284)(7.0000,14.803)(7.0000,16.9493)(7.0000,19.4069)(7.0000,22.2209)(7.0000,25.4429)(7.0000,29.132)(7.0000,33.3561)(7.0000,38.1926)(7.0000,43.7304)(7.0000,50.0712)(7.0000,57.3313)(7.0000,65.6442)(7.0000,75.1624)(7.0000,86.0608)(7.0000,98.5393)(7.0000,112.827)(7.0000,129.187)(7.0000,147.919)(7.0000,169.366)(7.0000,193.924)(7.0000,222.042)(7.0000,254.238)(7.0000,291.101)(7.0000,333.31)(7.0001,381.639)(7.0011,436.976)(7.0052,500.336)(7.0220,572.883)(7.0797,655.95)(7.1625,751.06)(7.2276,859.962)(7.2737,984.654)(7.2885,1127.43)(7.2640,1290.9)(7.2288,1478.08)(7.2738,1692.39)(7.3503,1937.78)(7.3089,2218.76)(7.2385,2540.47)(7.1791,2908.83)(7.1055,3330.6)(7.0564,3813.53)(7.0314,4366.48)(7.0191,4999.61)(7.0124,5724.54)(7.0082,6554.58)(7.0084,7504.97)(7.0100,8593.17)(7.0073,9839.15)(7.0035,11265.8)(7.0024,12899.3)(7.0022,14769.7)(7.0013,16911.2)(7.0009,19363.3)(7.0015,22170.9)(7.0008,25385.6)(7.0001,29066.5)(7.0000,33281)(7.0000,38106.7)(7.0000,43632)(7.0000,49958.6)(7.0000,57202.4)(7.0000,65496.6)(7.0000,74993.4)
  (7.0000,74993.4)(7.0000,65496.6)(7.0000,57202.4)(7.0000,49958.6)(7.0000,43632)(7.0000,38106.7)(7.0000,33281)(6.9999,29066.5)(6.9992,25385.6)(6.9985,22170.9)(6.9991,19363.3)(6.9987,16911.2)(6.9978,14769.7)(6.9976,12899.3)(6.9965,11265.8)(6.9927,9839.15)(6.9900,8593.17)(6.9916,7504.97)(6.9918,6554.58)(6.9876,5724.54)(6.9809,4999.61)(6.9686,4366.48)(6.9436,3813.53)(6.8945,3330.6)(6.8209,2908.83)(6.7615,2540.47)(6.6911,2218.76)(6.6497,1937.78)(6.7262,1692.39)(6.7712,1478.08)(6.7360,1290.9)(6.7115,1127.43)(6.7263,984.654)(6.7724,859.962)(6.8375,751.06)(6.9203,655.95)(6.9780,572.883)(6.9948,500.336)(6.9989,436.976)(6.9999,381.639)(7.0000,333.31)(7.0000,291.101)(7.0000,254.238)(7.0000,222.042)(7.0000,193.924)(7.0000,169.366)(7.0000,147.919)(7.0000,129.187)(7.0000,112.827)(7.0000,98.5393)(7.0000,86.0608)(7.0000,75.1624)(7.0000,65.6442)(7.0000,57.3313)(7.0000,50.0712)(7.0000,43.7304)(7.0000,38.1926)(7.0000,33.3561)(7.0000,29.132)(7.0000,25.4429)(7.0000,22.2209)(7.0000,19.4069)(7.0000,16.9493)(7.0000,14.803)(7.0000,12.9284)(7.0000,11.2912)(7.0000,9.86133)(7.0000,8.61254)(7.0000,7.52189)(7.0000,6.56935)(7.0000,5.73744)(7.0000,5.01088)(7.0000,4.37632)(7.0000,3.82213)(7.0000,3.33811)(7.0000,2.91539)(7.0000,2.5462)(7.0000,2.22376)(7.0000,1.94215)(7.0000,1.69621)
} --cycle;
\addplot[cFive, line width=1.3pt] coordinates {(6.660,1602.51)(7.340,1602.51)};
\addplot[only marks, mark=*, mark size=1.5pt, cFive] coordinates {(7,1602.51)};
\node[font=\footnotesize, text=cFive, fill=white, fill opacity=0.85, text opacity=1, rounded corners=1pt, inner sep=1.3pt] at (axis cs:7,168828) {100.4$\times$};
\node[anchor=north, font=\footnotesize, text=cFive, yshift=-1pt] at (axis cs:7,0.811141) {Geogram};
\end{axis}
\end{tikzpicture}

%% file: diagrams/comparison_pairwise_table.tex
% Pairwise-boolean median wall-clock, geometric-mean slowdown vs trueform, and
% validity (watertight solid; signed volume and area match) per library, over the 1000-pair corpus.
\begin{tabular*}{\columnwidth}{@{\extracolsep{\fill}}lrrr@{}}
\toprule
library & median (ms) & geomean $\times$ & valid \\
\midrule
trueform & $15.7$   & $1.0\times$   & $1000/1000$ \\
MeshLib  & $86.4$   & $5.5\times$   & $999/1000$  \\
Manifold & $118.0$  & $7.6\times$   & $1000/1000$ \\
EMBER    & $143.8$  & $10.4\times$  & $1000/1000$ \\
Cherchi  & $416.9$  & $28.7\times$  & $1000/1000$ \\
CGAL     & $493.6$  & $32.1\times$  & $999/1000$  \\
Geogram  & $1644.3$ & $100.4\times$ & $1000/1000$ \\
\bottomrule
\end{tabular*}

%% file: diagrams/comparison_nary.tex
% AUTO-GENERATED by experimentation/csg_comparison/make_nary_figure.py --- do not edit.
% Median N-ary union runtime relative to trueform vs operand count N (best of K,
% Option-A timing, 40-set corpus). CGAL excluded: pairwise accumulation too slow
% to complete the sweep. Re-run the generator to refresh from nary_results.csv.
\begin{tikzpicture}
\begin{axis}[
  width=\columnwidth, height=6.0cm,
  xmode=log, log basis x=2, ymode=log, log basis y=10,
  xlabel={operand count $N$},
  ylabel={median runtime relative to trueform ($\times$)},
  xtick={4,16,64}, xticklabels={4,16,64},
  xmin=3.5, xmax=72,
  ymin=0.6, ymax=972,
  grid=both, grid style={black!8},
  tick label style={font=\footnotesize},
  label style={font=\footnotesize},
  legend cell align=left,
  legend columns=3,
  legend style={font=\footnotesize, draw=black!20,
                at={(0.5,1.03)}, anchor=south,
                /tikz/every even column/.append style={column sep=1.2em}},
]
% trueform: the 1x baseline (dashed), first legend entry -> 6 entries, 3x2
\addplot[tfink, dashed, line width=1.0pt] coordinates {(4,1)(64,1)};
\addlegendentry{trueform}
\addplot[cThree, mark=*, mark size=1.5pt, line width=1.3pt] coordinates {(4,11.7)(16,14.3)(64,14.3)};
\addlegendentry{Manifold}
\addplot[cTwo, mark=*, mark size=1.5pt, line width=1.3pt] coordinates {(4,51)(16,69.2)(64,43.7)};
\addlegendentry{Cherchi}
\addplot[cOne, mark=*, mark size=1.5pt, line width=1.3pt] coordinates {(4,38.6)(16,45.9)(64,138)};
\addlegendentry{MeshLib}
\addplot[cFive, mark=*, mark size=1.5pt, line width=1.3pt] coordinates {(4,113)(16,213)(64,648)};
\addlegendentry{Geogram}
\addplot[cSix, dashed, mark=*, mark options={solid}, mark size=1.5pt, line width=1.3pt] coordinates {(4,10.1)(16,14.3)(64,13.9)};
\addlegendentry{EMBER}
\end{axis}
\end{tikzpicture}

%% file: diagrams/comparison_nary_table.tex
% AUTO-GENERATED by experimentation/csg_comparison/make_nary_figure.py --- do not edit.
\begin{tabular*}{\columnwidth}{@{\extracolsep{\fill}}l rrr c rrr@{}}
\toprule
& \multicolumn{3}{c}{median time (ms)} & & \multicolumn{3}{c}{valid sets} \\
\cmidrule(lr){2-4}\cmidrule(lr){6-8}
method & $4$ & $16$ & $64$ & & $4$ & $16$ & $64$ \\
\midrule
trueform & 6 & 25 & 103 & & $40/40$ & $40/40$ & $40/40$ \\
EMBER & 61 & 352 & 1{,}429 & & $40/40$ & $40/40$ & $40/40$ \\
Manifold & 71 & 354 & 1{,}467 & & $40/40$ & $40/40$ & $40/40$ \\
Cherchi & 310 & 1{,}711 & 4{,}495 & & $40/40$ & $40/40$ & $40/40$ \\
MeshLib & 234 & 1{,}134 & 14{,}158 & & $40/40$ & $38/40$ & $37/40$ \\
Geogram & 688 & 5{,}265 & 66{,}637 & & $40/40$ & $40/40$ & $40/40$ \\
\bottomrule
\end{tabular*}

%% file: diagrams/comparison_browser.tex
% AUTO-GENERATED by experimentation/csg_comparison/make_browser_violin.py --- do not edit.
% In-browser pairwise-boolean wall-clock per library (Chrome 148, M4 Max,
% same corpus as Section 4.3.1). y = time (ms, log); one violin per library.
\begin{tikzpicture}
\begin{axis}[
  width=\columnwidth, height=6.2cm,
  ymode=log, log basis y=10,
  ylabel={time per boolean (ms)},
  xmin=0.4, xmax=3.6,
  xtick={1,2,3},
  xticklabels={trueform,Manifold,three-bvh-csg},
  ymin=1.068, ymax=1.253e+05,
  grid=both, grid style={black!8},
  tick label style={font=\footnotesize},
  label style={font=\footnotesize},
  x tick label style={font=\footnotesize},
]
% --- trueform ---
\addplot[draw=tfink, fill=tfink, fill opacity=0.30, line width=0.5pt] coordinates {
  (1.0000,2.08117)(1.0000,2.35222)(1.0000,2.65857)(1.0001,3.00481)(1.0010,3.39615)(1.0039,3.83845)(1.0055,4.33836)(1.0061,4.90338)(1.0090,5.54198)(1.0096,6.26375)(1.0112,7.07953)(1.0250,8.00154)(1.0541,9.04364)(1.1062,10.2215)(1.1641,11.5527)(1.2054,13.0573)(1.2501,14.7578)(1.3092,16.6798)(1.3608,18.8522)(1.3852,21.3074)(1.3538,24.0824)(1.2799,27.2188)(1.2202,30.7638)(1.1823,34.7703)(1.1413,39.2987)(1.1005,44.4169)(1.0739,50.2016)(1.0668,56.7397)(1.0602,64.1294)(1.0464,72.4814)(1.0395,81.9212)(1.0352,92.5903)(1.0271,104.649)(1.0182,118.278)(1.0121,133.682)(1.0091,151.093)(1.0090,170.771)(1.0082,193.012)(1.0065,218.149)(1.0043,246.56)(1.0032,278.671)(1.0013,314.964)(1.0002,355.985)(1.0001,402.347)(1.0007,454.748)(1.0016,513.973)(1.0010,580.911)(1.0002,656.567)(1.0000,742.077)(1.0000,838.723)(1.0000,947.955)(1.0000,1071.41)(1.0000,1210.95)(1.0000,1368.66)(1.0000,1546.91)(1.0000,1748.38)(1.0000,1976.08)(1.0000,2233.44)(1.0000,2524.32)(1.0000,2853.08)(1.0000,3224.66)(1.0000,3644.63)(1.0000,4119.3)(1.0000,4655.78)(1.0000,5262.14)(1.0000,5947.46)(1.0000,6722.04)(1.0000,7597.51)(1.0000,8586.98)(1.0000,9705.33)(1.0000,10969.3)(1.0000,12397.9)(1.0000,14012.6)(1.0000,15837.6)(1.0000,17900.2)(1.0000,20231.5)(1.0000,22866.4)(1.0000,25844.4)(1.0000,29210.3)(1.0000,33014.6)
  (1.0000,33014.6)(1.0000,29210.3)(1.0000,25844.4)(1.0000,22866.4)(1.0000,20231.5)(1.0000,17900.2)(1.0000,15837.6)(1.0000,14012.6)(1.0000,12397.9)(1.0000,10969.3)(1.0000,9705.33)(1.0000,8586.98)(1.0000,7597.51)(1.0000,6722.04)(1.0000,5947.46)(1.0000,5262.14)(1.0000,4655.78)(1.0000,4119.3)(1.0000,3644.63)(1.0000,3224.66)(1.0000,2853.08)(1.0000,2524.32)(1.0000,2233.44)(1.0000,1976.08)(1.0000,1748.38)(1.0000,1546.91)(1.0000,1368.66)(1.0000,1210.95)(1.0000,1071.41)(1.0000,947.955)(1.0000,838.723)(1.0000,742.077)(0.9998,656.567)(0.9990,580.911)(0.9984,513.973)(0.9993,454.748)(0.9999,402.347)(0.9998,355.985)(0.9987,314.964)(0.9968,278.671)(0.9957,246.56)(0.9935,218.149)(0.9918,193.012)(0.9910,170.771)(0.9909,151.093)(0.9879,133.682)(0.9818,118.278)(0.9729,104.649)(0.9648,92.5903)(0.9605,81.9212)(0.9536,72.4814)(0.9398,64.1294)(0.9332,56.7397)(0.9261,50.2016)(0.8995,44.4169)(0.8587,39.2987)(0.8177,34.7703)(0.7798,30.7638)(0.7201,27.2188)(0.6462,24.0824)(0.6148,21.3074)(0.6392,18.8522)(0.6908,16.6798)(0.7499,14.7578)(0.7946,13.0573)(0.8359,11.5527)(0.8938,10.2215)(0.9459,9.04364)(0.9750,8.00154)(0.9888,7.07953)(0.9904,6.26375)(0.9910,5.54198)(0.9939,4.90338)(0.9945,4.33836)(0.9961,3.83845)(0.9990,3.39615)(0.9999,3.00481)(1.0000,2.65857)(1.0000,2.35222)(1.0000,2.08117)
} --cycle;
\addplot[tfink, line width=1.3pt] coordinates {(0.660,23.6449)(1.340,23.6449)};
\addplot[only marks, mark=*, mark size=1.5pt, tfink] coordinates {(1,23.6449)};
\node[font=\footnotesize, text=tfink] at (axis cs:1,49262.4) {1.0$\times$};
% --- Manifold ---
\addplot[draw=cThree, fill=cThree, fill opacity=0.30, line width=0.5pt] coordinates {
  (2.0000,2.08117)(2.0000,2.35222)(2.0000,2.65857)(2.0000,3.00481)(2.0000,3.39615)(2.0000,3.83845)(2.0000,4.33836)(2.0000,4.90338)(2.0000,5.54198)(2.0000,6.26375)(2.0000,7.07953)(2.0000,8.00154)(2.0000,9.04364)(2.0000,10.2215)(2.0000,11.5527)(2.0000,13.0573)(2.0000,14.7578)(2.0000,16.6798)(2.0000,18.8522)(2.0000,21.3074)(2.0000,24.0824)(2.0000,27.2188)(2.0000,30.7638)(2.0000,34.7703)(2.0000,39.2987)(2.0000,44.4169)(2.0000,50.2016)(2.0000,56.7397)(2.0000,64.1294)(2.0000,72.4814)(2.0000,81.9212)(2.0000,92.5903)(2.0020,104.649)(2.0242,118.278)(2.1072,133.682)(2.2177,151.093)(2.2698,170.771)(2.2919,193.012)(2.3140,218.149)(2.2965,246.56)(2.2323,278.671)(2.2359,314.964)(2.3626,355.985)(2.4000,402.347)(2.3172,454.748)(2.2290,513.973)(2.1477,580.911)(2.0950,656.567)(2.0486,742.077)(2.0151,838.723)(2.0024,947.955)(2.0002,1071.41)(2.0000,1210.95)(2.0000,1368.66)(2.0000,1546.91)(2.0000,1748.38)(2.0000,1976.08)(2.0000,2233.44)(2.0000,2524.32)(2.0000,2853.08)(2.0000,3224.66)(2.0000,3644.63)(2.0000,4119.3)(2.0000,4655.78)(2.0000,5262.14)(2.0000,5947.46)(2.0000,6722.04)(2.0000,7597.51)(2.0000,8586.98)(2.0000,9705.33)(2.0000,10969.3)(2.0000,12397.9)(2.0000,14012.6)(2.0000,15837.6)(2.0000,17900.2)(2.0000,20231.5)(2.0000,22866.4)(2.0000,25844.4)(2.0000,29210.3)(2.0000,33014.6)
  (2.0000,33014.6)(2.0000,29210.3)(2.0000,25844.4)(2.0000,22866.4)(2.0000,20231.5)(2.0000,17900.2)(2.0000,15837.6)(2.0000,14012.6)(2.0000,12397.9)(2.0000,10969.3)(2.0000,9705.33)(2.0000,8586.98)(2.0000,7597.51)(2.0000,6722.04)(2.0000,5947.46)(2.0000,5262.14)(2.0000,4655.78)(2.0000,4119.3)(2.0000,3644.63)(2.0000,3224.66)(2.0000,2853.08)(2.0000,2524.32)(2.0000,2233.44)(2.0000,1976.08)(2.0000,1748.38)(2.0000,1546.91)(2.0000,1368.66)(2.0000,1210.95)(1.9998,1071.41)(1.9976,947.955)(1.9849,838.723)(1.9514,742.077)(1.9050,656.567)(1.8523,580.911)(1.7710,513.973)(1.6828,454.748)(1.6000,402.347)(1.6374,355.985)(1.7641,314.964)(1.7677,278.671)(1.7035,246.56)(1.6860,218.149)(1.7081,193.012)(1.7302,170.771)(1.7823,151.093)(1.8928,133.682)(1.9758,118.278)(1.9980,104.649)(2.0000,92.5903)(2.0000,81.9212)(2.0000,72.4814)(2.0000,64.1294)(2.0000,56.7397)(2.0000,50.2016)(2.0000,44.4169)(2.0000,39.2987)(2.0000,34.7703)(2.0000,30.7638)(2.0000,27.2188)(2.0000,24.0824)(2.0000,21.3074)(2.0000,18.8522)(2.0000,16.6798)(2.0000,14.7578)(2.0000,13.0573)(2.0000,11.5527)(2.0000,10.2215)(2.0000,9.04364)(2.0000,8.00154)(2.0000,7.07953)(2.0000,6.26375)(2.0000,5.54198)(2.0000,4.90338)(2.0000,4.33836)(2.0000,3.83845)(2.0000,3.39615)(2.0000,3.00481)(2.0000,2.65857)(2.0000,2.35222)(2.0000,2.08117)
} --cycle;
\addplot[cThree, line width=1.3pt] coordinates {(1.660,297.023)(2.340,297.023)};
\addplot[only marks, mark=*, mark size=1.5pt, cThree] coordinates {(2,297.023)};
\node[font=\footnotesize, text=cThree] at (axis cs:2,49262.4) {12.6$\times$};
% --- three-bvh-csg ---
\addplot[draw=cFive, fill=cFive, fill opacity=0.30, line width=0.5pt] coordinates {
  (3.0000,2.08117)(3.0000,2.35222)(3.0000,2.65857)(3.0000,3.00481)(3.0000,3.39615)(3.0000,3.83845)(3.0000,4.33836)(3.0000,4.90338)(3.0000,5.54198)(3.0000,6.26375)(3.0000,7.07953)(3.0000,8.00154)(3.0000,9.04364)(3.0000,10.2215)(3.0000,11.5527)(3.0000,13.0573)(3.0000,14.7578)(3.0000,16.6798)(3.0000,18.8522)(3.0000,21.3074)(3.0000,24.0824)(3.0000,27.2188)(3.0000,30.7638)(3.0000,34.7703)(3.0000,39.2987)(3.0000,44.4169)(3.0000,50.2016)(3.0000,56.7397)(3.0000,64.1294)(3.0000,72.4814)(3.0000,81.9212)(3.0000,92.5903)(3.0000,104.649)(3.0000,118.278)(3.0000,133.682)(3.0000,151.093)(3.0000,170.771)(3.0000,193.012)(3.0004,218.149)(3.0033,246.56)(3.0162,278.671)(3.0479,314.964)(3.0975,355.985)(3.1529,402.347)(3.1933,454.748)(3.2014,513.973)(3.2024,580.911)(3.2256,656.567)(3.2497,742.077)(3.2557,838.723)(3.2680,947.955)(3.2973,1071.41)(3.3002,1210.95)(3.2577,1368.66)(3.1977,1546.91)(3.1500,1748.38)(3.1200,1976.08)(3.0933,2233.44)(3.0672,2524.32)(3.0485,2853.08)(3.0366,3224.66)(3.0286,3644.63)(3.0236,4119.3)(3.0191,4655.78)(3.0135,5262.14)(3.0109,5947.46)(3.0099,6722.04)(3.0074,7597.51)(3.0046,8586.98)(3.0028,9705.33)(3.0017,10969.3)(3.0008,12397.9)(3.0006,14012.6)(3.0012,15837.6)(3.0012,17900.2)(3.0005,20231.5)(3.0001,22866.4)(3.0000,25844.4)(3.0000,29210.3)(3.0000,33014.6)
  (3.0000,33014.6)(3.0000,29210.3)(3.0000,25844.4)(2.9999,22866.4)(2.9995,20231.5)(2.9988,17900.2)(2.9988,15837.6)(2.9994,14012.6)(2.9992,12397.9)(2.9983,10969.3)(2.9972,9705.33)(2.9954,8586.98)(2.9926,7597.51)(2.9901,6722.04)(2.9891,5947.46)(2.9865,5262.14)(2.9809,4655.78)(2.9764,4119.3)(2.9714,3644.63)(2.9634,3224.66)(2.9515,2853.08)(2.9328,2524.32)(2.9067,2233.44)(2.8800,1976.08)(2.8500,1748.38)(2.8023,1546.91)(2.7423,1368.66)(2.6998,1210.95)(2.7027,1071.41)(2.7320,947.955)(2.7443,838.723)(2.7503,742.077)(2.7744,656.567)(2.7976,580.911)(2.7986,513.973)(2.8067,454.748)(2.8471,402.347)(2.9025,355.985)(2.9521,314.964)(2.9838,278.671)(2.9967,246.56)(2.9996,218.149)(3.0000,193.012)(3.0000,170.771)(3.0000,151.093)(3.0000,133.682)(3.0000,118.278)(3.0000,104.649)(3.0000,92.5903)(3.0000,81.9212)(3.0000,72.4814)(3.0000,64.1294)(3.0000,56.7397)(3.0000,50.2016)(3.0000,44.4169)(3.0000,39.2987)(3.0000,34.7703)(3.0000,30.7638)(3.0000,27.2188)(3.0000,24.0824)(3.0000,21.3074)(3.0000,18.8522)(3.0000,16.6798)(3.0000,14.7578)(3.0000,13.0573)(3.0000,11.5527)(3.0000,10.2215)(3.0000,9.04364)(3.0000,8.00154)(3.0000,7.07953)(3.0000,6.26375)(3.0000,5.54198)(3.0000,4.90338)(3.0000,4.33836)(3.0000,3.83845)(3.0000,3.39615)(3.0000,3.00481)(3.0000,2.65857)(3.0000,2.35222)(3.0000,2.08117)
} --cycle;
\addplot[cFive, line width=1.3pt] coordinates {(2.660,973.637)(3.340,973.637)};
\addplot[only marks, mark=*, mark size=1.5pt, cFive] coordinates {(3,973.637)};
\node[font=\footnotesize, text=cFive] at (axis cs:3,49262.4) {41.2$\times$};
\end{axis}
\end{tikzpicture}